\documentclass[iop]{emulateapj}
\usepackage{natbib}
\usepackage{xcolor}
\usepackage{verbatim}
\usepackage{booktabs}
\usepackage{enumitem}
\usepackage{hyperref}
\usepackage[utf8]{inputenc}
\usepackage{muni}

\def\apjl{ApJL}               
\def\apjs{ApJS}               
\def\aap{A\&A}                

\shorttitle{Ram Pressure Stripping of the LMC}
\shortauthors{Salem et al.}

\begin{document}

\title{Ram Pressure Stripping of the Large Magellanic Cloud's Disk as a Probe of the Milky Way's Circumgalactic Medium}

\author{Munier Salem\altaffilmark{1}, Gurtina Besla\altaffilmark{2}, Greg Bryan\altaffilmark{1}, Mary Putman\altaffilmark{1}, Roeland P. van der Marel\altaffilmark{3}, Stephanie Tonnesen\altaffilmark{4}}  
\altaffiltext{1}{Department of Astronomy, Columbia University, 550 West 120th Street, New York, NY, 10027, USA}
\altaffiltext{2}{Department of Astronomy, University of Arizona, 933 North Cherry Avenue, Tucson, AZ, 85721, USA}
\altaffiltext{3}{Space Telescope Science Institute, 3700 San Martin Drive, Baltimore, MD, 21218, USA}
\altaffiltext{4}{Carnegie Observatories, 813 Santa Barbara St, Pasadena, CA 91101}
\email{gbesla@email.arizona.edu}

\begin{abstract}
Recent observations have constrained the orbit and structure of the Large Magellanic Cloud (LMC), implying a well-constrained pericentric passage about the Milky Way (MW) $\sim 50$ Myr ago. In this scenario, the LMC's gaseous disk has recently experienced stripping, suggesting the current extent of its HI disk directly probes the medium in which it is moving.  From the observed stellar and HI distributions of the system we find evidence of a truncated gas profile along the windward ``leading edge' of the LMC disk, despite a far more extended stellar component. We explore the implications of this ram pressure stripping signature, using both analytic prescriptions and full 3-dimensional hydrodynamic simulations of the LMC. Our simulations subject the system to a headwind whose velocity components correspond directly to the recent orbital history of the LMC. We vary the density of this headwind, using a variety of sampled parameters for a $\beta$-profile for a theoretical MW circumgalactic medium (CGM), comparing the resulting HI morphology directly to observations of the LMC HI and stellar components. This model can match the radial extent of the LMC's leading (windward) edge only in scenarios where the MW CGM density at pericentric passage is $n_p(R = 48.2 \pm 5\;{\rm kpc}) = 1.1^{+.44}_{-.45} \times 10^{-4} \; {\rm cm}^{-3}$. The implied pericentric density proves insensitive to both the broader CGM structure and temperature profile, thus providing a model-independent constraint on the local gas density. This result imposes an important constraint on the density profile of the MW's CGM, and thus the total baryon content of the MW. From our work, assuming a $\beta$-profile valid to $\sim r_{\rm vir}$, we infer a total diffuse CGM mass $M(300 \;\rm{kpc}) = 2.6 \pm 1.4 \times 10^{10} M_\odot$ or approximately 15\% of a $10^{12} M_\odot$ MW's baryonic mass budget.
\end{abstract}


\section{Introduction}
The Circumgalactic Medium (CGM) of the Milky Way (MW) represents a gaseous, multiphase plasma surrounding the Galactic plane and extending out to the virial radius. Though substantially more rarefied than the interstellar-medium (ISM), the density of plasma within this vast volume remains relatively uncertain. Studies in X-ray have detected CGM material $\sim10^6$ K, though the total mass, and whether or not it accounts for a majority of baryons in the MW remains in dispute \citep{Wang1993,Wang2005,Gupta2012,Mathur2012,Wang2012, Fang2013,Miller2013,Miller2015,Anderson2010}. Absorption studies in UV of $L \sim L_\star$ galaxies at low-redshift have likewise found evidence for a large fraction of galactic baryons residing in the CGM \citep{Tumlinson2011,Peeples2014}, a majority of which may be in the form of diffuse gas at surprisingly low temperatures $<10^5$ K \citep{Werk2014, Tumlinson2013}.

The CGM represents an important store of fuel for star formation (SF) as well as a tracer of inward flows from the pristine intergalactic medium (IGM) and outward flows of enriched material from the star-forming disk plane \citep[e.g.][]{Binney1977,vandeVoort2011,Cowie1995,Dave2012,Fraternali2013,Suresh2015}. Thus characterizing its composition aids in a comprehensive understanding of galaxy formation. For a recent review of the many components of the CGM, see \cite{Putman2012}.

Hydrodynamic probes of the MW CGM gas structure provide a powerful compliment to observational studies in absorption and emission. Modeling ionization states relies on assumptions regarding the gas' temperature and sources of photoionization; a single waveband study cannot probe the broad range of gas phases spanned by the CGM in all its forms. In contrast, studies that exploit ram pressure stripping (RPS) probe diffuse gas of all temperatures and ionization states, and the stripping dynamics are insensitive to the oncoming wind's temperature \cite[e.g.][find only a slight dependence on the Mach number of the flow]{Roediger2005}. Authors such as \cite{Grcevich2009} and \cite{Gatto2013} considered a host of dwarf spheroidals orbiting the MW, most devoid of HI, to place preliminary bounds on the CGM's gas profile. In this paper we consider instead the Large Magellanic Cloud (LMC), a relatively massive, late-type dwarf, with well-studied HI and stellar disks and well-constrained orbital properties.

The Magellanic Clouds (MCs) move through the MW at $\sim 50$ kpc from the Galactic Center at velocities $\sim 300$ km/s, with a Leading Arm (LA) and trailing Magellanic Stream (MS) of gas strewn across much of the southern sky. The interaction of this complex with the CGM likely alters the appearance and dynamics of these gaseous components. The LMC's proper motion vector implies that any CGM headwind would most directly impact the disk's northeastern edge, referred to hereafter as its ``leading edge''. The HI profile here truncates abruptly at $R \approx 6.2$ kpc from the kinematic center of the LMC (see Section \ref{sec:lmc-model}), in contrast to other quadrants of the LMC \citep[e.g.][]{Putman2003}. Despite the absence of gas, the stellar profile continues uninterrupted well beyond this radius \citep{vanderMarel2001}, which would rule out a tidal explanation for this HI truncation (as tides would truncate both gas and stars). The leading edge is characterized by multiple HI velocity components, the faster among them being possibly extra-planar \citep{Luks1992,Nidever2008,Nidever2010}, which point towards a strong interaction between the LMC's ISM and the ambient material it passes through. HI column and star formation are both more concentrated in the southeastern portion of the LMC disk; an entire ``supershell'' of denser gas and rapid star formation exists at the leading edge of the LMC disk, perhaps due to a bow-shock \citep{deBoer1998,Murali2000}, and regions devoid of gas exist despite the presence of young stars \citep{Indu2011}. Likewise the MS and LA are likely significantly altered by this ambient medium. The stream exhibits strong $\rm H\alpha$ emission \citep{Bland-Hawthorn2007,Weiner1996} and high velocity clouds (HVCs) kinematically associated with the whole complex consistently feature head-tail structures well aligned with the orbital movement of the Magellanic Complex at large \citep{Putman2011,For2014}.

Precise proper motion (PM) measurements of MC stars in the past decade have substantially improved our knowledge of the LMC's PM vector \citep{Kallivayalil2006,Piatek2008} and enhanced our understanding of the LMC's orientation and internal rotation and the dynamical relationship between the MCs \citep{Kallivayalil2006,vanderMarel2014}. These results imply that the LMC's kinetic energy is on par with its binding energy to the MW, leading to a picture in which the LMC passed through its last perigalacticon less than 100 Myr ago and is either on its first infall or else on a highly eccentric orbit whose previous perigalacticon occurred a substantial fraction of a Hubble time ago \citep[hereafter B07]{Besla2007}. The most recent epoch of PM measurements slightly reduced the implied LMC proper motion velocity, but still confirms this highly eccentric, possibly first infall scenario \citep[hereafter K13]{Kallivayalil2013}.

A detailed study of ram pressure stripping's effect on the LMC disk gas' extent was last undertaken by \cite{Mastropietro2005} who followed the LMC's orbital history within an N-body/SPH simulation of both the MW and LMC's dark and baryonic components. Their work suggested a MW gaseous halo density of $\approx 8 \times 10^{-4}$ cm$^{-3}$ at 50 kpc from the Galactic center. However, they used a substantially lower-energy LMC orbit, which allowed for multiple pericentric passages of comparable depth into the MW potential, and a much longer interaction time between the LMC gaseous disk and the MW CGM, during which time the headwind compressed and heated the LMC disk material, allowing for slow, continual stripping down to the present observed truncation radius --- a decidedly delocalized probe of this ambient material. The slower orbit also enhanced the role of tidal interactions between the MW and LMC. \cite{Mastropietro2009} used the updated orbital scenario of B07 but focused on the effects of compression-induced star formation and did not vary the halo density or consider how the leading edge infers a MW diffuse CGM density.
\cite{Besla2012} explored a toy simulation of the evolution of the gaseous Magellanic System wherein a ram pressure acceleration term is applied uniformaly to all gas particles. They find that the structure of the LMC's gas disk should be very sensitive to the properties of the ambient medium through which the galaxy is moving.   

The new, longer period, eccentric orbits of B07 and K13 no longer allow for slow, continual stripping mechanisms, and suggest a scenario in which the current HI truncation radius along the leading edge may be set by fast stripping, well described by the model of \cite{Gunn1972} (hereafter G72), having occurred during the LMC's recent pericentric passage of unprecedented speed and depth within the MW potential. This points to a picture where the LMC's current HI extent along the leading edge may provide a direct probe of the MW CGM density, likely localized to simply a measurement of this density at the LMC's recent perigalacticon at $R =  48.2 \pm 2.5$ kpc from the Galactic Center. Corroborating this notion and deriving a bound for this density is the primary focus of this paper.

This paper is organized as follows. In Section \ref{sec:model} we introduce an analytic model for the LMC internal structure and dynamics, the LMC's motion through the MW halo and the structure of the MW CGM. In Section \ref{sec:analytic} we use this setup along with an analytic description of RPS to construct a toy model through which we arrive at a localized density constraint at $r \sim 50$ kpc. In Section \ref{sec:sims} we then explore this process in greater detail via three-dimensional hydrodynamic simulations of the LMC passing through the MW halo. Section \ref{sec:error-anal} synthesizes the analytic and simulated results together along with an enumeration of relevant uncertainties to form an accurate diffuse gas density bound. Section \ref{sec:discussion} discusses the implications of our work for the total CGM mass budget and how our work informs the current understanding of the Magellanic Stream's (MS's) formation. Section \ref{sec:summary} summarizes our main findings.

\section{Theoretical Model for RPS of the LMC}
\label{sec:model}
In this section, we outline a model framework that will be applied consistently to all analytic and simulation work in this paper. Sections \ref{sec:lmc-model} -- \ref{sec:orbit-model} outline our prescriptions for the LMC internal structure and mass components, the MW diffuse gaseous halo, and the LMC's center of mass (COM) motion through the MW, respectively.

\begin{table}[tb]
\small
\begin{center}
	\begin{tabular}[b]{ll}
	\toprule		
	\multicolumn{2}{l}{\textbf{Static DM Potential} } 			\\
	\multicolumn{2}{l}{Spherical Profile \citep{Burkert1995}} 	\\
	$\rho_0	$	& $3.4 \times 10^{-24}$ g/cm$^3$		\\
	$r_0$		& $3$ kpc							\\
	$M(100 \; {\rm kpc})$		&	$5 \times 10^{10} M_\odot$ \\
	\midrule
	\multicolumn{2}{l}{\textbf{Static Stellar Potential} }		\\
	\multicolumn{2}{l}{Plummer-Kuzmin Disk} 				\\
	\multicolumn{2}{l}{\cite{Miyamoto1975}} 				\\
	$M_{\star}$	& $2.7 \times 10^{9} $ $M_\odot$		\\
	$a_{\star}$	& $1.7$ kpc						\\
	$b_{\star}$	& $.34$ kpc						\\
	\midrule
	\multicolumn{2}{l}{\textbf{Initial Gas Distribution}} 		\\
	\multicolumn{2}{l}{Exponential Disk} 					\\
	\multicolumn{2}{l}{\cite{Tonnesen2009}} 				\\
	$M_{\rm gas}$	& $5 \times 10^{8} \; M_\odot$			\\
	$a_{\rm gas}$	& $1.7$ kpc						\\
	$b_{\rm gas}$	& $.34$ kpc						\\
	$M_{\rm Tot}$	& $7.2 \times 10^{8} \; M_\odot$		\\
	\midrule
	\multicolumn{2}{l}{\textbf{Dynamical Mass}} 		\\
	$M_{\rm dyn}(8.7 \; {\rm kpc})$	& $1.4 \times 10^{10} M_\odot$ \\ 
	\bottomrule
	\end{tabular}
\end{center}
\small {\textbf{Table 1:}} Parameters for our LMC model, which consists of static DM and stellar potentials and a self-gravitating gas disk with an initially exponential profile. For this simple setup, all profiles share a common kinematic center and angular momentum axis, and the orbits are completely circularized. Here the dynamical mass includes all components (stars, gas and DM).
\label{tab:ICs}
\end{table}

%
\subsection{Model for the LMC}
\label{sec:lmc-model}
We begin by describing our model for the LMC's baryonic and DM distributions. Table 1 summarizes the important parameters.

For the stellar and gaseous disks of the LMC we follow the setup of \cite{Roediger2006} and \cite{Tonnesen2009}, hereafter T09. We assume the stellar and gaseous disks are circularized, share a common kinematic center, and are aligned in their rotation axes. In reality, the present day stellar distribution is intrinsically asymmetric and elliptical, and the position angle of progressively fainter isophotes of the stellar distribution rotate by $\sim 90^\circ$ \citep{vanderMarel2001}. However, these simplifications should not substantially alter our RPS dynamics, which is mostly governed by the stellar and gaseous surface densities at $R \sim 6$ kpc and to a lesser extent the circular velocity of the HI, related to the total enclosed mass at these radii. \cite{Kim1998}, hereafter K98, studied the HI velocity field to find an upper limit on the dynamical mass within a radius of 4 kpc of $M(4 \; {\rm kpc}) = 3.5 \times 10^{9} M_\odot$. From their HI surface maps, they also inferred a total disk mass of $2.5 \times 10^{9} M_\odot$ and a gaseous component of $5 \times 10^{8} M_\odot$ within the same radius. We use these values to constrain the normalization of our profiles. For the stars we employ a static Plummer-Kuzmin disk \citep{Miyamoto1975},
\begin{equation}
\Phi_{\star}(R,z) = GM_\star \left[R^2 + \left(a_{\star} + \sqrt{z^2 + b_{\star}^2} \right)^{2} \right]^{-1/2}
\label{eq:stellar}
\end{equation}
with mass $M_\star= 2.7 \times 10^{9} M_\odot$, radial scale length $a_{\star} = 1.7 \; {\rm kpc}$ and vertical scale height $b_{\star} = .34 \; {\rm kpc}$. These scale radii are motivated by \cite{vanderMarel2001}. For the gas disk, we adopt the profile of T09, 
\begin{equation}
\rho_{\rm gas}(R,z) = \frac{M_{\rm gas}}{2 \pi a^2_{\rm gas} b_{\rm gas}} 0.5^2 {\rm sech}\left( \frac{R}{a_{\rm gas}} \right) {\rm sech}\left( \frac{z}{b_{\rm gas}} \right)
\label{eq:gas}
\end{equation}
where $a_{\rm gas}$ and $b_{\rm gas}$ are the radial and vertical scale heights of the disk and $M_{\rm gas}$ is a normalization related to the total mass of the disk by $M_{\rm T,gas} \approx 1.44 M_{\rm gas}$. We consider only profiles that roughly obey the constraint of K98: $M_{\rm gas}( 4 \; {\rm kpc}) = 5 \times 10^8 M_\odot$. For the majority of simulations, we then set the gaseous scale radii equal to that of the stellar distribution, and thus $a_{\rm gas } = 1.7 \; {\rm kpc}$, $b_{\rm gas} = .34 \; {\rm kpc}$ and $M_{\rm gas} = 5 \times 10^8 M_\odot$. This ``characteristic mass'' is identical to the mass within 4 kpc by coincidence; the total enclosed mass of this disk model to large radii $M_{T} \approx 1.44 M_{\rm gas} = 7.2 \times 10^{8} M_\odot$. Figure \ref{fig:obs-profile} compares this model of the LMC gaseous disk to the observed HI profile, as found from averages of radial profiles in four distinct quadrants of the system. The observed HI column was taken from the Parkes HI Survey~\citep[HIPASS, see][]{Putman2003}. This comparison shows our model provides an excellent match to the central HI column of the observed system, and matches the radial falloff in two quadrants of the system (northwest and southeast, ignoring the heavily star forming 30 Doradus region). In the southwest quadrant, leading into the Magellanic Bridge, the observed column predictably rises above the radial falloff. The worst agreement is beyond 4 kpc on the leading edge (northeast). This is itself evidence of a role for ram pressure stripping, especially in light of the uninterrupted presence of stars beyond the HI truncation in this quadrant. 

Equating our model's surface density to the observed HI surface density ignores other phases of the ISM. However, dense molecular clouds behave more akin to the stellar component in a ram pressure stripping scenario, and thus could be thought of as a small error in the stellar disk mass. And while we do not expect ionized HII to be a large contribution to the gas profile by mass, we do account for HII in our simulations, crudely, via a density cut at $\rho = .03$ cm$^{-3}$ \citep[see, for example,][]{Rahmati2013}. 

This model also produces a total gas mass and HI diameter in line with the tight relation found for a survey of late-type dwarfs by \cite{Swaters2002}. The faster falloff along the leading edge today is not in line with this relation, which further suggests it is not a long-lived remnant of a pre-MW LMC but rather induced by RPS, RP compression of contours or RP-induced star formation and feedback. Our model is on the lower mass end of all models that obey both the Swaters relation and the mass constraint of K98. Thus in section \ref{sec:gas-rich} we consider a more gas-rich initial disk profile as well.

\stdFullFig{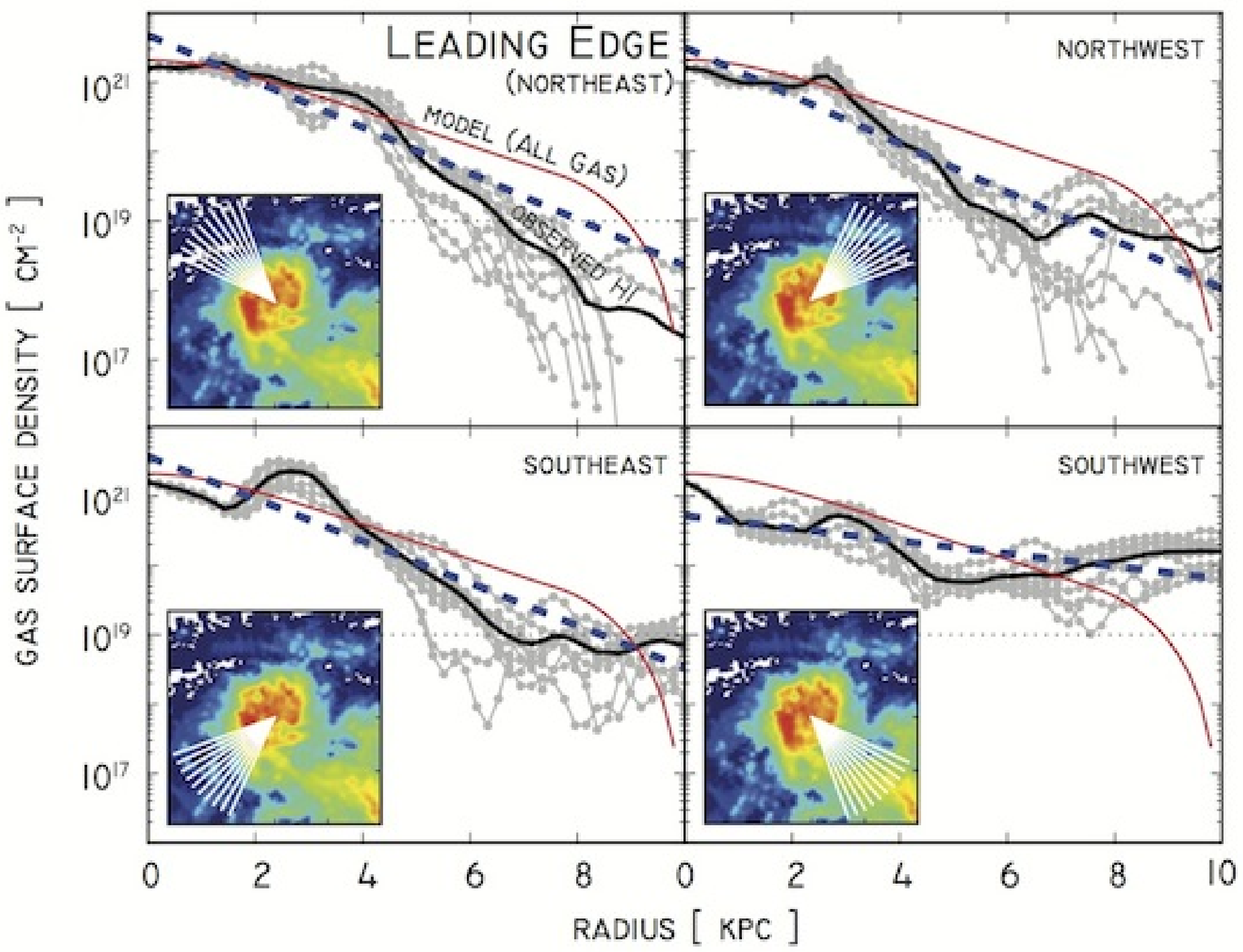}{Comparison of the observed HI  column density from the Parkes HI Survey~\citep[HIPASS, see][]{Putman2003} to our model's gaseous surface density (red line). Here we plot surface density versus radial distance from the LMC kinematic center in the plane of the LMC disk. The inset shows the observed HI surface density map from which our profiles are derived. To map the observed two-dimensional line-of-sight (LOS) column's position to a radius in the disk plane, we assumed the gas's height above the plane was zero and made use of the coordinate systems and transformations described in Section \ref{sec:coordinates}. Next we sampled the observed profile along a dozen rays, equally spaced in angle from $30^\circ$ to $60^\circ$ in each quadrant (grey lines with dots, white lines on the inset) and then averaged these radial samplings into an observed radial profile (thick black line). The central surface density of our model matches the observed profile well, and the radial falloff is in agreement in the northwest and southeast quadrants. The leading edge (northeast) shows a stronger radial falloff beyond 4 kpc, itself evidence of ram pressure stripping, as the stellar component does not show any such curtailment here. Figure \ref{fig:HI} shows the observed stellar profile.}{fig:obs-profile}{.7}

Our LMC disk sits within a large DM halo, which we treat as a spherical potential whose center is matched with those of the stellar and gaseous distributions. Inferring the circular velocity and dynamical mass of the LMC is not a straightforward process. Measurements of the rotation curve from different stellar and gaseous populations yield disparate results, convoluted further by imprecise knowledge of the system's orientation and ellipticity \citep[section 2.5 for a review]{vanderMarel2009}. \cite{Olsen2011} found the HI rotation curve peaks at $87 \pm 5$ km/s beyond $2.4 \pm .1$ kpc. More recently, \cite[][hereafter vdM14]{vanderMarel2014} took 3D stellar proper motions and consistently fit the bulk motion, orientation and rotation curve of the LMC disk, assuming a flat, circularized stellar disk, to find a flat rotation curve of $91.7 \pm 18.8$ km/s, corresponding to a dynamical mass of $M(8.7\;{\rm kpc}) = 1.7 \times 10^{10} M_\odot$. Our model produces a rotation curve of $80$ km/s, which is consistent with vdM14, though admittedly on the lower end of their error space. Our dynamical mass is thus $\approx 1.4 \times 10^{10} M_\odot$ (includes gas, stars and DM). We also perform a few simulations with peaks of 90 and 100 km/s, though we find no significant effect from this change. We emphasize that an accurate rotation speed is of secondary importance to our results, since where RPS truncates the disk gas does not depend strongly on this value (see Section \ref{sec:analytic}). For the DM, following T09, we employ the static, spherical model of  \cite{Burkert1995}, 
\begin{equation}
\rho_{\rm DM}(r) = \rho_0 r_0^3 \left[ ( r + r_0 ) ( r^2 + r_0^2 ) \right]^{-1}
\label{eq:DM}
\end{equation}
with characteristic density and radius $\rho_0 = 3.4 \times 10^{-24}$ g/cm$^3$ and $r = 3$ kpc, values chosen to produce a rotation curve peaked at 80 km/s and also to closely follow the equivalent NFW profile with concentration parameter $c = 10$. Figure \ref{fig:ICs} shows this rotation curve also broken down into mass components, with comparisons to the constraints of K98, vdM14, \cite{Olsen2011}.

In our simulations, as a practical consideration, we implemented a hydrostatic LMC gas halo (LGH) profile, whose pressure declines towards the box edge. This setup mitigated problems with our initial headwind setup. Our inclusion of this LMC gas halo is otherwise immaterial since the headwind washes this small gas component away well before the LMC's gaseous disk gets stripped. The low density, hydrostatic gas distribution obeys
\begin{eqnarray}
T_{\rm LGH} &=& \frac{GM_{\rm DM}(R)\mu m_p}{3k_br}		\\
n_{\rm LGH} &=& n_{LGH0} \left(\frac{T}{T_{LGH0}}\right)^{-1}\left(\frac{r}{r_{LGH0}}\right)^{-3}
\end{eqnarray}
where $r_{LGH0} = 2$ kpc and $n_{LGH0} = 1.8 \times 10^{-3}$ cm$^{-3}$ were chosen to provide ample pressure support near the LMC disk but a very low pressure, low density ambient medium near where the simulation wind first propagates into the domain. $T_{\rm LGH0} = T_{\rm LGH}(r_{\rm LGH0})$, and these expressions lead to a total LMC gaseous halo mass of roughly $5 \times 10^{6} M_{\odot}$, or roughly 1\% the LMC gaseous disk mass. An appreciably higher fraction of baryons could reside in this halo without affecting our results, since the material, spread over the halo's large volume, would still be very diffuse, and thus easy to unbind long before pericentric passage. 

\stdFig{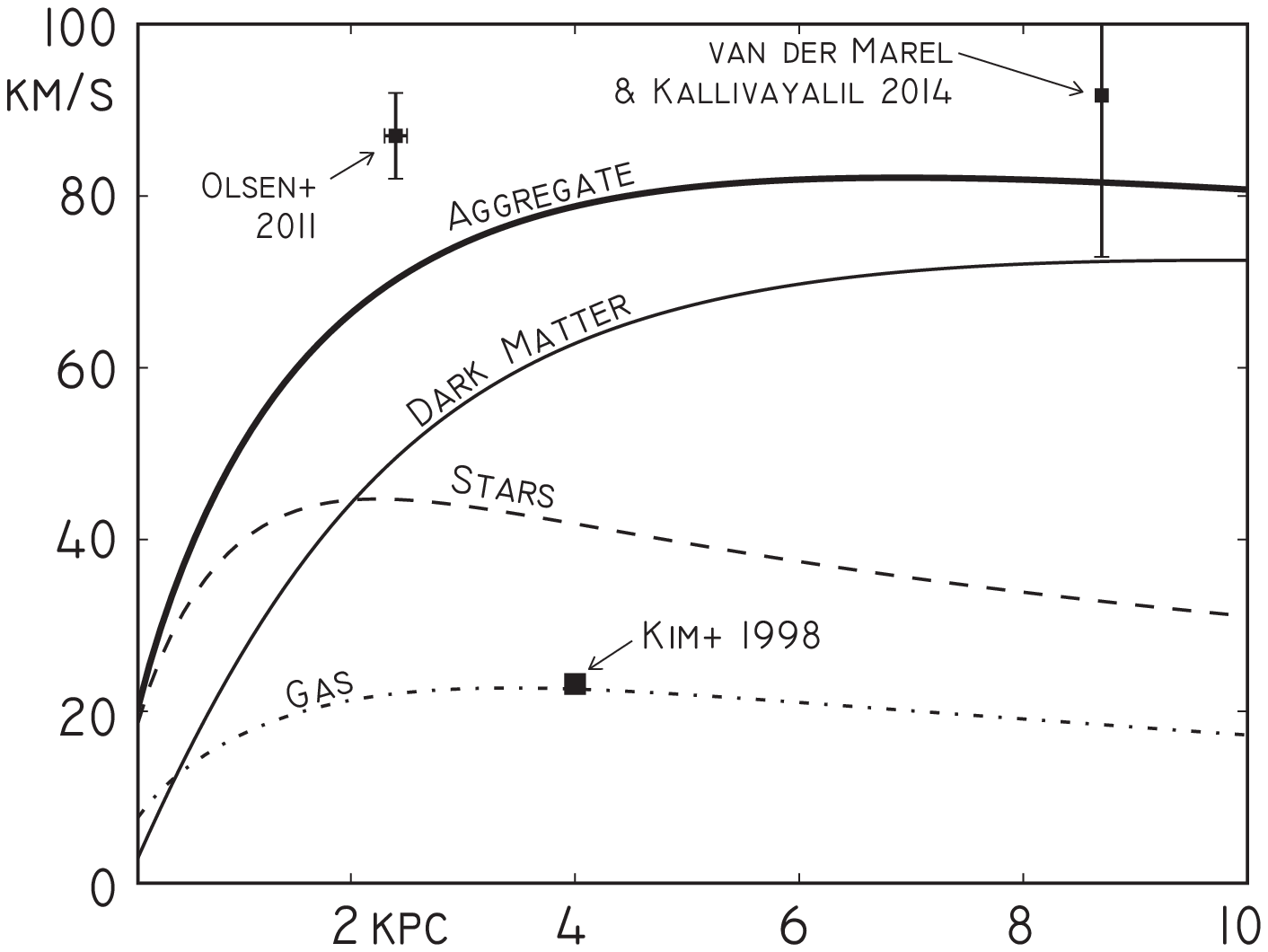}{Our model LMC's rotation curve, broken down by mass components. The net rotation curve peaks at 80 km/s, on the lower end of current observations. Observationally inferred circular velocities are over-plotted with errors: vdM14 estimated the amplitude of the circular velocity from stellar kinematics, whereas \cite{Olsen2011} did so via HI. \cite{Kim1998} also used the HI velocity field along with the luminosity of the baryonic components to find an HI mass within $r = 4$ kpc. We emphasize that correctly modeling the stellar and gaseous surface densities at the observed HI truncation radius of $\approx 6$ kpc is the chief concern for our RPS study of the LMC disk, whereas the rotation curve profile in interior regions is of limited importance. }{fig:ICs}{.55}

%
\subsection{Model for the MW Diffuse CGM}
\label{sec:halo-model}
For the extended, diffuse phases of the circumgalactic medium, we adopt a $\beta$-model in density~\citep{Makino1998}, motivated by observations of the hydrostatic gas filling large X-ray clusters,
\begin{equation}
n(r) = n_0\left[1+ \left( \frac{r}{r_c} \right)^2  \right]^{-3\beta/2} \;,
\label{eq:beta}
\end{equation}
where $n_0$ and $r_c$ describe a ``core'' density and radius within which the profile is roughly flat, and $\beta$ is a power law decay constant applicable at larger radii $r > r_c$.
 \cite{Miller2013}, hereafter MB13, fit a $\beta$-profile to the distribution of $\sim 10^{6}$ K diffuse halo gas to match OVI and OVII absorption line  strengths for 29 X-ray sources (primarily AGN). They found best fit parameters for the MW,
\begin{eqnarray}
n_0	&=&	0.46^{+0.74}_{-.35} \; {\rm cm^{-3}}	\nonumber \\
r_c 	&=&	0.35^{+0.29}_{-.27} \; {\rm kpc}		\nonumber \\
\beta	&=&	0.71^{+0.13}_{-.14} \nonumber
\end{eqnarray}
which suggest the MW core region ends at radii well below our region of interest ($r \sim 50$ kpc). Thus for our LMC this profile can be approximated as 
\begin{equation}
n(r) \approx \frac{n_0 r_c^{3\beta}}{r^{3\beta}}
\label{eq:beta-approx}
\end{equation}
to good precision --- i.e. our model is more or less a simple power law and thus a function of two important variables: an amplitude $n_0 r_c^{3\beta}$, and a falloff $\beta$. Extrapolated mass estimates for the CGM will depend most sensitively on $\beta$. Section \ref{sec:param-choices} details how we chose various parameters in our suite of simulations.

To completely specify the thermodynamic state of our CGM model we'll need not only this $\beta$-profile density and our universal ideal gas law equation of state, but also a temperature profile. The choice of temperature profile is of small significance, since simulations both with and without gas cooling have found the stripping dynamics strongly insensitive to the headwind's temperature and pressure \citep[e.g. T09,][]{Roediger2005}. Again appealing to cluster observations, we choose to model the MW halo temperature with \citep{Makino1998}
\begin{equation}
k_b T(r) = \gamma \frac{G\mu m_p M(r)}{3r}
\label{eq:halo-temp}
\end{equation}
where $\gamma \approx 1.5$ is found in numerical simulations. For the mass of the MW, we follow the classic NFW profile \citep{Navarro1997} with $M_{200} = 10^{12} M_\odot$ and concentration parameter $c = 12$. 

%
\subsection{Model for LMC Orbit}
\label{sec:orbit-model}

Our work relies on a model of the LMC's orbital history to produce the CGM headwind's time-evolving density and 3D velocity.  Such an orbit was generated via a backwards-orbit integration for the LMC/MW system using methods described in B07. These calculations treat the MW as an NFW profile and the LMC as a Plummer profile and ignore the LMC's major companion, the Small Magellanic Cloud (SMC). But such simplifications will not have a huge impact on the accuracy for the most critical past 100 Myr surrounding pericentric passage \citep{Kallivayalil2013,Besla2007}. 

The initial conditions for the LMC's motion are the latest proper motion measurements of \cite{Kallivayalil2013} (hereafter K13). For our fiducial orbital scenario, we adopt an orbit with the pericentric velocity 340 km/s at a distance of 48.1 kpc from the Galactic Center. This fiducial orbit was calculated assuming a MW with $M_{\rm vir} = 1\times10^{12} M_\odot$, $c_{\rm vir} = 9.86$ and $R_{\rm vir} = 261$ kpc with $M_{\rm MW, disk} = 6.5 \times 10^{10} M_\odot$, as in K13. Our fiducial model involves an LMC Plummer sphere with mass $1.8 \times 10^{11} M_\odot$ with dynamical friction calculated also as in K13. This dynamical friction implies that uncertainy in the MW/LMC mass ratio results in uncertainty in the LMC's orbital history, which we explore in Section \ref{sec:error-anal}.

As described in the introduction to this paper, the LMC's orbital period is an appreciable fraction of a Hubble time, implying either a first-infall scenario or an orbit in which the last pericentric passage occurred much farther from the Galactic Center. Thus its recent pericentric passage with a speed of $\approx 340$ km/s involved a ram pressure headwind of unprecedented strength for the LMC. A plot of this evolving ram pressure appears in Figure \ref{fig:wind}.

\subsection{Defining the Truncation Radius of the LMC's Leading Edge}
To compute the observed leading-edge truncation radius, we took the HIPASS HI column density map and re-projected the HI column into our LMC line-of-sight (LOS) point-of-view (POV) (see Sections \ref{sec:coordinates} and \ref{sec:sim-results}). We then extended a dozen rays radially from the kinematic center of the LMC out to the leading edge, which we define as the second quadrant of a cartesian frame centered on the LMC kinematic center where the $\hat{y}$-axis is tangent to the line of constant declination passing through the origin (see Section \ref{sec:coordinates}). Table 2 summarizes position and orientation values relevant to the present work. We then find the radial distance at which the column drops below $10^{19}$ cm$^{-2}$ for the final time (Section \ref{sec:sim-results} justifies this choice). Finally, we fit an ellipse to these values whose axis-ratio is constrained to correspond to a circle in the LMC disk plane, minimizing the sum of the square of the residuals (SSR). The best fit ellipse has a corresponding radius of $6.2$ kpc in the LMC disk plane. The uncertainty in this value greatly exceeds any number implied by the goodness of fit of our constrained ellipse or concerns over the LMC disk plane orientation. Rather, this uncertainty is set by the flocculent nature of the observed HI distribution, with super bubbles and large whorls induced by star formation, feedback and the ram pressure dynamics itself. By eye, we judge a conservative (large) standard deviation of this value to be $\sim .25$ kpc.

\section{Inferred MW CGM Density from Analytic Arguments}
\label{sec:analytic}

In this section we take the model outlined in Section \ref{sec:model} and use analytic methods to demonstrate how RPS considerations coupled with the LMC's observed leading edge truncation radius, $R_t$, can lead to tight constraints on the MW CGM gas density at $r \sim 50$ kpc. Section \ref{sec:error-anal} embarks on a systematic enumeration of our model uncertainties, which we ignore in the present section for the sake of clarity.


G72 explored how the ram pressure experienced by matter plunging into a denser cluster environment can alter the extent of a galaxy's gas disk by equating the ram pressure headwind experienced by the galaxy with the restoring pressure of the disk, the latter being proportional to both the gaseous and stellar surface densities,
\begin{equation}
\rho v^2 = 2 \pi G (\Sigma_\star(R) + \alpha \Sigma_{\rm g} ) \Sigma_{\rm g}(R) \; ,
\label{eq:GunnGott}
\end{equation}
where $\rho$ is the physical density of the ambient medium the galaxy passes through, in our case, the MW's diffuse, gaseous halo; $v$ is the speed of the galaxy's motion relative to this gas; $\Sigma_\star(R)$ is the stellar surface density at cylindrical radius $R$ from the galaxy's center; $\Sigma_{\rm g}(R)$ is the surface density of disk gas, and $\alpha \in [0,1]$ represents the degree to which the gas experiences its own self-gravity during stripping. For MW-sized systems falling into cluster environments -- the scenario explored in a majority of the RPS literature -- $\alpha = 0$ has been employed, since the gas fraction is quite low in these systems. However, for gas-rich late-type dwarves like the LMC, this restoring force cannot be ignored. We choose $\alpha = 1.0$, i.e. full self-gravity, a choice well-supported by analysis in Section \ref{sec:gas-rich}. 

According to this model, (hereafter the GG model), regions of the disk where the ram pressure exceeds the gravitational restoring force would experience a rapid loss of gas. For a simple exponential disk profile, this restoring pressure falls off monotonically, implying there exists a truncation radius beyond which gas is completely removed. This simple model thus provides a map between the wind pressure and a truncation radius. For real systems where there are considerations of the stripping timescale, nearby extraplanar material, and the effects of viewing angle, the truncation radius predicted by GG, $R_{\rm GG}$, may not match the observed truncation radius, $R_t$, precisely. Thus for our simulation work, we will consider an extended model
\begin{equation}
\log_{10}( n ) = \log_{10}\left[n_{\rm GG}(R_t)\right] + D
\label{eq:offset}
\end{equation}
where 
\begin{equation}
n_{\rm GG}(R_t) = \frac{2 \pi G}{\mu m_p} \frac{\Sigma_{\rm gas}(R_t)[\Sigma_\star(R_t) + \Sigma_{\rm gas}(R_t)]}{v^2_{\rm p}} \; .
\end{equation}
Here $n$ is the actual CGM halo density, whereas $n_{\rm GG}$ is the prediction of the GG model. $D$ is an offset term meant to capture any observed discrepancies between the two. In the present section, where we explore the model's capabilities using simple analytic calculations, we set $D=0$. In section \ref{sec:error-anal}, we use data from a large suite of 3D hydrodynamic simulations to better describe the offset term and its effect on the implied MW CGM density.

The LMC's velocity during its most recent pericentric passage is well constrained (K13), as is its gaseous and stellar surface density profiles. Within the GG model, this leaves only the ambient CGM's density, $\rho$, as a free parameter, which given the rapid stripping timescales of R05 and T09, we may interpret as the density at perigalacticon. Thus GG together with the LMC's observed HI and stellar extent imply a localized measurement of MW CGM density at LMC perigalacticon.

To explore this approach, we sampled the $\beta$-profile (Equation \ref{eq:beta}), selected a fiducial LMC orbital history (defined in Section \ref{sec:orbit-model}), used the implied halo density at perigalacticon and orbital speed to find the peak ram pressure, and then applied the GG model of Equation \ref{eq:GunnGott} to find a predicted truncation radius for the LMC disk's leading edge. We then explored which CGM models matched the observed HI truncation radius.

5,000 $\beta$-profiles were randomly selected, drawing the three parameters from independent, uniform distributions: the core density $n_0 \in [ 0 ,1 ]$ cm$^{-3}$, $r_c \in [0,1.2]$ kpc and $\beta \in [0,1.4]$. The choice of these bounds were guided by MB13 (see Figure \ref{fig:analytic-params}). We selected the fiducial orbit of K13, which at perigalacticon had $r_p = 48.1$ kpc and $v_p = 340.$ km/s. In section \ref{sec:error-anal}, we explore how the spread in allowed orbits affects our results. With this speed and distance from the Galactic Center, Equation \ref{eq:beta} provides a mapping to the density at perigalacticon, $n_p$, and peak ram pressure, $P_p$, for each halo model. We then apply GG via Equation \ref{eq:GunnGott}, solving numerically for the gaseous disk's truncation radius, $R_{\rm GG}$, assuming the LMC gas and stellar distribution follow the model outlined in Section \ref{sec:lmc-model}, via Equation1 and 2. Thus for each halo model, there exists an implied truncation radius, $R_{\rm GG}$, which can be compared to the observed value.

\begin{table*}[tbp]
\small
\renewcommand{\tabcolsep}{12 pt}
\begin{center}
\begin{tabular}{lllll}
\toprule
Quantity		&	Value		& Unit		&	Source				& Description			\\	
\midrule
$\alpha_0$  	&	$78.76$		& deg		&	vdM14, Tab. 1, col. 3		& CM position in celestial 	\\
$\delta_0$ 	&	$-69.19$		& deg		&	vdM14, Tab. 1, col. 3 	&  coords				\\
$D_0$ 		&	$50.1$		& kpc		&	\cite{Freedman2001}		&  					\\
			&				&			&						&					\\
$v_x$ 		&	$-453.8$		& km/s		&	K13 					& Orbital velocity in the	\\
$v_y$	 	&	$54.4$		& km/s		&	K13 					& LOS frame (Sun at rest) \\
$v_z$	 	&	$-262.2$		& km/s		&	K13 					& 					\\
			&				&			&						&					\\
$\Theta$	 	&	$139$		& deg		&	vdM02 				& Disk pos/inc angles and \\
$i$	 		&	$34.7$		& deg		&	vdM02 				& angular change rate	\\
$di/dt$ 		&	$0$			& deg/Gyr		&	vdM14 				&				 	\\
			&				&			&						&					\\
$\vsys$  		&	$262.20$ 		& km/s 		&	(calculated) 			& Recession velocity 	\\
$v_t$ 		&	$457.05$		& km/s		&						&  Transverse speed		\\
$\Theta_t$ 	&	$83.16$  		& deg		&						&  Position angle of $v_t$ 	\\
\bottomrule
\label{tab:position-params}
\end{tabular}
\end{center}
 \small {\textbf{Table 2:} Summary of LMC COM position and velocity as well as the disk plane's orientation and solid body motion. The top three quantities describe the LMC COM's position, with RA and DEC ($\alpha_0$ and $\beta_0$) and the distance from the Sun to the LMC, $D_0$. The next three quantities describe the COM's proper motion in the LOS frame (See Section \ref{sec:coordinates}). The third set describe the LMC disk plane's orientation on the sky via the position angle of the line of nodes $\Theta$, the plane's inclination angle with respect to the sky plane $i$, and the rate of change of this inclination angle $di/dt$ which alters the appearance of internal motion within the LMC (see Section \ref{sec:sim-results}). The final set of quantities uses Equation \ref{eq:celestial-velocity} to translate the COM into a form amicable to the velocity corrections applied in Section \ref{sec:sim-results}.  }
\end{table*}

\stdFullFig{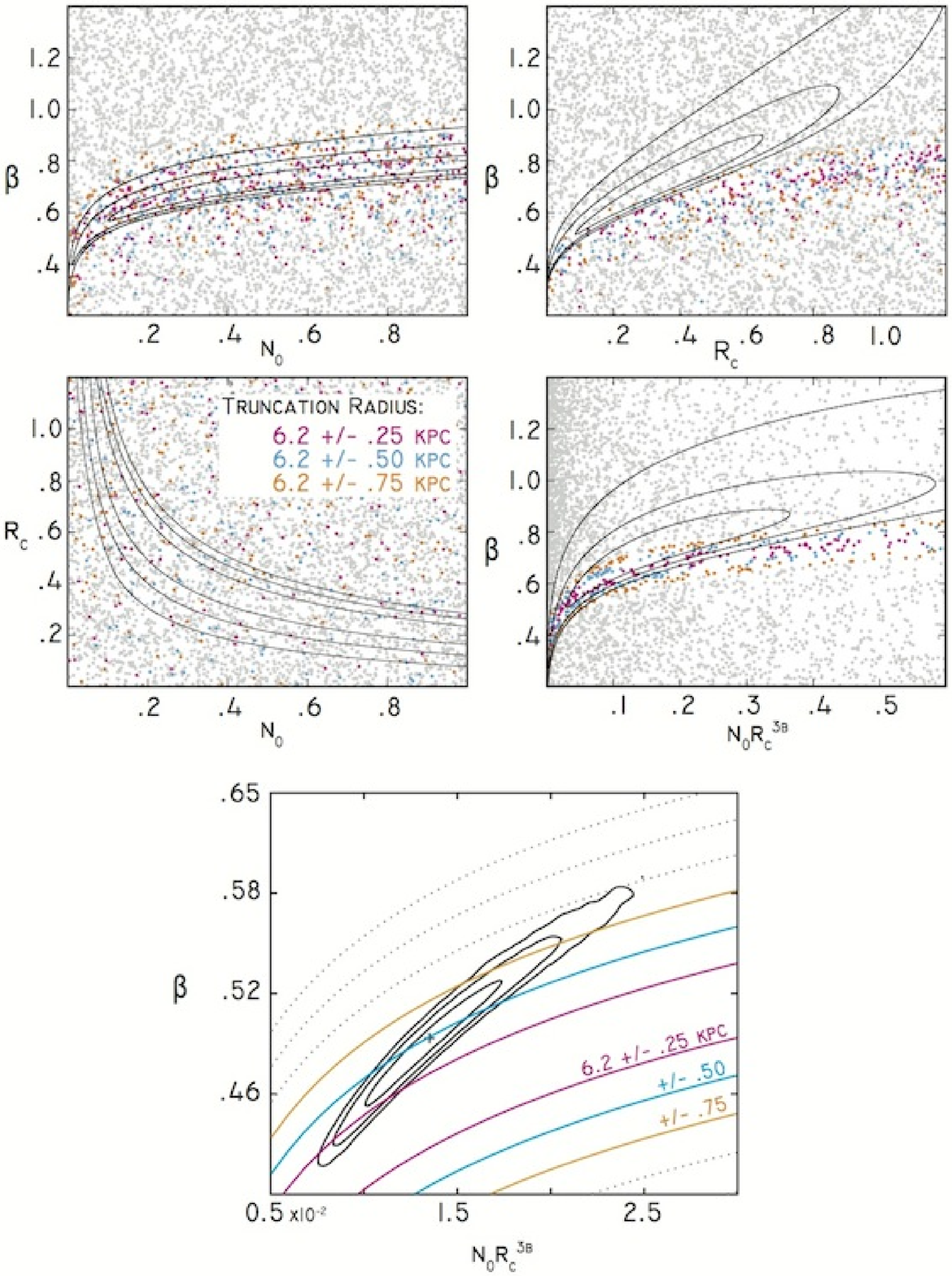}{Parameter space for a $\beta$-profile model of the MW's diffuse CGM. The core density, $n_0$, core radius, $r_c$ and exponential falloff, $\beta$, completely specify the profile, described by Equation \ref{eq:beta}. The \textbf{scatter plots} show 5,000 model choices, randomly selected from independent, uniform distributions of the three parameters. The GG model (Equation \ref{eq:GunnGott}) was then used along with our LMC model's stellar and gaseous distributions to determine a predicted RPS truncation radius of the LMC's leading edge. Models where this value is within 1-$\sigma$ of the observed radius ($6.2 \pm .25$ kpc) are shown in magenta, models within 2-$\sigma$ in cyan and 3-$\sigma$ in orange. These cuts separate cleanly only in the bottom-right panel, where we've plotted the ``parameters'' $\beta$ vs. $n_0 r_c^{3\beta}$, motivated by the simplified Equation \ref{eq:beta-approx}, valid for $r/r_c \gg 1$. Over-plotted on each panel, we show the 1-, 2- and 3-$\sigma$ confidence intervals for the constraints found by MB13 (their Figure 4), from X-Ray emission and quasar absorption lines. Towards the end of our investigations, MB15 published tighter constraints on this parameter space, shown in the \textbf{bottom contour plot}, with our analytic results again over-plotted. Figure \ref{fig:sim-results} reproduces this bottom panel, but with the results of our 3D hydrodynamic simulations and full error analysis folded in.}{fig:analytic-params}{.75}

Figure \ref{fig:analytic-params} shows the distribution of our 5,000 halo models in parameter space, both for the full $\beta$-profile, in terms of $n_0$, $r_c$ and $\beta$ and for the approximation of Equation \ref{eq:beta-approx}, valid at LMC distances and described by $\beta$ and $n_0 r_c^{3 \beta}$. We highlight models that successfully fall within 1-, 2- and 3-$\sigma$ of the observed truncation radius. The top four panels of Figure \ref{fig:analytic-params} show that this constraint places a strong upper limit on $\beta$, for reasonable core density choices (governed by $n_0$ and $r_c$), and for a fixed core profile, selects a narrow range of allowed $\beta$ exponents. This reflects how much leverage our measurement at $\sim 50$ kpc from the Galactic Center has, which strongly constrains any simple exponential halo profile's falloff. In contrast, the LMC's orbit does not afford us a strong probe of the innermost region of the CGM, as evidenced by the nearly uniform spread in $n_0$ - $r_c$ space. The $\beta$ - $n_0r_c^{3\beta}$ space of the approximation in Equation \ref{eq:beta-approx} provides the cleanest separation, and selects out the narrowest range of $\beta$ for a given core density. Also over-plotted in the top four panels of Figure \ref{fig:analytic-params} are the 1-, 2- and 3-$\sigma$ contours from MB13's work analyzing direct observations of $\sim10^{6}$ K gas. These results suggest our approach predicts consistently lower $\beta$-values (i.e. higher density at large radii) than their results, though their is good agreement for the smallest $r_c$ and $n_0$ core parameters shown. Towards the end of our investigations, \cite{Miller2015} (hereafter MB15) published tighter constraints on the allowed halo parameters, narrowing the results to $\beta$-profiles with small cores ($n_0 r_c^{3\beta} \sim .01$). The final panel of Figure \ref{fig:analytic-params} plots the agreement of our analytic RPS model with the observed truncation radius with MB15's updated error space superimposed. We again find rough agreement, though the 1-sigma contours just barely fail to overlap. Averaging the density at LMC pericenter from these samples, we find $n(48 \; {\rm kpc}) \approx 8 \times 10^{-5} \; {\rm cm}^{-3}$. Note our results in these plots represent only an analytic toy model. We reproduce this panel in Figure \ref{fig:sim-results} with the results of our hydrodynamical simulations and full error analysis.


This crude approach applies best to a smooth, azimuthally symmetric LMC disk of low gas-fraction, facing a strong, constant, head-on wind (aligned with the disk's angular momentum axis). In reality, a wind of variable strength arrives at a varying angle to an LMC disk riddled with large scale inhomogeneities and a significant velocity dispersion. On a localized level, the GG model itself comes into question, as the details of momentum transfer between the wind and gas involve shock fronts propagating in front of and behind the contact point between the wind and disk, making this as much a thermal process as a mechanical one. In addition, the flocculent, multiphase ISM is threaded with magnetic fields and infused with photons and energetic particles, complicating notions of its tensile strength and role as a monolithic impediment to the wind's motion. Since pericentric passage happens quickly, and has occurred less than 100 Myr ago, the validity of using the ram pressure at this epoch depends on the timescale of RPS. The GG argument's only (implicit) timescale is the crossing time of the wind through the galaxy, set by the windspeed and the disk's scale height, which may play no role in setting the actual stripping timescale.

Despite this, many numerical simulations have confirmed GG's applicability in a host of stripping scenarios. In particular,~\cite{Roediger2005}, hereafter R05, simulated ram pressure in two dimensions, and looked across many wind strengths and vertical disk scale heights, and found the timescale of fast stripping to occur $\sim 20 -200$ Myr, followed by longer, slower phases of stripping, the final of which was driven by viscous processes. The results were only strongly dependent on the wind properties (not the disk scale height), most notably its ram pressure, and in good agreement with GG. In their followup,~\cite{Roediger2006}, hereafter R06, performed fully three-dimensional runs, showing that disk orientation does not strongly alter the stripping extent except for extreme angles (beyond $\sim60^\circ$). Beyond this, T09 explored stripping for a host of ram pressures both with and without radiative cooling, and found agreement to within 10\% of GG. These explorations suggest the model of Equation \ref{eq:GunnGott} is a powerful, predictive tool relevant to our studies. In particular, the LMC's recent orbital history involves a disk galaxy having passed through a period of strong, peak ram pressure over the last $\sim100$ Myr, with an orientation angle between wind and disk vector well within the $\sim60^\circ$ limit of R06.

\section{Inferred  MW CGM Density from Simulations }
\label{sec:sims}

Motivated by the analytic proof-of-concept in Section \ref{sec:analytic}, we now present three-dimensional, global galaxy simulations of an LMC-analog isolated disk galaxy with stellar, gaseous and dark matter components, subjected to a time-varying headwind meant to mimic the diffuse CGM of the MW as the dwarf swoops through its recent pericentric passage. The chief goal of this work is to corroborate and extend the model above, though other useful information will arise from these investigations, including a look at how RPS influences the LMC's HI velocity structure, and an upper bound on how much HI RPS can redistribute from the LMC disk to the MS. Sections \ref{sec:implementation} through \ref{sec:param-choices} describe the numerical implementation, wind setup and halo model selection. Some of this material is presented mostly for the sake of reproducibility, and the casual reader may wish to skip details, particularly Section \ref{sec:coordinates}. Section \ref{sec:sim-results} provides the results of these simulations, building towards a thorough understanding between the CGM density and the LMC's leading edge truncation radius.

%
\subsection{Numerical Implementation}
\label{sec:implementation}

For the present work, we employed the Eulerian hydrodynamics code Enzo, described in~\cite{Bryan2014}. Our runs are fully three-dimensional and make use of Enzo's robust ZEUS hydro method~\citep{Stone1992}. Self gravity of the gas is also included.

Our LMC sits within a $60$ kpc-wide cubic box, with a $256^3$ base grid resolution. One of Enzo's main strengths is adaptive mesh refinement (AMR), which uniformly resolves the entire simulation region on a course grid but provides higher resolution, ``refined'' sub-grids as needed in regions where the dynamics grow complex. In our work, gas density a factor of four above the background density triggers higher mesh refinement, on up to three additional levels, refining by a factor of two each time. This leads to an effective resolution of 30 pc throughout the entire disk, or less than a tenth of the disk's vertical scale height.

We use a ``wind tunnel'' setup for our simulations, working in a frame where the LMC sits at rest within our simulation domain, and the MW Halo thus becomes a ``headwind'', produced from the boundary of the simulation box. This setup reduces computational costs in a few ways: first it guarantees the densest, most-refined disk material is at rest in our simulation frame, which prevents relative motion of this gas from limiting the simulation's time stepping; and second it alleviates the need to simulate the entire volume of the Milky Way halo through which the LMC moves. We place our galaxy $20$ kpc in each dimension from the corner of our box designated as the origin, where the halo wind first propagates inward. Our initial distribution of baryons follows the idealized LMC model outlined in Section \ref{sec:lmc-model} and is consistent with a three-dimensional, time-dependent extension of the analytic analysis of Section \ref{sec:analytic}. Our gaseous disk is evolved in isolation, before any ram pressure headwind arrives, for 1 Gyr, to allow transient oscillations in the structure to die away. We then launch the wind to mimic the LMC's orbit, beginning 1 Gyr before the present day, long before the ram pressure's impact grows significant.

Our simulations include self-gravitating gas, representing the atomic and ionized components of the LMC's ISM and the MW gaseous halo. We elect to shut off radiative cooling and star formation and feedback in this simulation. This choice was motivated by an acknowledgement that the current generation of feedback models fail to produce an accurately pressurized ISM \citep[although see][]{Hopkins2014}, and thus inclusion of this physics may lead to an ISM with erroneously low tensile strength, thereby overestimating its susceptibility to stripping. Stars and dark matter are included in a static sense: we compute the gravitational acceleration caused by the stellar disk and LMC DM halo distributions discussed in Section \ref{sec:lmc-model} and include these time-invariant, spatially varying acceleration vectors into the simulation. The potential of the MW and  forces associated with our accelerating, wind-tunnel reference frame are ignored, as our system's dynamics of interest lie well within the likely LMC tidal radius, which vdM14 found to be $r = 22.3 \pm 5.2$ kpc. To cleanly separate the role of RPS from other LMC-specific dynamics, we do not include the SMC, which has a strong gravitational influence on the LMC.

%
\subsection{Coordinate Systems}
\label{sec:coordinates}

Following the work of \cite{vanderMarel2001} and \cite{vanderMarel2002} (hereafter vdM02), we define the following four important Cartesian reference frames used in the present work.

\paragraph{Line of Sight (LOS) Frame: }  A 3-D cartesian coordinate system whose origin is the ``center'' of the LMC. For our simplistic model of the LMC, this corresponds to the center of its circular gaseous disk's initial distribution; its static, circular stellar potential; and its static, spherically symmetric dark matter potential. The $\hat{x}$-axis lies anti-parallel to the right ascension (RA, $\alpha$, at the origin of the system), the $\hat{y}$-axis parallel to the declination ($\delta$) and the $\hat{z}$-axis pointed towards the solar neighborhood (i.e. the observer). From this frame we can easily collapse our 3-D data into 2-D projections of the LMC that can be overlaid on and compared to real observations with relative accuracy.

\paragraph{LMC Frame: } Another frame first devised in \cite{vanderMarel2001} centered on the LMC, with the vertical $\hat{z}'$-axis aligned \emph{antiparallel} to the galaxy's angular momentum vector, $\textbf{L}$. In the present model, $\textbf{L}$ is perfectly aligned for the stars, gas and DM. This new frame is defined in terms of two rotations out of the LOS frame: first a counterclockwise rotation about the $\hat{z}$ (line-of-sight) axis by an angle $\theta$, and second a clockwise rotation about the new $\hat{x}'$-axis by an angle $i$. Here $i$ corresponds to the inclination angle of the LMC disk out of the LOS plane: the plane on the sky passing through the LMC center whose normal vector is aligned with the line-of-sight. And $\theta$ corresponds to the angle between the horizontal LOS axis and the line of nodes, which represents the intersection of the LMC disk plane and the LOS $x$-$y$ plane. For these angles, we use the results of vdM02 obtained from analysis of carbon star kinematics, setting $i = 34.7^\circ$ and $\Theta = \theta - 90 = 139.9^\circ$.

\paragraph{Simulation (SIM) Frame: } Working in the LMC frame is an attractive option, since height above and below the disk plane is aligned with the $\hat{z}'$-axis and all of the disk gas' circular motion is in the $x'$-$y'$ plane. This not only simplifies how \verb|enzo| generates the initial conditions but also aids in the solver's accuracy, since angular momentum is better preserved when rotational motion is aligned with the grid axis.
However, one downside of the frame as defined by \cite{vanderMarel2001} is that the headwind experienced by the LMC would have velocity components that switch sign over the course of the LMC's orbit. This means a wind being blown in from one boundary face of our simulation box would eventually switch to being blown in from the opposite boundary face --- a complication that presents a formidable software challenge. We can easily circumvent this issue by rotating the LMC frame $100^\circ$ clockwise about the $\hat{z}'$-axis, retaining the LMC frame's attractive features. All our present simulations were run in this new SIM frame.

\paragraph{Galactic Frame: } This last frame of interest is the only one not centered on the LMC but rather on the solar neighborhood, and where the LMC orbit data from \cite{Besla2012} is most naturally defined. Here the $\hat{z}^{\rm MW}$-axis is aligned with the MW's angular momentum vector, and the $\hat{x}^{\rm MW}$ axis points away from Sgr A*. Here the following rotation matrix pulls a vector in LOS coordinates into this frame

\begin{equation}
	\left[ \begin{array}{c} x^{\rm MW} \\ y^{\rm MW} \\ z^{\rm MW} \end{array} \right] = 
	\left[ \begin{array}{lll}
			0.11638 	& 	-0.98270	& 	-0.14410	\\
			0.57132	&	-0.05244	&	0.81905	\\
			-0.81243	&	-0.17765	&	0.55533
	\end{array} \right] \cdot
	\left[ \begin{array}{c} x \\ y \\ z \end{array} \right] \; .
\end{equation}

For the sake of both clarity land reproducibility, Table 3 displays important vector quantities used in this work, instantiated in our four coordinate frames.

\begin{table*}[tb]\begin{tiny}\begin{center}\begin{tabular}{l|llll}
Quant					& LMC					& SIM					& LOS					& Galactic					\\ \hline \\
$\hat{x}_{\rm LMC}$			& \vvb{1.}{0.}{0.}			& \vv{-0.174}{-0.985}{0.0}		& \vv{-0.656}{-0.755}{0.}		& \vv{0.665}{-0.335}{0.667}	\\ \\
$\hat{y}_{\rm LMC}$			& \vvb{ 0.}{1.}{0.}			& \vv{0.985}{-0.174}{0.0}		& \vv{0.620}{-0.539}{-0.569}	& \vv{0.684}{-0.083}{-0.724}	\\ \\
$\hat{z}_{\rm LMC}$			& \vvb{0.}{0.}{1.}			& \vv{0.}{0. }{1.}				& \vv{0.430}{-0.373}{0.822}	& \vv{0.299}{0.939}{0.174}	\\ \\ \hline \\
$\hat{x}_{\rm LOS}$	(WEST)	& \vv{-0.656}{0.620}{0.430}	& \vv{0.725}{0.538}{0.430}	&\vvb{1.}{0.}{0.}				& \vv{0.116}{0.571}{-0.812}	\\ \\
$\hat{y}_{\rm LOS}$	(NORTH)	& \vv{-0.755}{-0.539}{-0.373}	& \vv{-0.400}{0.837}{-0.373}	&\vvb{ 0.}{1.}{0.}			& \vv{-0.983}{-0.052}{-0.177}	\\ \\
$\hat{z}_{\rm LOS}$	(-AWAY)	&  \vv{0.}{-0.569}{0.822}		& \vv{-0.561}{0.099}{0.822}  	&\vvb{0.}{0.}{1.}				&  \vv{-0.144}{0.819}{0.555}	\\ \\ \hline \\
Angle between disk \\and $x$-$y$-plane	
						& $0.0^\circ$				& $0.0^\circ$				& $34.7^\circ$				& $76.3^\circ$				\\ \\ \hline \\
$\textbf{L}$				&  \vvb{0.}{0.}{-1.}			& \vv{0.}{0.}{-1.}			& \vv{-0.430}{0.373}{-0.822}	& \vv{-0.299}{-0.938}{-0.174}	\\ \\ \hline \\
$\textbf{v}_{\rm Sun}$		&  \vv{-72.0}{-18.6}{240.3}		& \vv{-5.85}{74.1}{240.3}		& \vv{138.9}{-25.4}{208.2}		& \vvb{11.1}{251.2}{7.25}*		\\ \\ 
$\textbf{v}_{\rm LMC,unc}$
						& \vv{256.7}{-161.7}{-430.8}	& \vv{-203.8}{-224.7}{-430.8}	& \vv{-453.8}{54.4}{-262.2} 	& \vvb{-68.5}{-476.8}{213.4}**	\\ \\ 
$\textbf{v}_{\rm LMC}= $\\$\;\;\;\;\; \textbf{v}_{\rm LMC, unc} + \textbf{v}_{\rm Sun}$ 
						& \vv{184.6}{-180.3}{-190.5}	& \vv{-209.6}{-150.5}{-190.5}	& \vvb{-314.8}{29.03}{-53.951}	& \vv{-57.4}{-225.6}{220.7}	\\ \\ \hline \\
Angle: $\textbf{v}_{\rm LMC}$, $\textbf{L}$ 
						& $53.571^\circ$			&  $53.571^\circ$			&  $53.571^\circ$			&  $53.571^\circ$			\\  \\
Angle: $\textbf{v}_{\rm wind}$, $\textbf{L}$
						& $126.429^\circ$			& $126.429^\circ$ 			& $126.429^\circ$ 			& $126.429^\circ$ 			\\ \\ 
\end{tabular}\end{center}
* See discussion in Section 4.1 of \cite{vanderMarel2012}, ** Value from K13 \\
\end{tiny} \\
{ \small \textbf{Table 3:} Various vector quantities expressed in our four cartesian coordinate frames defined in Section \ref{sec:coordinates}. Bold quantities represent observed inputs or quantities expressed in their most natural coordinate frame. Here $\textbf{L}$ denotes the LMC disk's angular momentum unit vector. $\textbf{v}_{\rm Sun}$ is the Sun's velocity with respect to the Galactic Center (GC). $\textbf{v}_{\rm LMC, unc}$ is motion of the LMC COM with respect to the Sun, whereas the ``corrected'' $\textbf{v}_{\rm LMC}$ is relative to the GC. This last quantity is the most useful in constructing our ram pressure headwind, which is denoted here at present day as $\textbf{v}_{\rm wind} = -\textbf{v}_{\rm LMC}$. }
\label{tab:coords}
\end{table*}

%
\subsection{Simulation Boundary Wind}
\label{sec:wind}

\stdFig{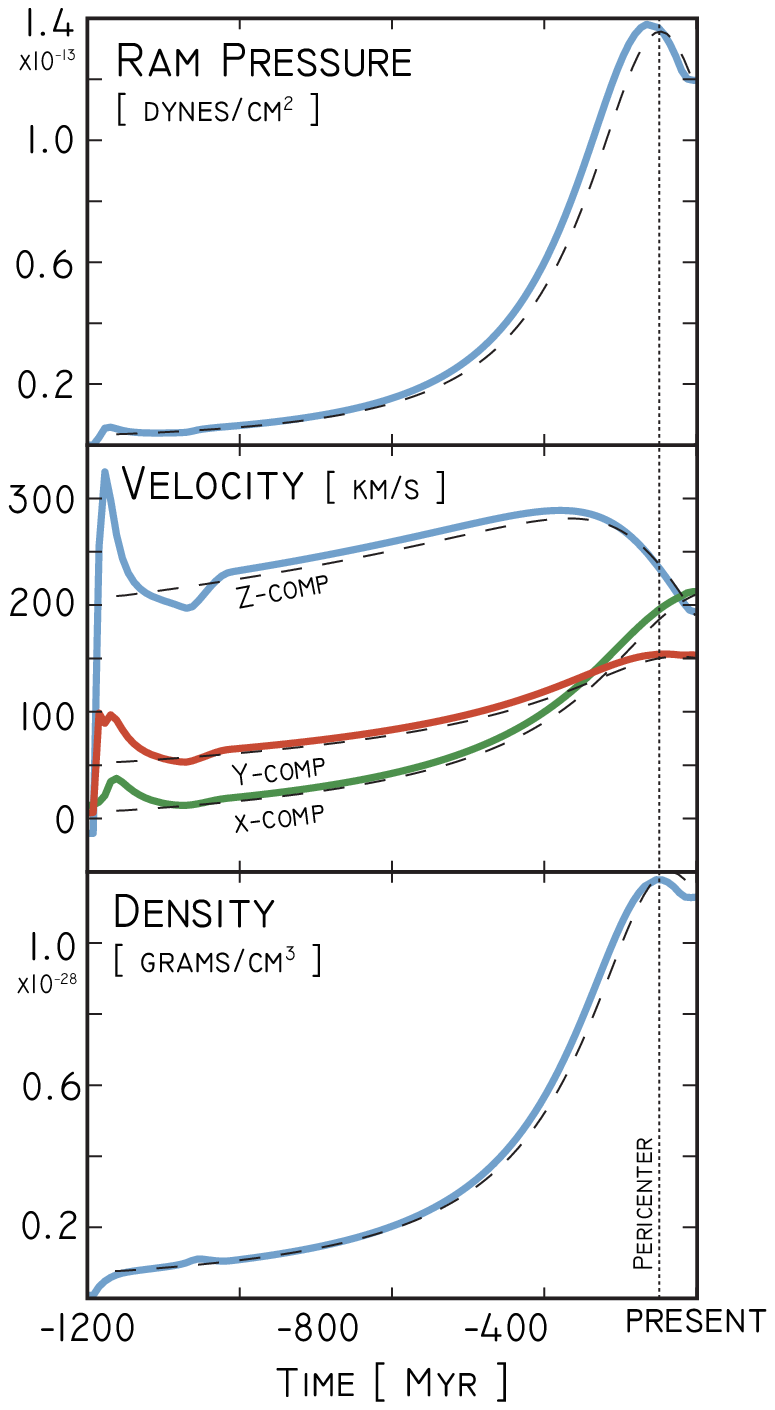}{Comparison of intended wind within our simulation box (dashed lines) to the wind actually measured within the simulation over time (solid, colored lines). This wind depends on both the LMC orbit and chosen MW gaseous halo model. Shown here is our fiducial, M-1.2, simulation. The simulation measurement shown here was taken halfway between the corner of the simulation box where the wind is introduced and the center of the LMC. We also tested the fidelity of the wind as it propagated further into the simulation domain by running a simulation devoid of an LMC, with similarly well-matched results.}{fig:wind}{1.0}

Our simulated LMC stands still in our simulation box, with its angular momentum axis aligned with the cartesian grid. To capture the hydrodynamical effects of its orbit through the MW, we launch a wind into the box from three orthogonal, adjacent sides of the domain. The opposite three sides are set to ``outflow'' conditions, to allow the downstream gas to escape the domain. The density, temperature and three-dimensional velocity components of this wind evolve over time to mimic the headwind experienced by an orbiting LMC in its final Gyr before the present.

We take our fiducial LMC orbit from K13 and assume the orientation of the LMC angular momentum axis is constant in a Galactic frame. Neither this particular orbital scenario nor this constant orientation assumption will strongly influence our results (see Section \ref{sec:error-anal}). This orbital data supplies us with the time-evolving three dimensional headwind velocity and the evolving radial distance of the LMC from the MW center which we can transform into our simulation's frame of reference with the aid of Section \ref{sec:coordinates}. Equations \ref{eq:beta} and \ref{eq:halo-temp} map the radial distance to the wind's density and temperature, thus completing its description. 

Our simulated wind is an orchestrated attempt to make fluid inflow from three orthogonal boundary surfaces of our domain to mimic gas in hydrostatic equilibrium within the MW's gravitational potential, as experienced by a dwarf system whose local rest frame is non-inertial. Thus achieving a good match between model and reality throughout our simulation box is a non-trivial matter that needs to be verified ``experimentally''. Figure \ref{fig:wind} shows how this wind evolves over time for our ``fiducial'' simulation \verb|M-1.2| (see Section \ref{sec:param-choices}). The figure displays not only the input wind given to our simulation software but also a measurement of the wind probed directly in our simulation box halfway between the box's origin and the LMC kinematic center, upstream of the galaxy. This probe demonstrates excellent fidelity between the theoretical wind evolution and the wind actually experienced by the simulated disk. The initially wide disparity between modeled and simulated velocity components signifies the presence of an initial ``adjustment'' shock, and does not affect our results since the ram pressure at this early stage is far too weak to alter the disk's gas structure. The fidelity of the propagating wind was also verified for larger distances from the box edge in a simulation devoid of an LMC itself. Figure \ref{fig:wind} also illustrates how the ram pressure is far stronger at the LMC's recent pericentric passage than at earlier times in its orbital evolution, suggesting its current HI extent is a \emph{localized} probe of the hydrodynamics at $r_p \approx 48$ kpc. 

The changing components of the wind's three dimensional velocity in Figure \ref{fig:wind} demonstrate how the orientation between the LMC disk and the headwind changes measurably over time. \cite{Roediger2006} simulated ram pressure stripping for a variety of wind-disk inclination angles, and showed that the gas disk rapidly strips to the truncation radius predicted analytically by Gunn \& Gott in all but the the most edge-on orientations. We explore the role of wind orientation angle further in Section \ref{sec:sim-results}.

\subsection{Simulation CGM Model Selection}
\label{sec:param-choices}
We seek to constrain the CGM density at radii corresponding to the LMC's recent pericentric passage at $r \approx 48$ kpc using the current observed HI disk extent. As discussed in Section \ref{sec:wind}, ram pressure is strongest at the LMC's most recent pericentric passage, suggesting that the resulting HI truncation radius will be mostly sensitive to the local halo density at pericenter, $n_p = n( r \approx 48 \;{\rm kpc})$. It is this quantity we systematically vary in our simulations. 

However, our simulations must also address what role the broader CGM density profile plays for a fixed pericentric density. If the Gunn \& Gott picture of ram pressure stripping holds precisely, and the process occurs instantaneously, then indeed the density at exactly pericenter, for a given orbital model, is all that matters. But if the stripping time scale, chaotic gas dynamics, oscillations within the disk, and the history of the ram pressure faced by the disk influence the final outcome, the results may become sensitive to the broader picture of gas in the halo.

While our CGM density model has three parameters, $n_0$, $r_c$ and $\beta$, as explained in Section \ref{sec:halo-model}, our LMC effectively experiences a simpler profile with two parameters: an amplitude $n_0 r_c^{3 \beta}$ and an exponential falloff $\beta$. Thus while the chief goal of our parameter search will be to understand the role of $n_p$, a secondary goal will be to explore a few combinations of $n_0 r_c^{3 \beta}$ and $\beta$ for a range of pericentric densities.

We chose three representative amplitudes: $n_0 r_c^{3 \beta} \in \{ .01, .079, .5 \}$ cm$^{-3}$ kpc$^3$, designated as lo-, mid- and hi-core simulations, respectively. The mid-core runs were given $n_0 = .46$ cm$^{-3}$ and $r_c = .35$ kpc, corresponding to the maximal likelihood values of MB13. The other two pairs of $n_0$ and $r_c$ were chosen to roughly span two orders of magnitude in the core constant. For each of these three scenarios, we then selected $\beta$ chosen to fix $n_p$ to a wide range of values from $\sim 10^{-7}$ - $10^{-2}$ cm$^{-3}$, with a denser sampling around the optimal parameter choices for matching the observed truncation radius. 

Table 4 summarizes these simulation choices, grouped by core constant. The runs follow a standard naming convention with a capital letter denoting the core constant level followed by a dash and the halo gas density at LMC pericenter in units of $10^{-4}$ cm$^{-3}$. For instance, the medium-core run with $n_0 r_c^{3\beta} \approx .079$ and a pericentric halo density of $1.2 \times 10^{-4}$ cm$^{-3}$ is denoted \verb|M-1.2|; the high-core density run with pericentric density $7.96 \times 10^{-5}$ is labeled \verb|H-.8|; etc. A few miscellaneous runs were also thrown in with widely varying halo parameters. Finally, we explored the role of the LMC wind orientation by performing a series of runs with varying halo profiles where the wind was launched \emph{face-on} at the galaxy, along its angular momentum axis, denoted with the prefix \verb|FL-|, \verb|FM-| and \verb|FH-| for low, medium and high core halos, and some runs at a fixed 30-degree inclination angle for the wind, denoted \verb|IL-|, \verb|IM-| and \verb|IH-|. Figure \ref{fig:sim-params} also shows each simulation choice graphically in parameter space, atop the same analytic models of Figure \ref{fig:analytic-params}. From this overlay, it's clear that our densest sampling of models feature systematically lower $\beta$'s (and thus, systematically higher pericentric densities) than the optimal values predicted by the toy model in Section \ref{sec:analytic}. We discuss and quantify this offset in Section \ref{sec:error-anal}.

The simulations described here all make use of our standard LMC model, described in Section \ref{sec:lmc-model}. While GG predicts that the choice of DM halo has a negligible effect on the gas stripping radius, our choice of stellar and gaseous disks, in particular their surface densities as a function of radius, will strongly impact results. We consider uncertainties in the stellar disk via our analytic GG model in Section \ref{sec:error-anal}. For the gaseous component, we ran a smaller suite of simulations with a ``gas-rich'' LMC disk, with the results and implications discussed in Section \ref{sec:gas-rich}.

\stdFullFig{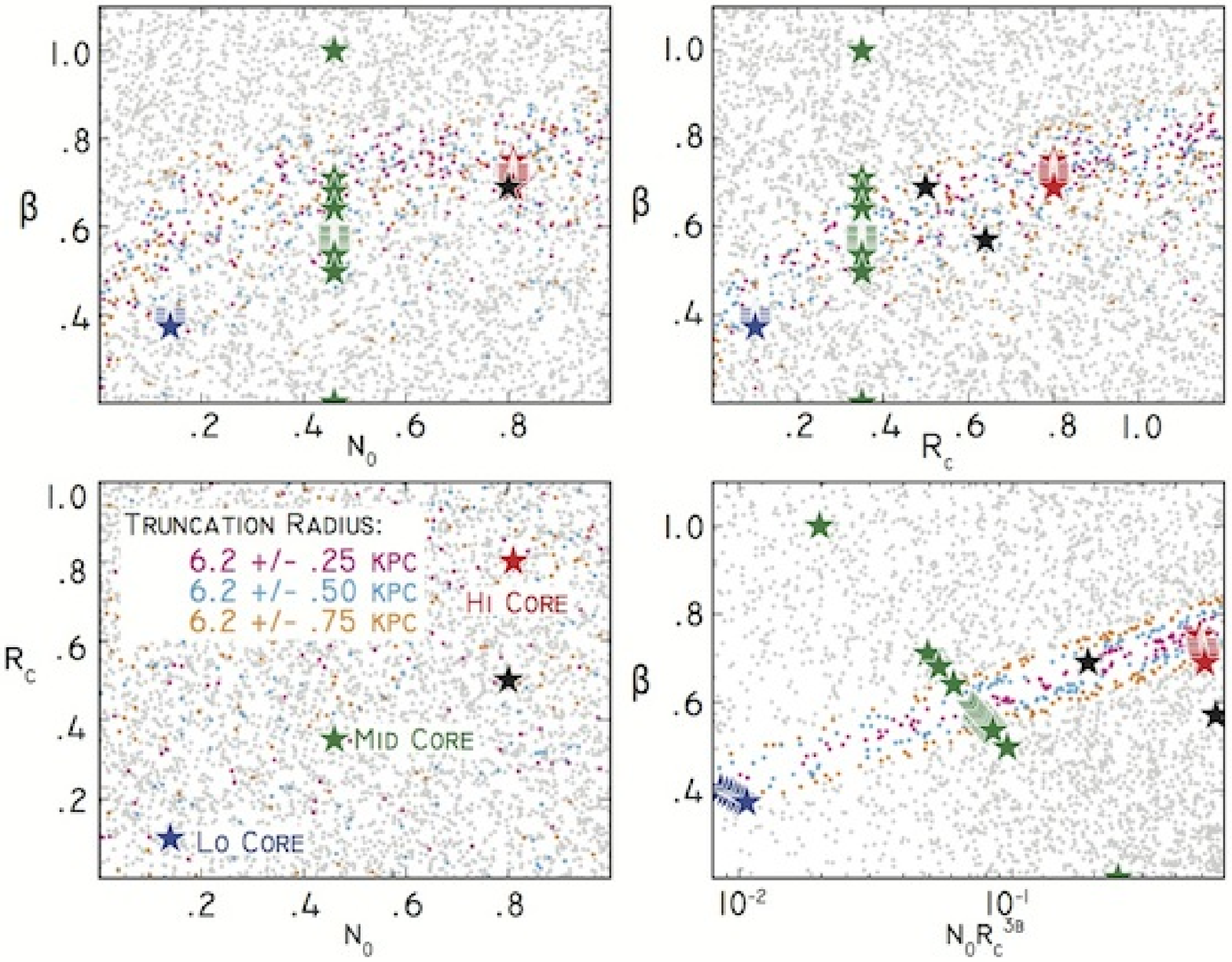}{Our halo model's parameter space, as in Figure \ref{fig:analytic-params}, but now with our 44 three-dimensional, hydrodynamic simulations over-plotted as stars. For the majority of runs, the halo's core parameters, $n_0$ and $r_c$, were held at three distinct, fixed combinations designated ``lo core'' (blue stars), ``mid core'' (green stars) and ``hi core'' (red stars). For each of these core choices, $\beta$, the exponential falloff, was incrementally varied to explore a range of MW halo densities at the LMC's recent perigalacticon, $n_p$. A broad range of $\beta$'s were also explored for the mid-core case, and a few miscellaneous parameter choices were also explored (black stars, some beyond the plotted range). Beneath these stars, the analytic models of Section \ref{sec:analytic} are also shown.}{fig:sim-params}{.7}

\begin{table*}[p]
\tiny
\renewcommand{\tabcolsep}{11pt}
\begin{center}\begin{tabular}{lllllllll}

\multicolumn{9}{c}{\textbf{Lo Core} (blue) --- $n_0 r_c^{3\beta} \approx .01$} \\
\hline \hline 
$n_0$	& $r_c$	& $\beta$		& $P_p$			& Total Mass	& $n_p$	& $R_{\rm GG}$	& $R_t$	& Name		\\
\hline 
0.139	& 0.100     & 0.403		& 9.04e-14		& 1.18e+10       & 7.95e-05		& 6.09				& 6.56        			& L-0.8		\\
0.139	& 0.100     & 0.396		& 1.03e-13		& 1.37e+10       & 9.06e-05		& 5.97				& 6.39        			& L-0.9		\\
0.139	& 0.100     & 0.391		& 1.13e-13		& 1.52e+10       & 9.93e-05		& 5.89				& 6.17        			& L-1.0		\\
0.139	& 0.100	& 0.385		& 1.26e-13		& 1.73e+10	& 1.11e-04		& 5.78				& 6.05				& L-1.1		\\
0.139 	& 0.100	& 0.381		& 1.36e-13		& 1.88e+10	& 1.20e-04		& 5.72				& 5.98				& L-1.2		\\
0.139	& 0.100	& 0.376		& 1.49e-13		& 2.09e+10	& 1.31e-04		& 5.63				& 5.81				& L-1.3		\\
0.139	& 0.100	& 0.372		& 1.61e-13		& 2.28e+10	& 1.41e-04		& 5.57				& 5.71				& L-1.4		\\ \\

\multicolumn{9}{c}{\textbf{Mid Core} (green) --- $n_0 r_c^{3\beta} \approx .079$ } \\ 
\hline \hline
$n_0$	& $r_c$	& $\beta$		& $P_p$			& Total Mass	& $n_p$			& $R_{\rm GG}$		& $R_t$				& Name		\\ 
\hline 
0.46 		& 0.35 	& 1.00 		& 2.01e-16 		& 1.89e+07 	& 1.77e-07 		& 12.6 				& No Effect			& VHiBeta		\\
0.46 		& 0.35 	& 0.71 		& 1.46e-14 		& 1.05e+09 	& 1.29e-05 		& 7.82 				& 6.69				& M-0.13		\\ 
0.460	& 0.350	& 0.680		& 2.27e-14		& 1.68e+09	& 2.00e-05		& 7.40				& 6.72				& M-0.2		\\
0.46 		& 0.35 	& 0.64 		& 4.10e-14 		& 3.22e+09 	& 3.61e-05 		& 6.83				& 6.83				& M-0.36		\\

0.460	& 0.350	& 0.591		& 8.45e-14		& 7.22e+09	& 7.45e-05		& 6.15				& 6.47				& M-0.75		\\
0.460	& 0.350     & 0.586		& 9.10e-14		& 7.85e+09       & 8.02e-05		& 6.08				& 6.56        			& M-0.8		\\
0.460	& 0.350     & 0.578		& 1.02e-13		& 8.97e+09       & 9.03e-05		& 5.97				& 6.56       			& M-0.9		\\
0.460	& 0.350	& 0.571		& 1.14e-13		& 1.01e+10	& 1.00e-04		& 5.88				& 6.49				& M-1.0 		\\

0.46		& 0.35 	& 0.57 		& 1.15e-13 		& 1.02e+10 	& 1.02e-04 		& 5.87				& 6.37				& M-1.0-II		\\
0.460	& 0.350	& 0.565		& 1.24e-13		& 1.11e+10	& 1.09e-04		& 5.80				& 6.43				& M-1.1		\\
\textbf{0.46}& \textbf{0.35} & \textbf{0.559} & \textbf{1.36e-13}& \textbf{1.23e+10}&\textbf{1.20e-04}	& \textbf{5.72}& \textbf{6.26}& \textbf{Fiducial (M-1.2)}		\\
0.46		& 0.35 	& 0.553 		& 1.48e-13		& 1.36e+10	& 1.31e-04		& 5.64				& 6.13				& M-1.3		\\

0.46		& 0.35 	& 0.548		& 1.60e-13		& 1.48e+10	& 1.41e-04		& 5.57				& 6.02				& M-1.4		\\
0.46		& 0.35	& 0.544		& 1.69e-13		& 1.59e+10	& 1.49e-04		& 5.52				& 6.18				& M-1.5		\\
0.46		& 0.35	& 0.539		& 1.82e-13		& 1.73e+10	& 1.61e-04		& 5.45				& 5.64				& M-1.6		\\
0.46		& 0.35	& 0.535		& 1.93e-13		& 1.85e+10	& 1.70e-04		& 5.40				& 5.89				& M-1.7		\\
 
0.460	& 0.350	& 0.497		& 3.39e-13		& 3.52e+10	& 2.99e-04		& 4.90				& 4.89				& M-3.0		\\
0.46 		& 0.35	& 0.20 		& 2.74e-11 		& 6.34e+12 	& 2.40e-02 		& 1.26 				& Destroyed			& VLoBeta	\\ \\

\multicolumn{9}{c}{\textbf{Hi Core}  (red) --- $n_0 r_c^{3\beta} \approx .5$} \\
\hline \hline 
$n_0$	& $r_c$	& $\beta$		& $P_p$			& Total Mass	& $n_p$			& $R_{\rm GG}$		& $R_t$ 				& Name		\\
\hline 
0.810	& 0.800     & 0.751		& 9.01e-14		& 6.23e+09       & 7.96e-05		& 6.09				& 6.19        			& H-0.8		\\
0.810	& 0.800     & 0.732		& 1.14e-13		& 7.98e+09       & 1.00e-04		& 5.88				& 6.12        			& H-1.0		\\
0.810	& 0.800	& 0.725		& 1.24e-13		& 8.75e+09	& 1.09e-04		& 5.80				& 6.30				& H-1.1		\\
0.810	& 0.800	& 0.718		& 1.35e-13		& 9.60e+09	& 1.19e-04		& 5.72				& 6.10				& H-1.2		\\
0.810	& 0.800	& 0.711		& 1.47e-13		& 1.05e+10	& 1.30e-04		& 5.64				& 6.21				& H-1.3		\\
0.810	& 0.800	& 0.705		& 1.59e-13		& 1.14e+10	& 1.40e-04		& 5.57				& 6.19				& H-1.4		\\
0.810	& 0.800     & 0.699		& 1.71e-13		& 1.24e+10       & 1.51e-04		& 5.51				& 6.09        			& H-1.5		\\
0.810	& 0.800     & 0.689		& 1.93e-13		& 1.41e+10       & 1.70e-04		& 5.40				& 6.04        			& H-1.7		\\ \\

\multicolumn{9}{c}{\textbf{Exploratory} (purple)} \\ 
\hline \hline 
$n_0$	& $r_c$	& $\beta$		& $P_p$			& Total Mass	& $n_p$			& $R_{\rm GG}$		& $R_t$ 				& Name		\\
\hline 
1.20		& 0.64 	& 0.57 		& 8.44e-13 		& 7.50e+10 	& 7.44e-04 		& 4.12 				& 4.10				& MaxWind	\\
0.80 		& 0.50 	& 0.69 		& 7.12e-14 		& 5.22e+09 	& 6.29e-05 		& 6.31 				& 6.42				& T-6		\\
1.50 		& 1.50 	& 0.97 		& 7.03e-14 		& 5.41e+09 	& 6.22e-05 		& 6.32				& 6.49				& T-6-X	         \\ \\

\multicolumn{9}{c}{\textbf{Face-On Wind} (yellow dots)} \\
\hline \hline 
$n_0$	& $r_c$	& $\beta$		& $P_p$			& Total Mass	& $n_p$			& $R_{\rm GG}$		& $R_t$ 				& Name		\\
\hline 
0.139      & 0.100      & 0.381		& 1.36e-13		& 1.88e+10	& 1.20e-04		& 5.72				& 6.44      				& FL-1.2		\\
0.460      & 0.350      & 0.586		& 9.10e-14		& 7.85e+09	& 8.02e-05		& 6.08				& 7.06      				& FM-0.8		\\
0.460      & 0.350      & 0.559		& 1.36e-13		& 1.23e+10	& 1.20e-04		& 5.72				& 6.91       			& FM-1.2		\\
0.460      & 0.350      & 0.535		& 1.93e-13		& 1.85e+10	& 1.70e-04		& 5.40				& 6.34       			& FM-1.7		\\ 
0.810      & 0.800      & 0.718		& 1.35e-13		& 9.60e+09	& 1.19e-04		& 5.72				& 6.88      				& FH-1.2		\\ \\

\multicolumn{9}{c}{\textbf{$30^{\circ}$ Inclined Wind} (Shown in Figure \ref{fig:inclined}) } \\
\hline \hline
$n_0$	& $r_c$	& $\beta$		& $P_p$			& Total Mass	& $n_p$			& $R_{\rm GG}$		& $R_t$ 				& Name		\\
\hline 
0.139      & 0.100      & 0.381		& 1.36e-13		& 1.88e+10	& 1.20e-04		& 5.72				& 6.03      				& IL-1.2		\\
0.460      & 0.350      & 0.559		& 1.36e-13		& 1.23e+10	& 1.20e-04		& 5.72				& 6.22   	   			& IM-1.2		\\
0.810      & 0.800      & 0.718		& 1.35e-13		& 9.60e+09	& 1.19e-04		& 5.72				& 6.26     				& IH-1.2		\\ \\

\label{tab:all-sims}
\end{tabular}\end{center}

{ \small \textbf{Table 4:} Our 44 wind tunnel simulations. The top three groups (L-, M-, and H- runs) each have a different, fixed core halo density (determined by $n_0$ and $r_c$) with various $\beta$ chosen to explore a range of pericentric densities. The fourth group are runs with miscellaneous halo parameter choices. The fifth group is a set of runs where the orbital wind was delivered entirely face on to the galaxy. The first three columns, $n_0$ [ cm$^{-3}$ ], $r_c$ [ kpc ] and $\beta$, denote the MW halo parameter choices that inform the LMC wind density. The next three columns list important implications of these parameter choices: $P_{\rm ram}$, the peak (pericentric) ram pressure experienced by the LMC; the total mass of the MW halo, if we were to extrapolate the $\beta$-profile to large radii ($M(r \gtrsim r_{\rm vir})$); and $n_{\rm p}$, the MW halo density at the LMC's pericentric passage. $R_{\rm GG}$ is the truncation radius of the LMC disk inferred from the pericentric ram pressure and the initial disk profile. $R_t$ is the truncation radius for the observable HI inferred from the simulations (see Section \ref{sec:sim-results}). The final column indicates the name of the run (see Section \ref{sec:param-choices} for naming conventions).}
\end{table*}

\subsection{Simulation Results}
\label{sec:sim-results}

The following analysis made use of \verb|yt| \citep{Turk2011} for simulation data reduction, \verb|Astropy| \citep{Astropy2013} for observed data reduction and map projections, and \verb|emcee| \citep{Foreman2013} for MCMC analysis, all community-developed, Python packages.

\subsubsection{HI Column Distribution and Inferred Truncation Radius}
We begin by describing our fiducial simulation, \verb|M-1.2|, a mid-core halo run with $n_p = 1.2 \times 10^{-4}$ cm$^{-3}$, which produced one of the most successful matches between the simulated and observed HI truncation radius on the disk's leading edge. The LMC gaseous disk is, at first, completely unperturbed by the low density headwind the galaxy experiences 1 Gyr ago, at our run's start. At this point, the truncation radius predicted by Gunn \& Gott, $R_{\rm GG}$, lies well beyond the disk's initial extent. The LMC then slowly descends further into the MW halo, picking up speed and facing an increasingly stronger ram pressure, an effect compounded by the increasing CGM density as the LMC travels towards pericenter of its orbit. The theoretical $R_{\rm GG}$ soon shrinks to become smaller than the LMC's initial extent. At this point, from face-on projections of the gaseous disk, we see material torn off the disk. The stripping is not entirely smooth, but rather episodic. The disk's extent often lies a substantial fraction of a kiloparsec beyond $R_{\rm GG}$, defying the expected stripping. Then, on a timescale of a few Myr, this disk material beyond $R_{\rm GG}$ fragments globally, and disintegrates, beginning at the leading-edge. Any remaining large-scale asymmetries are evened out within one rotation period of the LMC disk. Material pulled beyond the disk plane is first ``folded back'' before flying off the disk. As discussed below, this often enhances the appearance, from our viewing perspective on Earth, of an asymmetric HI column distribution. From slices of gas density, we see that material unbound from the disk often lingers behind the LMC disk, shielded from the wind and still bound to the system, though at unobservably low column.

\stdFullFig{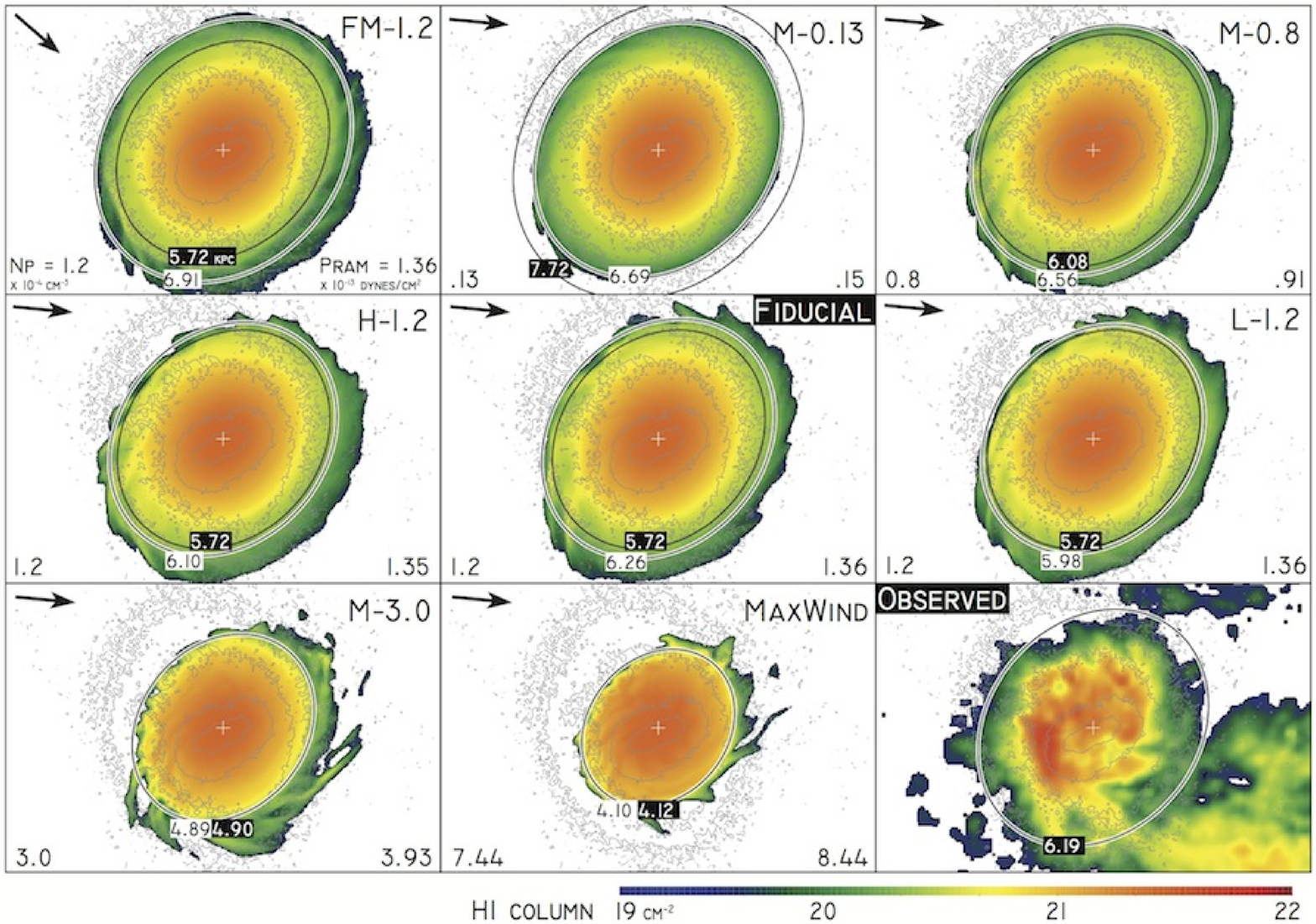}{Present Day \textbf{HI column} distribution for 8 representative simulations compared against the observed HI column (bottom right panel). Also overlaid in \textbf{grey contours} is the observed stellar distribution from 2MASS \cite{vanderMarel2001}. The name of the run, which can be matched to Table \ref{tab:all-sims}, is displayed in the upper-right corner. The pericentric MW gaseous halo density, $n_p$ in each simulation is displayed in the bottom right of each panel in units of $10^{-4} \; {\rm cm}^{-3}$, with the corresponding peak ram pressure experienced at this point in the orbit displayed in the bottom right, also in cgs. These two quantities tend to increase from top to bottom in this figure. The observed truncation radius, found from fitting the leading edge of the disk (upper left quarter) where the HI column falls to $10^{19}$ cm$^{-2}$, is displayed as black ellipse with a white outline, whose axis ratio is constrained to match the LMC disk orientation found in vdM02. The corresponding truncation radius predicted by Gunn \& Gott is also plotted as a thinner \textbf{black ellipse}. The \textbf{black arrow} in the upper left of each panel indicates the direction of the headwind experienced by the LMC at present day. Finally, the \textbf{white cross} denotes the LMC kinematic center as traced by the HI}{fig:HI}{.43}

Figure \ref{fig:HI} provides mock observations from a subset of representative simulations compared to the observed HIPASS HI column map in the LOS point of view (LOS POV). For the simulations this LOS POV is a cartesian reduction (sum) performed with \verb|yt| of physical density into surface density along equally spaced rays throughout the domain orientated along the LOS $\hat{z}$-axis, as defined in Table 3. The resulting 2D surface density map is centered on the LMC kinematic center, with a north-vector aligned with the LOS $\hat{y}$-axis. For the observed HI column, we interpolated the data onto an orthographic projection with \verb|astropy|'s World Coordinate System (WCS) centered on the LMC kinematic center. In this projection, $\hat{x}$- and $\hat{y}$-vectors are tangent to the lines of constant RA and DEC passing through this center, the former anti-aligned with RA so it increases left-to-right. Distances are then mapped to kpc via the measured distance to the LMC (see Table \ref{tab:position-params}) and by approximating the map as locally flat. Thus the LOS POV for both the simulations and observations roughly corresponds to 2D maps in the LOS frame defined in Section \ref{sec:coordinates}. The error introduced by this mapping at the observed $R_t \approx 6$ kpc is $\mathcal{O}(\theta^3) \sim  ({6 \; \rm kpc} / {50 \;  \rm kpc})^3 \approx  .2 \% $.

The central panel of Figure \ref{fig:HI} shows a mock HI observation of our fiducial simulation, \verb|M-1.2| at present day, $\approx 50$ Myr past the LMC's recent perigalacticon. The \emph{observed} stellar distribution is also overlaid  \citep{vanderMarel2001} as contours on the panel. To produce these plots, we tagged our simulation's gas disk with a color field, to identify where disk material would end up in the simulation. We then eliminated any cells where this disk gas had fallen below .03 cm$^{-3}$, in density, as this material is likely highly ionized \citep{Rahmati2013}. We then used \verb|yt| to produce a column density projection of the surviving HI along the LOS axis between the solar neighborhood and the LMC's kinematic center. This vector is labeled $\hat{z}_{\rm LOS}$ in Table 3, and was found in the simulation frame, SIM, with the coordinate transformation matrices described in Section \ref{sec:coordinates}. A planar projection was used for simplicity, though the opening angle of the observed LMC disk is  $\theta \approx \tan^{-1}(5 \; {\rm kpc} / 50 \; {\rm kpc} ) = 5.7^\circ$, and so using non-converging column rays introduces errors $\sim.1$ kpc. We then cut out any region of the projection where the total HI column falls below $10^{19}$ cm$^{-2}$. A similar cut was made to the observed HI distribution, shown in the bottom-right panel of Figure \ref{fig:HI}. The present direction of the headwind experienced by the LMC (in a frame where the LMC is at rest) is shown in the upper-left corner of the panel.

Figure \ref{fig:HI} also includes a black ellipse on each simulation panel to denote the present day truncation radius predicted by the GG model, $R_{\rm GG}$. To provide a direct numerical comparison to this, we take the ``leading quadrant'' of the LMC disk, defined from $90^\circ$ - $180^\circ$ in the LOS projection frame, and search along rays equally spaced in angle for the last point where the HI column drops below $10^{19}$ cm$^{-3}$. We then fit a ellipse to these points, by minimizing the sum of the square of the radial residuals to each point. This ellipse is shown in black with a white outline. Both the fitted and Gunn/Gott ellipses have orientations and axis ratios constrained to match a circle in the LMC frame as would be seen from the LOS POV.

Analysis across all simulations indicated that this measure of truncation radius is systematically smaller (and closer to the prediction of GG) than measurements taken from face-on projections of the LMC disk (as seen from a fictional observer directly above the LMC disk plane). This discrepancy vanishes if we use a higher column threshold, such as $10^{20}$ cm$^{-3}$, providing evidence that the difference is caused by extraplanar, ``folded back'' material that is less visible from the LOS POV. Further, the systematic difference between this measurement taken from the LOS view and the face-on view is substantially wider than the scatter across the simulations. Thus this is an important feature of our simulations completely unexplained by the analytic GG model that will have a measurable impact on our inferred diffuse CGM density. 

Visual inspection of this gas beyond the predicted GG radius but within $R_t$ reveals a rippled structure to its column density, suggesting the gas has been destabilized by the wind, and is perhaps en route to becoming liberated from the disk. Visual inspection also reveals that the trailing (bottom-right) edge of the disk has gas well beyond the fitted truncation radius. This asymmetry in the HI morphology is both a product of viewing perspective (folded back material we can see on the trailing edge and not the leading edge) and an actual discrepancy in the truncation radius on the leading and trailing edges, as seen in face-on projections of the disk. Bear in mind that the simulated stellar distribution's iso-surface-density contours are themselves ellipses of the same axis ratio and orientation as those plotted in Figure \ref{fig:HI}. Thus we indeed see an offset in observable HI on the leading and trailing edges unexplained by the simple GG model. Notably, the discrepancy disappears for the highest density runs shown here (e.g. \verb|Maxwind|).

An important point to make regarding these mock observations is the absence of high-column gas beyond the disk, particularly in the direction of the Magellanic Bridge (bottom-right). This stands in stark contrast to the observed HI profile, where a significant quantity of gas lies in the bridge region, which current models attribute to the work of tidal forces from the LMC-SMC interaction \citep{Besla2010}. The absence of high column gas beyond the disk plane is a feature present across our runs, along all sight-lines.

The runs highlighted in Figure \ref{fig:HI} were chosen to visually illustrate some points that further analysis across time and across the broader suite of 42 simulations corroborates. First, we note the obvious, by considering the top-right, central, and bottom-left panels, i.e. runs $M-0.8$, $M-1.2$ and $M-3.0$, which feature the same central MW gaseous halo density but decreasing $\beta$, and by extension larger $n_p$ and higher ram pressure: consistent with the model of GG, these runs feature progressively smaller truncation radii.

Next, we note the central row shows runs \verb|H-1.2|, \verb|M-1.2| (fiducial), and \verb|L-1.2|, which all exhibit identical pericentric densities but decreasing $\beta$ (i.e., from left to right these MW gas halos have slower density falloffs). From visual inspection, the runs appear more or less identical, and the discrepancy in their fitted truncation radii do not follow any trend with $\beta$. These anecdotes suggest a point we strengthen later: that the LMC's present day HI truncation radius is a clean probe of local density, $n_p = n(R \approx 48\; {\rm kpc})$, i.e. the maximal ram pressure sets the value, and is independent of the broader halo profile the system has orbited through. These simulations thus bolster a key assumption of Section \ref{sec:analytic}.

Another comparison worth highlighting involves the top-left panel and the central fiducial panel. The former shows a face-on wind simulation, \verb|FM-1.2|, where we launched the wind directly along the $\hat{z}$-axis of the LMC disk plane, eliminating any asymmetry between how different quadrants of the disk plane experience the headwind. This run is otherwise identical (in wind speed, density, temperature and evolution) to the fiducial \verb|M-1.2| run. The first important point to make is the inferred truncation radius of 6.91 kpc is substantially larger than any of the values in the central row of Figure \ref{fig:HI}, despite all these runs having identical $n_p$, putting it in even further disagreement with the GG model's prediction. The second important point involves asymmetry in the observed HI column: despite the symmetric pummeling of all edges of the disk, from the LOS POV the gas still appears substantially farther from the LMC kinematic center on the trailing edge of the disk. This is the clearest example yet of extra-planar, ``folded back'' material influencing the observed HI distribution. It further highlights the importance of viewing perspective. From visual inspection we see this gas is at substantially lower column (though still above our cut) and exhibits a wavy, unstable structure rife with filaments slowly breaking away from the disk.

Finally, we call attention to the top-center panel, run \verb|M-0.13| where the typical relation between $R_{\rm GG}$ and $R_t$ has been reversed: the simulated truncation of the mock HI observation lies \emph{within} the GG model's prediction. This is simply a product of our density cut choice: before the wind has even gained enough strength to induce any stripping of LMC disk material, gas below our physical density cut of $.03$ cm$^{-3}$ dominates at large radii. Whatever high density gas does go into our mock density projection is then removed by the surface density cut at $10^{19}$ cm$^{-3}$. Indeed, choosing a cut of $10^{18}$ restores agreement, though the probed radius exhibits an even higher degree of scatter across the simulations and disagrees even further with the prediction of the GG model. This issue motivates our choice of a higher column cutoff. However, if we choose too high a surface density cut --- say, $10^{20}$ cm$^{-2}$ --- the observed truncation radius of the \emph{initial} HI mock observation is smaller than the present observed extent of the LMC, rendering our predictions useless. Thus $10^{19}$ cm$^{-2}$ is optimal for our purposes. Varying the physical density cut from $.03$ cm$^{-3}$ likewise impacts the results, though this value is motivated by studies of the ISM and not an arbitrary choice.

\subsubsection{Radial Profiles of HI Column}

\stdFullFig{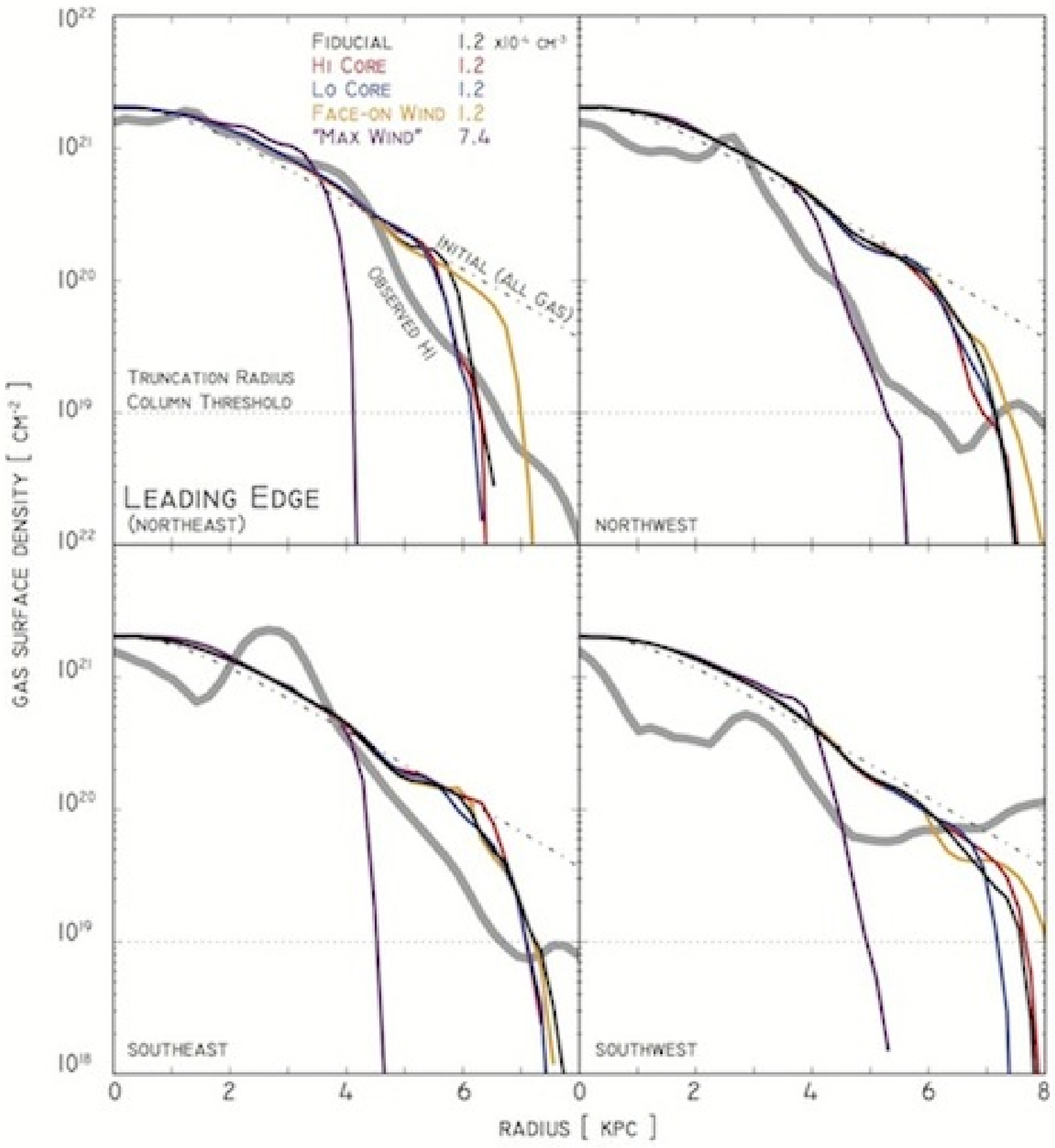}{Averaged profile of surface density in the four quadrants of the LMC, computed as in Figure \ref{fig:obs-profile}. Black lines shows our fiducial simulation, whereas red and blue lines show hi- and lo-core halo density models with an identical pericentric density of $1.2 \times 10^{-4}$ cm$^{-3}$. A thick grey line denotes the same average profile found from the observed HI column map. Also shown are the higher-density MaxWind simulation (purple) and a face-on wind run, FM-1.2 (orange).}{fig:surf-profiles}{.7}

Figure \ref{fig:surf-profiles} quantifies the surface density profiles of Figure \ref{fig:HI} by sampling the mock-HI surface density along a dozen rays from the LMC disk center out towards the edge in each disk quadrant (clockwise from top-left: the leading edge, southeast, towards the bridge/stream and northwest). The 2D distance along these rays is then mapped to a radial distance from the LMC kinematic center assuming all material lies in the disk plane, and then these profiles are added together into an average surface density profile along the leading edge. We performed this analysis in an identical fashion for both the simulations and the observed LMC HI distribution. From this we find the runs with $n_p = 1.2 \times 10^{-4}$ cm$^{-3}$ (regardless of hi/mid/lo core model) produce the most successful match to the observed surface density profile along the leading edge. The agreement among these three simulations further bolsters our argument that RPS is a clean probe of $n_p$, rather than the broader halo profile. The face-on wind model at the same pericentric density meanwhile produces consistently higher HI surface density, consistent with Figure \ref{fig:HI}. Also shown is a substantially higher density \verb|MaxWind| model, which truncates the disk at much smaller radius. 

We should emphasize here that our goal was to match the observed truncation radius, and not the broader HI profile. For the former goal, we considered where the column dropped to $10^{-19}$ cm$^{-2}$, a probe of compression and oblation of gas by ram pressure stripping. The broader profile however ought to be sensitive to the details of compression-induced star formation, as HI may be processed into stars, outflows and ionized gas, dropping the observed column deeper into the disk. This could perhaps even account for the steeper falloff of the leading edge, beginning at $r \sim 4$ kpc, which our simulations do a poor job recreating. 

Beyond the leading edge, we find the simulated gas disk extends further, hitting the column threshold of $10^{-19}$ cm$^{-2}$ beyond 7 kpc in the remaining three quadrants. Comparing to the observed profile, the southeast quadrant (bottom-left) features the closest match, ignoring the 30-Doradus anomaly in the inner profile. In the southeast quadrant leading in to the MS and bridge, the simulated profile fails to reproduce the high column at large radii seen in the observed system, which is expected since interactions with the SMC are not simulated.

\subsubsection{Time Evolution of Leading Edge Truncation Radius}
\stdFullFig{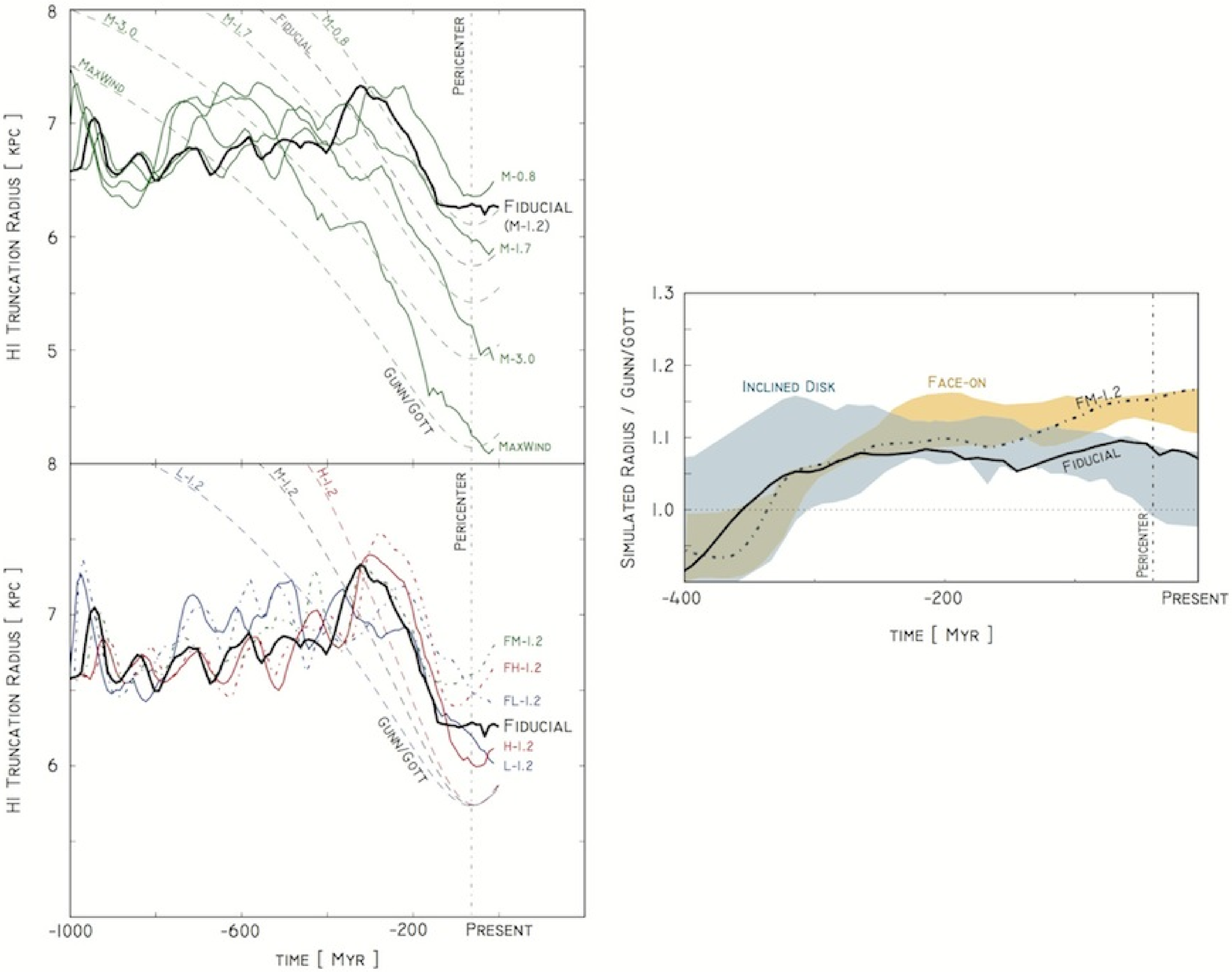}{The top left panel shows the evolution over time of the leading edge truncation radius, found from LOS column maps, in our simulations. The thick black lines denotes our fiducial simulation, with a pericentric density $n_p = 1.2 \times 10^{-4}$ cm$^{-3}$, while green solid lines show a variety of other pericentric densities. Dashed lines show the instantaneous stripping radius predicted by the GG model. The bottom-left panel shows the same format as the top plot, but for models with \emph{identical} pericentric halo densities but different exponential falloffs. The red curves are for the hi core, concentrated model, whereas the blue curves are for lo core, extended model. Also shown as dash-dot lines are so-called ``face-on'' runs, where the wind was launched perpendicular to the galaxy's disk plane, rather than as it would be experienced by an orbiting LMC. Finally the right plot shows the ratio of the simulated truncation radius to that predicted by GG. Blue and gold swatches denote the range spanned by all full-orbit- and face-on-wind runs, with the fiducial halo models highlighted in black.}{fig:time-radius}{.5}

We next explored the time-evolving nature of the truncation radius, found from mock LOS HI column maps such as in Figure \ref{fig:HI}, by again fitting the radius at which the gas column density drops to $10^{19}$ cm$^{-2}$ and then fitting an ellipse whose shape comes from assuming circular contours in the LMC disk plane. Figure \ref{fig:time-radius} shows how this fitted truncation radius evolves over the final gigayear of the LMC's orbit, across a suite of simulations, and compares this to the radius predicted by the GG model assuming instantaneous stripping. 

The top-left panel of Figure \ref{fig:time-radius} plots the fiducial simulation along with a host of higher and lower density halo models, all from the ``mid-core'' set (i.e. MW gas halos with the same core radius and density, but different exponential falloffs). All runs exhibit oscillations at early times, with amplitudes $\sim10\%$ of the HI truncation disk radius. These oscillations are leftover from our initial conditions, and are similar to those found in runs devoid of radiative cooling in T09. When the wind becomes sufficiently strong as to alter the observed truncation radius, there is an initial rise in the truncation radius. This is due to vertical compression of the disk, which causes more gas to lie above our $\rho = .03 \; {\rm cm}^{-3}$ density cut --- since similar plots that did not make such a cut do not exhibit the same spike. From there, the truncation radius shrinks rapidly as the LMC approaches pericenter, its slope in the radius-time graph consistent with the analytic GG prediction, though its observed value is about $5\%$ larger than the model predicts for mid-density runs (such as the fiducial case). Some of the runs, the fiducial run among them, then pull up further from the GG prediction near present day, which we discuss more while describing the next panel.

The bottom-left panel of Figure \ref{fig:time-radius} shows six runs with an \emph{identical} MW halo density at pericentric passage. An instantaneous application of the GG model at pericenter would thus predict an identical truncation radius across these models. Indeed runs with three different central halo densities and disparate $\beta$s --- \verb|H-1.2|, \verb|Fiducial|, and \verb|L-1.2|, in red, black and blue, respectively --- do not show a significant difference in the inferred present-day truncation radius. This result illustrates that the LMC's present day truncation radius is indeed a \emph{localized} probe of the MW gaseous halo's density. 

Figure \ref{fig:time-radius}'s bottom-left panel also plots hi-, mid- and lo- core runs at the fiducial $n_p$ for simulations with a face-on wind. While the early evolution for all three core models agrees well, the inferred truncation radius departs sharply from the full-orbit winds at late times, with the truncation radius rising rapidly in two of the three face-on wind runs. Although we do not explore the cause of this disparity in detail, it seems the symmetric nature of the wind perpendicular to the galaxy plane has sustained and enhanced, rather than damped-down oscillations in the inferred HI truncation radius.

The final, right panel of Figure \ref{fig:time-radius} plots the ratio of the simulated truncation radius to that predicted by GG, to quantify the disparity. Blue and gold shaded regions show the bounds within which all of our simulations lie, across various core concentration and pericentric density choices, for the full-orbit and face-on winds respectively. From this we find that the full-orbit simulations produce a truncation radius $\sim 5\%$ higher than GG across all runs, with a comparatively small scatter, whereas the face-on wind leads to radii over 10\% higher than the analytic model. These oscillations have a period of roughly 200 Myr, whereas the period of epicyclic oscillations in the disk is $\approx 450$ Myr at 6 kpc.

\stdFullFig{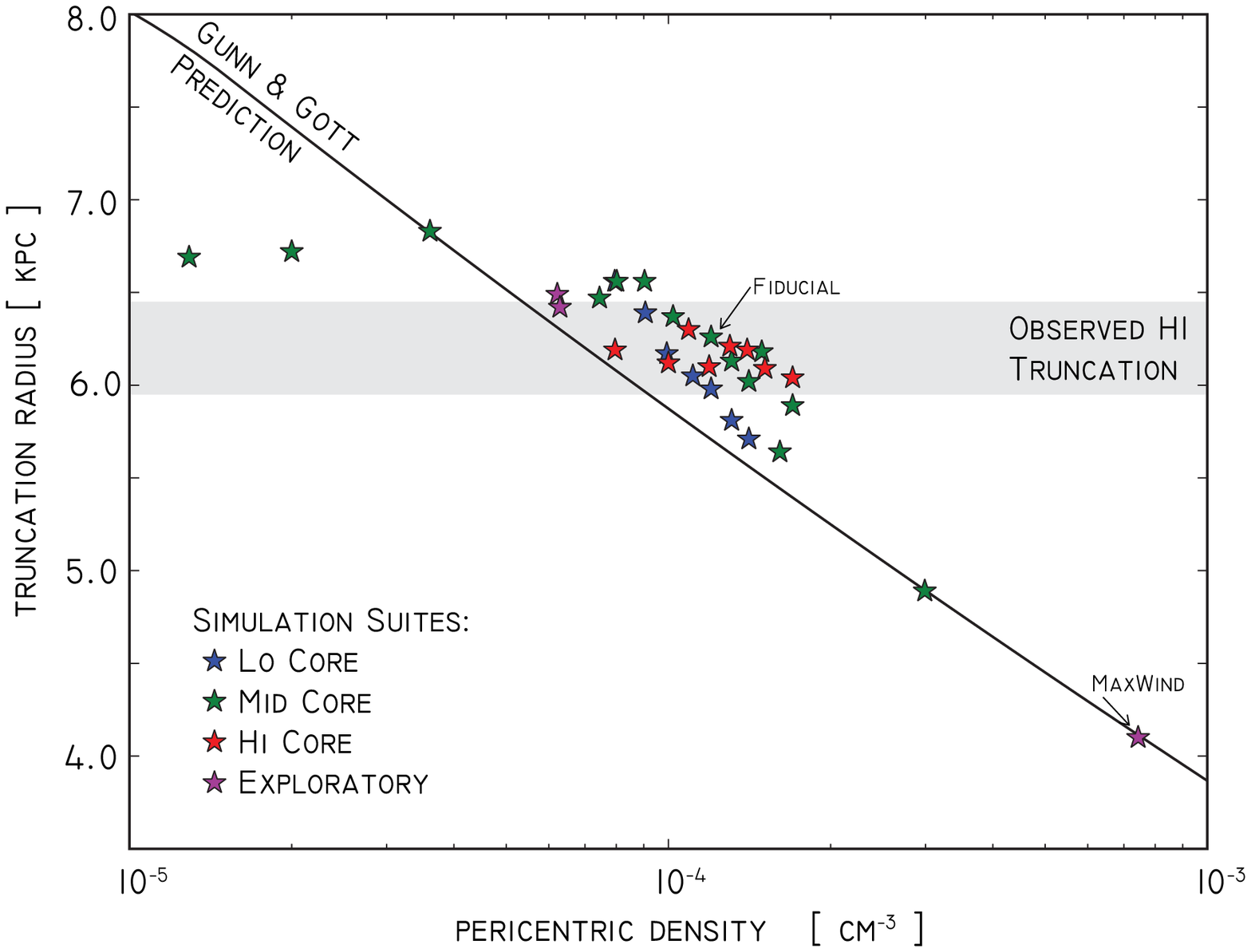}{The leading edge truncation radius, as inferred from mock HI observations such as in Figure \ref{fig:HI}, as a function of each simulation's pericentric MW gaseous halo density, $n_p$, for our entire suite of simulations using our standard LMC model described in Section \ref{sec:lmc-model}. Stars denote each simulation, with orbit models spanning two orders of magnitude in $n_p$, with a denser sampling where the simulation's results well matched the observed leading edge truncation radius (horizontal dashed line, with grey region denoting a $3-\sigma$ confidence region). In addition to sampling across pericentric densities, we also explored whether or not the broader halo profile impacted the observed radius, with the lo-, mid-, and hi-core models, displayed here in blue, green and red respectively (see Section \ref{sec:param-choices}). Although trends within and among these three classes of halo model are evident here, they exist within the general scatter, suggesting the observed truncation radius is indeed a clean probe of the local MW gaseous halo density at $r = 48$ kpc from the Galactic center.}{fig:n-r}{.8}

\subsubsection{Comparisons between GG toy model and simulations}

To move towards a quantitative prediction for the MW's diffuse CGM density, we next sought to quantify the inferred HI truncation radius across our entire suite of simulations. For each run, we produced the mock HI observations, as in Figure \ref{fig:HI} and once again fit a constrained ellipse to the column map where the surface density dropped to $10^{19}$ cm$^{-2}$. Figure \ref{fig:n-r} shows the results of this effort, plotting inferred truncation radius against each orbit model's pericentric MW halo density, $n_p$. From this we found a relation whose trend was well matched to the analytic prediction of GG, though near the observed truncation radius the distribution was consistently offset by $\sim 5\%$, just larger than the scatter in the results. Our ``mid-core'' halo models sampled roughly two orders of magnitude in $n_p$, with a dense sampling around the $1.2 \times 10^{-4}$ cm$^{-3}$ fiducial run, where the results best matched the observed truncation radius. The simulations break from the GG model at low density, since for our mock HI observations the density cut at $.03$ cm$^{-3}$ truncates even our initial disk near 7 kpc. In addition to sampling across $n_p$, we ran numerous simulations with more concentrated core densities (hi-core, red stars) and less concentrated (lo-core, blue stars). Although small trends are evident within and among these three model classes, they exist within the general scatter, suggesting the LMC HI truncation radius along the leading edge is indeed insensitive to the broader halo gas profile, and is thus a clean probe of the gas density at $\approx 48$ kpc. This conclusion stands in stark contrast to the orbital scenario explored by \cite{Mastropietro2009} whose lower-speed orbital scenario suggested the stripping process probed the broader gas profile over a longer duration of the LMC's orbit. A pair of our exploratory runs (T-6 and T-6X, purple stars) really serve to emphasize this locality result: these two models were both chosen to produce an $n_p$ that would truncate the disk at exactly 6 kpc, as predicted by the GG toy model. But they were chosen from opposite ends of the $\beta$-model parameter space. Despite this, they appear nearly overlapping in Figure \ref{fig:n-r}, with a separation much smaller than the overall scatter across our runs.

\stdFig{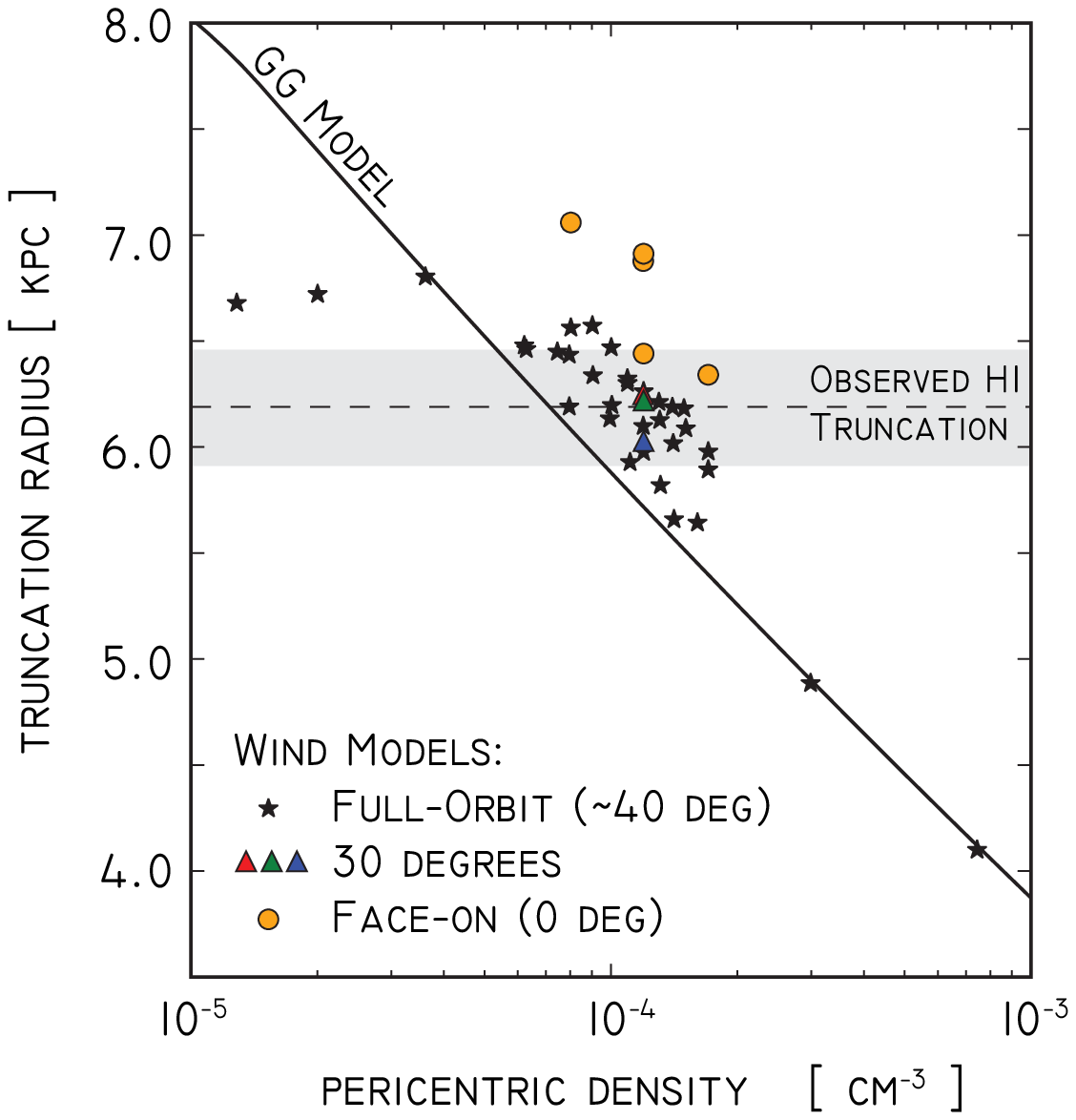}{A comparison of runs by the angle between the wind and disk plane's normal vector, $\psi$, plotting quantities as in Figure \ref{fig:n-r}. Black stars denote runs where the wind velocity mimics the orbit of the LMC, with $\psi(t) \sim 20^{\circ} \to 50^{\circ}$, with a pericentric value $\sim 40^{\circ}$. Yellow circles denote face-on runs, with $\psi = 0$ at all times, even as the wind speed and density vary as in the full-orbit runs. Finally, triangles show three runs at an intermediate, fixed $\psi = 30^{\circ}$. Blue, green and red markers denote IL-1.2, IM-1.2, and IH-1.2, for our three core density models with $n_p = 1.2 \times 10^{-4}$ cm$^{-3}$. While we find the face-on wind produces consistently higher inferred truncation radii at present day (due to the oscillations observed in Figure \ref{fig:time-radius}), both the 30-degree and full-orbit ($\psi_p \approx 40^{\circ}$) runs show excellent agreement in $R_t$, well within the scatter of our simulations. Thus we conclude the exact nature of disk orientation is of minimal importance, so long as the wind is not face-on.}{fig:inclined}{.7}

The right panel of Figure \ref{fig:time-radius} suggests that the wind's angle of approach (which we denote $\psi$ for brevity) might have a measurable impact on the implied MW halo density. Here $\psi(t)$ is the angle between disk plane's normal vector and the wind velocity vector.  In the majority of our simulations this quantity varies over time as the LMC orbits the MW. We assume the angular momentum vector is stable for the final 200 Myr of the orbit \citep[consistent with][]{Besla2012}, when the ram pressure ramps up to where it measurably impacts the disk gas. Under this scenario $\psi$ begins at $20^{\circ}$, climbs past $40^{\circ}$ at pericenter, and finishes off at the observed present value of about $50^{\circ}$. For our face-on simulations, by construction, $\psi(t) = 0$, even as the wind's speed and density evolve as in the full-orbit runs. These face-on runs showed the $\psi = 0$ runs have a systematically higher offset from the GG model, due to larger oscillations in the disk radius post-stripping than the full-orbit $\psi_p \sim 40^{\circ}$ runs. This suggests the disk's instantaneous truncation radius at present day may be sensitive to this angle. Thus we were led to perform a few simulations with a constant intermediate wind orientation of $\psi = 30^{\circ}$. We ran three simulations, all at the fiducial $n_p = 1.2$ cm$^{-3}$, across our three core models. These inclined wind runs are designated \verb|IL-1.2|, \verb|IM-1.2| and \verb|IH-1.2|. Figure \ref{fig:inclined} plots these runs against both the full-orbit  and face-on simulations. We see that the $30^{\circ}$ inclined wind runs across the three core models agree almost precisely with their full-orbit counterparts. This suggests the symmetry of the face-on run leads to enhanced oscillatory behavior qualitatively distinct from runs at any intermediate inclinations, and that our results remain insensitive to the exact value of the wind angle, $\psi$.

\stdFullFig{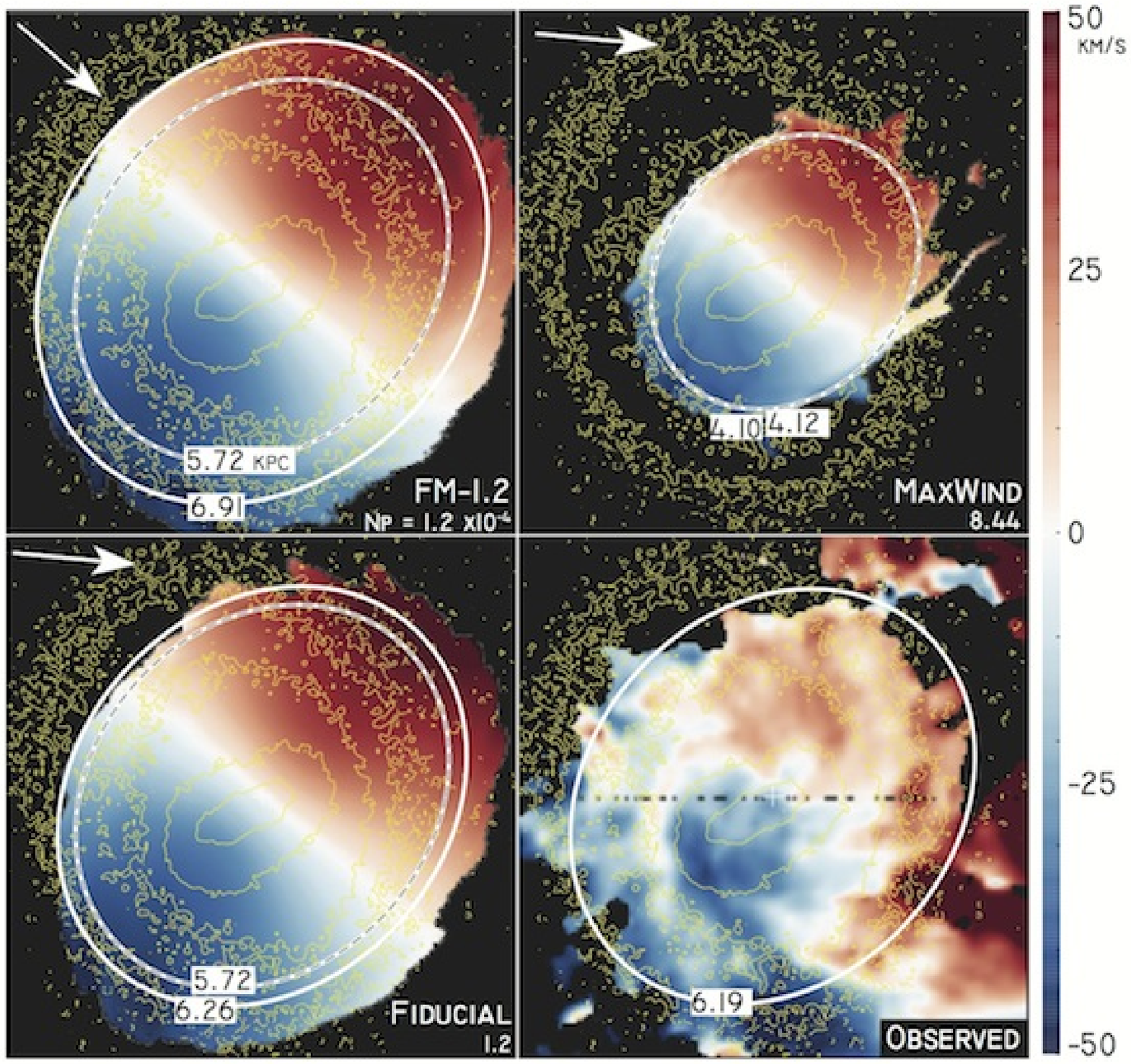}{Comparison of simulated HI radial velocity from mock-HI observations to the observed HI radial velocity (see text for explanation of proper motion corrections). Red and blue indicate motion away from and towards the observer, whereas white indicates gas with no radial velocity. Yellow contours again over plot the \emph{observed} stellar profile \citep{vanderMarel2001}, and white and dashed-line ellipses denote the truncation radius fitted to the HI column profile and predicted by the GG model, respectively. The simulations shown here are for the face-on and fiducial runs with $n_p = 1.2 \times 10^{-4}$ cm$^{-3}$ (top-left and bottom-left) and the ``max wind'' scenario with $n_p = 8.44 \times 10^{-4}$ cm$^{-3}$ (top-right).}{fig:LOS-velocity}{.6}

\subsubsection{Influence of Ram Pressure Stripping on Velocity Field of HI disk}
\label{sec:hi-velocity-map}

Beyond understanding how RPS alters the spatial distribution of the LMC's HI, our simulations allow us to study the gas's velocity structure. Figure \ref{fig:LOS-velocity} again provides mock HI observations, this time of the radial (LOS component of) velocity, as well as the observed HI radial velocity from \cite{Putman2003}. To form these images, we again sought to eliminate gas that is likely ionized, removing gas where $n < .03$ cm$^{-3}$. For each pixel we took a density-weighted average of $v_{\rm LOS}$, the component of the gas's velocity towards or away from the observer. Finally, we again eliminated from the map any region where the gas surface density fell below $10^{19}$ cm$^{-2}$. 

For the observed HI, the raw $v_{\rm LOS}$ data needed to be corrected for both the LMC's proper motion through space (relative to the Sun) and the disk plane's solid body motion about its center of mass. Only then could a fair comparison be made to our simulated disk which sits at rest in the ``wind tunnel''. To achieve this, we subtracted velocity corrections using machinery laid out in vdM02 and \cite{vanderMarel2001}, described in Appendix \ref{sec:velocity-transforms}. 

For reference, Figure \ref{fig:LOS-velocity} also plots the observed stellar distribution, the truncation radius found as in Figure \ref{fig:HI}, and the GG prediction for the truncation radius. Three simulations are shown: our fiducial (\verb|M-1.2|) run, a face-on wind run with an identical halo density model to the fiducial (\verb|FM-1.2|) and our highest density wind run, \verb|MaxWind|. For all three runs, regardless of wind direction or strength, the ram pressure fails to strongly perturb the basic gradient of a circularized, rotating disk. In contrast, the observed radial velocity in the disk plane is rife with substructure, most notably a swirl pattern with several arm-like structures. Because this observed velocity map's appearance depends on accurate parameters for the LMC's three-dimensional position, orientation and bulk motion, we varied these parameters systematically and found no choice of parameters that could severely diminish these swirls. We conclude the LMC disk plane indeed features a rich velocity structure unexplained by its proper motion, solid body rotation or circular motion. Our simulations also suggest that RPS cannot be responsible for this irregular motion to any appreciable extent.

\subsection{Implications of an LMC disk with a Higher Initial Gas Fraction}
\label{sec:gas-rich}

\begin{table*}[p]
\tiny
\renewcommand{\tabcolsep}{11pt}
\begin{center}\begin{tabular}{lllllllll}
\multicolumn{9}{c}{\textbf{Gas Rich Disk Simulations}} \\
\hline \hline 
$n_0$	& $r_c$	& $\beta$		& $P_p$			& Total Mass	& $n_p$			& $R_{\rm GG}$	& $R_T$	& Name		\\
\hline 
0.139	& 0.100	& 0.323		& 4.07e-13		& 6.46e+10	& 3.50e-04		& 6.63			& 7.32				& GL-3.5		\\
0.139	& 0.100	& 0.316		& 4.65e-13		& 7.66+10		& 4.06e-04		& 6.42			& 6.51				& GL-4.0		\\
0.139	& 0.100	& 0.309		& 5.23e-13		& 8.71+10		& 4.54e-04		& 6.21			& 6.76				& GL-4.5		\\
0.139	& 0.100	& 0.278		& 9.29e-13		& 1.69e+11	& 8.06e-04		& 5.32			& 5.75				& GL-8.0		\\ 
0.139	& 0.100	& 0.275		& 9.88e-13		& 1.80e+11	& 8.52e-04		& 5.23			& 5.36				& GL-8.5		\\ 
0.139	& 0.100	& 0.272		& 1.05e-12		& 1.92e+11	& 9.01e-04		& 5.15			& 5.21				& GL-9.0		\\ 
\hline
0.46		& 0.35 	& 0.559 		& 1.39e-13		& 1.25e+10	& 1.21e-04		& 8.39			& 8.73				& GM-1.2		\\
0.46		& 0.35 	& 0.486 		& 4.07e-13		& 4.25e+10	& 3.51e-04		& 6.62			& 7.35				& GM-3.5		\\
0.46		& 0.35	& 0.477		& 4.65e-13		& 4.96e+10	& 4.01e-04		& 6.41			& 7.01				& GM-4.0		\\
0.46		& 0.35	& 0.469		& 5.23e-13		& 5.69e+10	& 4.52e-04		& 6.22			& 6.83				& GM-4.5		\\ 
0.46		& 0.35	& 0.430		& 9.29e-13		& 1.11e+11	& 8.03e-04		& 5.32			& 5.88				& GM-8.0		\\ 
0.46		& 0.35	& 0.426		& 9.88e-13		& 1.19e+11	& 8.52e-04		& 5.23			& 5.99				& GM-8.5		\\ 
0.46		& 0.35	& 0.422		& 1.05e-12		& 1.28e+11	& 9.04e-04		& 5.14			& 5.88				& GM-9.0		\\ 
\hline 
0.810	& 0.800	& 0.630		& 4.07e-13		& 3.18e+10	& 3.52e-04		& 6.62			& 7.10				& GH-3.5		\\
0.810	& 0.800     & 0.620		& 4.65e-13		& 3.71e+10       & 4.03e-04		& 6.42			& 6.96        			& GH-4.0		\\
0.810	& 0.800     & 0.610		& 5.23e-13		& 4.27e+10       & 4.55e-04		& 6.23			& 6.91        			& GH-4.5		\\
0.810	& 0.800	& 0.563		& 9.29e-13		& 8.19e+10	& 8.02e-04		& 5.32			& 5.96				& GH-8.0		\\ 
0.810	& 0.800	& 0.558		& 9.88e-13		& 8.80e+10	& 8.52e-04		& 5.23			& 5.87				& GH-8.5		\\ 
0.810	& 0.800	& 0.554		& 1.05e-12		& 9.46e+10	& 9.06e-04		& 5.16			& 5.80				& GH-9.0		\\  \\
\label{tab:gas-rich-sims}
\end{tabular}\end{center}
{ \small \textbf{Table 5:} Our 19 gas-rich LMC disk model simulations (all prefixed with G). These runs are grouped by MW gaseous halo core density as in Table \ref{tab:all-sims} ( GL-, GM- and GH- runs) with various $\beta$ chosen to explore a range of pericentric densities. See Table \ref{tab:all-sims} for a description of column quantities.}
\end{table*}

\stdFullFig{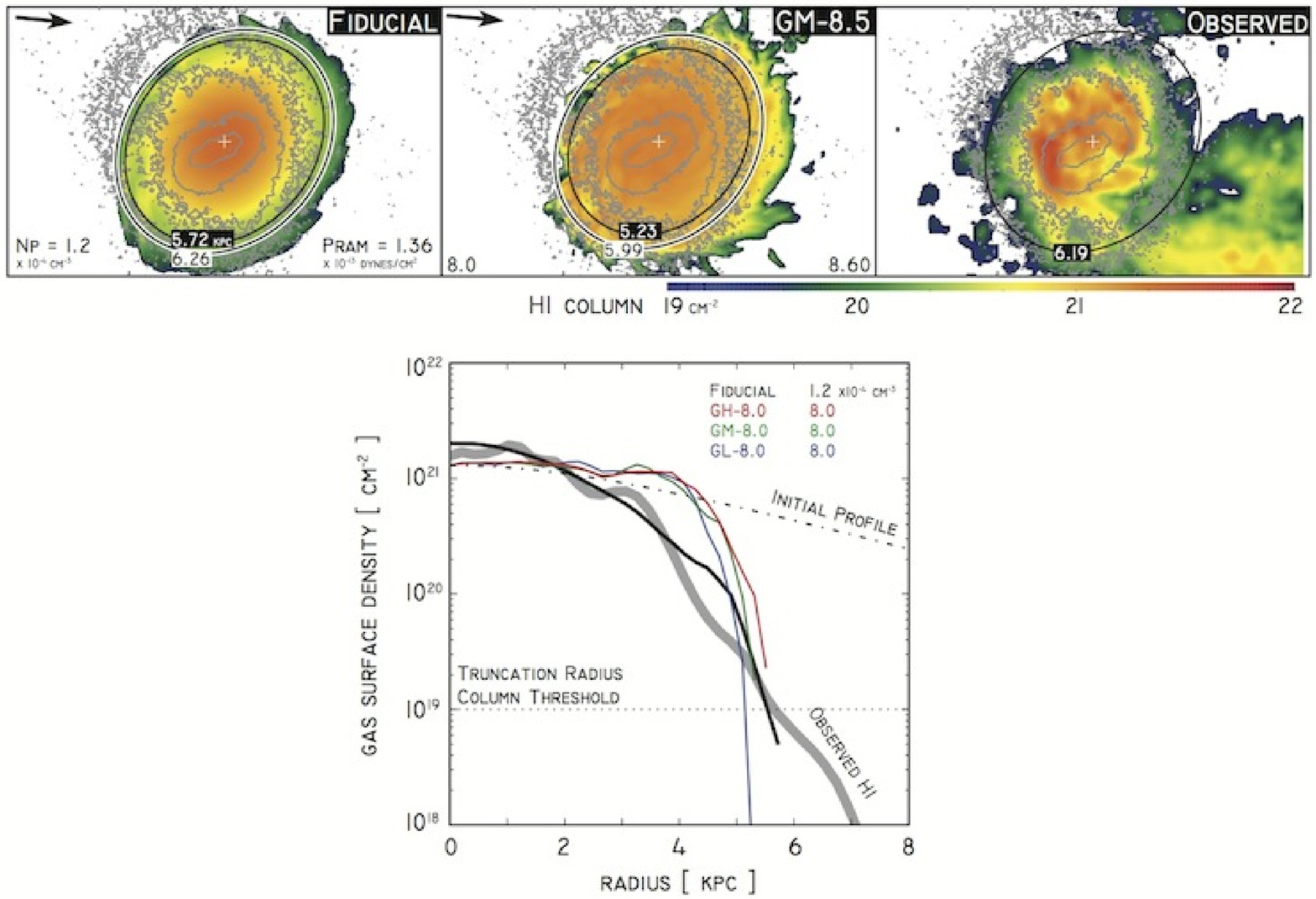}{Results for our gas-rich LMC disk model simulations, which began with a higher initial gas mass of $M_{\rm gas, tot} \approx 1.8 \times 10^{9} M_\odot$. \textbf{Top:} mock HI column map as in Figure \ref{fig:n-r} for the new gas-rich fiducial halo model run, GM-1.2 (center panel), compared to the fiducial simulation, M-1.2 (left), and observed HI distribution (right). Producing $R_t \sim 6$ kpc now requires a MW diffuse halo density at 50 kpc a full factor of 8 higher than for our standard disk model. This more violent scenario leads to a splotchier disk edge, which appears to match the observed profile better. However, ram pressure fails to alter the the gas column at intermediate radii, which remains noticeably higher than the observed system's profile, as the \textbf{bottom panel}'s plot of the leading edge surface density profile (as in Figure \ref{fig:surf-profiles}) shows. This panel also plots the fiducial simulation (black line) which features a falloff more in line with the observed system. These gas-rich results would require an unrealistically high amount of SF and feedback to bring this profile in line with current observations, suggesting such a pre-infall LMC HI disk is rather far-fetched.}{fig:gas-rich}{.6}

\stdFullFig{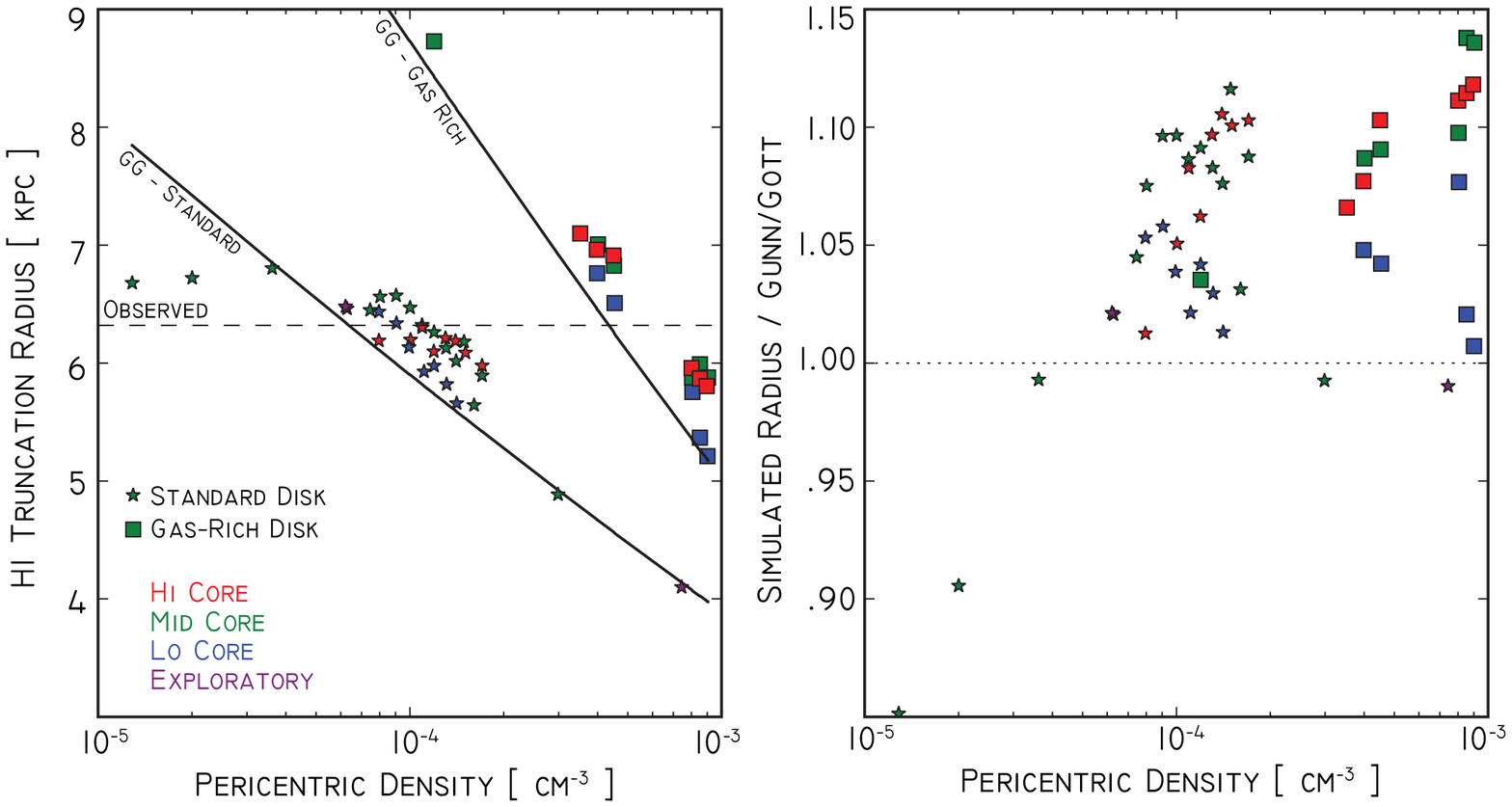}{A comparison of our gas-rich initial LMC disk simulations (squares) to our standard disk model simulations (stars). The \textbf{left panel} plots HI truncation radius, $R_t$, versus pericentric halo gas density, $n_p$, as in Figure \ref{fig:n-r}. The GG model is shown now for both sets of ICs (two black lines). The observed HI truncation radius is also shown as a dashed line. The systematic offset seen in the standard runs is again observed in the gas-rich set, with an equivalent offset in $n_p$.  From the simulations we find the gas-rich model requires a MW halo gas density $\approx 7$ times higher than the standard runs at $\sim 50$ kpc. The \textbf{right panel} takes a ratio of $R_t$ to the GG prediction across all simulations, with unity indicating a perfect match (dotted line). The gas rich runs feature an offset statistically consistent with the original runs. This agreement is strong evidence for the choice of $\alpha=1.0$ in Equation \ref{eq:GunnGott}, since models with a diminished gas self-gravity ($\alpha < 1.0$) provided a worse match with disparate offsets between gas-rich and standard runs. }{fig:gas-rich-n-r}{.7}

In this section we consider alternative models for the LMC's gaseous disk profile, and explore how they inform the implied CGM density. The LMC gas disk model chosen in Section \ref{sec:lmc-model} represents perhaps a lower-limit on the LMC disk's gas fraction prior to its recent pericentric passage. Alternative models that still obey the mass constraint of K98 are possible, producing gaseous and stellar disk properties (e.g. extent relations and gas fractions) consistent with observations of isolated late-type dwarf systems \citep{Swaters2002,Begum2005,Kreckel2011}. As the gas mass increases, the profiles do a consistently \emph{worse} job matching the LMC's present-day observed surface density profile, producing columns too low at its center with too shallow a radial falloff. However, ram pressure dynamics, including contour compression and enhanced SF can perhaps alleviate these issues at late times, especially since the headwind required to truncate the HI disk at its observed extent grows increasingly strong for these alternative models. The model we consider boosts the initial LMC gas mass by over a factor of two, to $\approx 10^{9} M_\odot$. Recent UV absorption line data coupled with \verb|Cloudy| modeling of ionized gas in the MS suggest the amount of gas stripped from the MCs may exceed their current HI mass \citep{Fox2014}, so this scenario is not outlandish.

To explore the scenario of an HI-rich initial LMC, we formulated a second set of initial conditions with an identical stellar and DM distribution, but with a gas disk whose scale radius was twice as extended as the standard run's scale radius (and thus twice the stars): $a_{\rm gas} = 3.4$ kpc. We then set the mass normalization constant, $M_{\rm gas} = 1.22 \times 10^{9} M_\odot$, which once-again ensured the dynamical mass within $R = 4$ kpc matched the constraint of K98. This result also fell on the relation of \cite{Swaters2002}, suggesting it as a plausible scenario for a gas-rich late-type dwarf galaxy before plunging into the MW halo and undergoing RPS. This produced an initial mock-HI surface density that was too low at the galaxy's center and much too high towards its edges in all directions, but this in itself does not preclude such a model, since the wind required to strip such an LMC to its present observed extent will be much higher density, and thus has the potential to dramatically reshape the entire HI profile.

We again performed a suite of simulations with MW gaseous halo models spanning a wide range of $n_p$, and also varied the halo's core density, as before. Table 5 summarizes these runs. Our $n_p$ choices are concentrated around where the predicted truncation radius would roughly match the present-day observed value $\sim 6$ kpc, but a run was also done at the fiducial density of $n_p = 1.2 \times 10^{-4} \; {cm^{-3}}$ for comparison. These runs are prefixed \verb|GL-|, \verb|GM-| and \verb|GH-|, with the first letter denoting their ``gas rich'' initial disk and the second their halo core model.

Figure \ref{fig:gas-rich} shows our canonical mock-HI observation as before for a representative gas-rich run, \verb|GM-8.5|, which produced a stripping radius in rough agreement with the standard disk's fiducial run and the observed distribution (also repeated in the new figure for comparison). Immediately evident is the higher gas column towards the disk edge, compared to the standard simulation and the observed map. This is seen most easily in the bottom panel of Figure \ref{fig:gas-rich}, where we again plot the averaged radial profile along the leading edge (as in Figure \ref{fig:surf-profiles}). In this sense, we find RPS cannot substantially alter the global gas profile, even when the wind has been ramped up to sufficient density to match the leading edge's extent. The rate of star formation required to alleviate this discrepancy is $\sim 10 M_\odot$/year, unrealistically high for the LMC, given observations. From visual inspection of the column maps that the gas-rich run's flocculent disk edge, shaped by the denser, more violent halo headwind, does a better job matching the observed profile. But here we caution that a more detailed treatment of ISM physics could easily reverse these fortunes. In all, the gas-rich runs provide a poor match to the observed HI content of today's LMC, and thus we deem this such high gas disk models to be unlikely. 

We can once again move towards an inferred halo density and compare to the standard runs by considering the whole suite of gas-rich simulations. Figure \ref{fig:gas-rich-n-r} accomplishes this, plotting the same derived quantities as in Figure \ref{fig:n-r}. This figure also compares both gas-rich and standard disk runs to the GG model's analytic prediction. The higher surface density gas disk shifts the GG model substantially, to give an inferred MW gaseous halo density $\sim 5$ times higher than the runs with the standard disk ICs. The simulated results follow this trend, once again exhibiting an offset from the analytic model and a scatter roughly equivalent to the original simulation suite. Figure \ref{fig:gas-rich-n-r} also plots the ratio of the simulated truncation radius, $R_t$, to the GG model, $R_{\rm GG}$. Here we see the gas-rich runs feature an offset from GG of $\approx 5\%$ on average, in good agreement with the standard runs. Note that this GG model includes the full self-gravity of the disk gas, typically ignored in RPS models of MW-sized systems with lower gas fractions. Including this self-gravity was captured by setting $\alpha = 1.0$ in Equation \ref{eq:GunnGott}. Other choices for $\alpha$, in particular the standard choice of $\alpha = 0$, led to larger offsets between the GG model and the simulations, most for the gas-rich runs. This suggests including the full self-gravity of the gas is the most accurate approach.

\section{Inferring the MW Gaseous Halo Density at LMC Distances}
\label{sec:error-anal}

The purpose of this section is to extend the analytic considerations of Section \ref{sec:analytic} with the simulation results of Section \ref{sec:sims} to produce a MW CGM density bounds implied from the truncation radius, $R_t$, of the LMC HI disk's leading edge. To do so, we first characterize the offset between the CGM density implied by the analytic GG model, and that found from simulations, as captured by the offset parameter $D$ in Equation \ref{eq:offset}. This together with a systematic enumeration of model uncertainties permits us to produce a quantitative prediction for $n_p( 48 \; {\rm kpc})$, the density of the MW diffuse gaseous halo at the distance of the LMC's most recent pericenter. We begin by describing the results of this effort, before going back to enumerate the various uncertainties we chose to model.

\subsection{A Density Constraint for the MW CGM at $r = 48$ kpc}

Using Equation \ref{eq:offset} together with samples from the error space of its various parameters (described below) we obtained an implied MW CGM density of
\begin{equation}
n_{\rm MW Halo}( R = 48.2 \pm 2.5 \; {\rm kpc} ) = 1.1^{+.44}_{-.45} \times 10^{-4} \; {\rm cm}^{-3} \; .
\label{eq:money-shot}
\end{equation}

\stdFullFig{./sim-results}{TOP: Distribution of important LMC RPS model params from our full model incorporating uncertainties described in Section \ref{sec:error-anal} including offset from GG model found from our suite of simulations. Left to right and top to bottom, these are the observed HI truncation radius (modeled as a gaussian, though this choice does not strongly influence results); MW gaseous halo density at LMC's pericenter, $n_p = n(r_p = 48.2 \pm 2.5 \; {\rm kpc})$; and ram pressure experienced by the LMC at pericenter. To compute this final value, we also sampled uniformly from $n_0 r_c^{3\beta} \in [.005,.03]$, motivated by MB15's error space. The average of this inferred halo mass is a factor of three higher than the value found in the toy model of Section \ref{sec:analytic}. BOTTOM: Colored lines show our model's span of the $\beta$-profile's parameter space, compared to the results of MB15. Our full model calibrated to the simulations has pulled down the model's exponential falloff, $\beta$ from our earlier toy model, inferring an average gaseous halo model denser than the results of MB15.}{fig:sim-results}{.7}

Figure \ref{fig:sim-results} displays the results of our full model graphically. The upper panels display the spread in the sampled $R_t$ and computed values $n_p$, $P_p$ and $M_{\rm MW gas halo}$. The first quantity was directly sampled from a normal distribution $R_t \sim N( 6.2 \; {\rm kpc} , (.25 \; {\rm kpc})^2)$. To judge the importance of this choice, we also used a uniform distribution with an identical standard deviation, which had only a small effect on the derived shapes and spreads in our other quantities. The pericentric MW gaseous halo density, $n_p$ (and corresponding ram pressures) found by our full model are substantially higher than the results of Section \ref{sec:analytic}, a reflection of the offset we found in our simulations, thus implying a more gas-rich MW halo at $\sim50$ kpc than the toy model suggests, by a factor of two. We can go a step further, albeit with healthy dose of extrapolation. Section \ref{sec:mass-estimate} extrapolates this density result to compute a total mass for the MW CGM, and compares to recent models based on studies in emission and absorption.

The final lower panel of Figure \ref{fig:sim-results} shows the agreement between our full model results and those of MB15 in parameter space of the $\beta$-model for the MW halo. From this we find, across halo core density, our model predicts a consistently shallower falloff in halo density (smaller $\beta$), implying a denser CGM profile. This degree of disagreement is hardly surprising: the hydrodynamic interaction of RPS involves the entire multiphase CGM, not merely the hot, X-ray emitting component. Finally, our RPS studies, as well as previous work \citep[e.g.][]{Roediger2005} shows that (supersonic) RPS is insensitive to the temperature of the ambient CGM, whereas X-ray models rely on thermodynamic assumptions for the gas. However, given these considerations, the degree of overlap we do find with MB15 suggests both probes are converging on an accurate picture for the MW's diffuse halo gas at intermediate radii.

\subsection{Sampling the Offset and Model Uncertainties}

To produce the above density bound, we needed to calibrate the offset term $D$ in Equation \ref{eq:offset} and then sample $D$ along with a host of other uncertain parameters. We finish this section by enumerating the sample distributions for these parameters, appealing to our full suite of simulations presented in Section \ref{sec:sim-results} and observed properties of the LMC.

\paragraph{Simulation Scatter (D):} The scatter observed in the simulated disk truncation radius is primarily due to oscillations in the disk extent, as seen in Figure \ref{fig:time-radius}, which we choose to model as simply random noise, as Figure \ref{fig:time-radius} indeed shows the phase of the oscillations shifts considerably across runs with no clear trend. Our data suggested the offsets in logspace did not have a strong trend with truncation radius (a t-test for a correlation coefficient produced a p-value of 33\%) and are consistent with being drawn from a normal distribution (a Shapiro-Wilkes test yielded a p-value of 13\%). We thus assume $\{D\} \sim N(B,\sigma_D^2)$, where $B$ is the mean of the offset and $\sigma_D^2$ its variance. We implicitly assume the offsets are independent of other changes in the LMC orbit and disk structure parameters. We then used \verb|emcee| \citep{Foreman2013} to find the joint distribution of $B$ and $\sigma_D$, from which to sample moving forward. This model is a quick way to capture uncertainties in the simulations that include everything from the aforementioned oscillations to the crude nature of mapping a 3D turbulent gas structure to a single truncation radius at a snapshot in time.

\paragraph{Orbital Uncertainties:} Our work has relied on a fiducial orbit taken from a large sample of backwards-orbit integration calculations for the LMC/MW system that makes use of the latest proper motion measurements of K13. The integrations treat the LMC as a Plummer profile within an NFW profile for the MW halo, completely ignoring the SMC's gravitational influence; these assumptions do not appreciably degrade the calculation's integrity during the past $\sim$100 Myr during which the LMC undergoes its most recent pericentric passage \citep{Kallivayalil2013,Besla2007}. 10,000 MC drawings were made for each combination of $M_{\rm MW} \in \{1,1.5,2\} \times 10^{12} M_\odot$ and $M_{\rm LMC} \in \{.3,.5,.8,1,5,8\} \times 10^{11} M_\odot$, though the inclusion of multiple galaxy masses enhances the spread in orbital parameters by less than $5\%$. From these samples we find $r_p = 48.1 \pm 2.5$ kpc, $v_p = 340 \pm 19$ km/s and $t_p = 46.4 \pm 8.5$ Myr, where $t_p$ indicates time since pericentric passage. Our numerous simulations show the observed truncation radius is roughly independent of the broader halo gas density profile, and thus a well localized probe of pericentric halo conditions. Thus the uncertainty in $r_p$, the distance of perigalacticon from Sgr A*, will not impact our density measurement but rather directly translates to uncertainty in where within the MW halo that measurement was made. In contrast, uncertainty in $v_p$ is very important to $n_p$, since the GG model depends quadratically on velocity. We thus sample this value from the orbit calculations when constructing our final measurement, along with the simulation scatter. To judge the importance of the timing errors, we took measurements of the simulated truncation radius both 20 Myr beyond and before the present day of our simulations. This introduced negligible scatter, and thus we ignore timing errors. 

\paragraph{Observed Truncation Radius ($R_t$):} Section \ref{sec:analytic} describes how we inferred our measured $R_t = 6.2 \pm .25$ kpc. To reiterate, our measurement of the radius of the LMC's observed leading edge ignores the flocculent nature of the HI profile, punctured by holes caused by star formation, accretion flows and other phenomena of the turbulent ISM. We thus sample $R_t$ from a normal distribution with $\sigma = .25$ kpc, a conservative (large) estimate for accuracy of the leading edge's position.
As previously stated, we have neglected radiative cooling and star formation in this analysis as T09 found agreement within 10\% of GG when accounting for a multiphase ISM. We further note that supergiant holes exist in the ISM of the LMC with sizes of on average 0.5 kpc \citep{Kim2003}. This size scale is encompassed within the uncertainty of the truncation radius of the LMC outer disk.   
In this analysis we have also assumed that the MW CGM is smooth.  If instead the LMC were primarily stripped by a clumpier medium, the clumps would need to be $\sim$20 kpc in size in order to explain the roughly uniform truncation of the LMC's leading edge (which had an initial disk radius of $\sim$10 kpc to match the current distribution of stars). Encounters with smaller, dense clumps may lead to a similar effect as a multiphase ISM, further motivating our listed truncation radius errors.  

\paragraph{Stellar Surface Density: } The GG model relies on the stellar surface density distribution, which we treat as an exponential disk (see Equation \ref{eq:stellar}) with characteristic mass $M_{\star}$, radial scale radius $a_{\star}$. \cite{Roediger2005} found the vertical structure and thickness of the disk did not impact the stripping dynamics, and thus we ignore uncertainty in $b_{\star}$, the scale height of the physical density profile. The surface density is proportional to the stellar number density distribution. \cite{vanderMarel2001b} (Section 3.1) fit an exponential falloff to this number density to find a scale radius $r_d = 1.44^\circ$. Starting with the data in their Figure 4 we used \verb|emcee| to find a standard deviation in this value of $\approx.04^{\circ}$. This combined with the known 5\% uncertainty in distance to the LMC leads to a combined error in our stellar scale radius of $6\%$, or $a_\star = 1.7 \pm .1$ kpc. Uncertainty in the total stellar mass is dominated by uncertainty in the stellar mass-to-light ratio, which vdM02 gives as $M/L = 0.9 \pm 0.2$, and thus a $22\%$ error, or $M_\star = 2.7 \pm .6 \times 10^9 M_\odot$. We include an estimation of both these uncertainties by sampling from independent Gaussian profiles for both these variables. Sampling from these distributions holding all other variables fixed allows us to gauge how sensitive our results are to these uncertainties. Using the observed truncation radius of $6.2$ kpc and the analytic Gunn / Gott prescription, we find the uncertainty in the scale radius introduces a $4\%$ uncertainty in the implied pericentric halo density, $n_p$, while the uncertainty in mass adds $23\%$. The latter is unsurprising, since in the Gunn/Gott model $n_p \propto \Sigma_\star \propto M_\star$.

\paragraph{Disk Orientation: } The LMC's disk orientation (position and inclination angle) can potentially introduce errors on both the simulated and observed end of our considerations. Regarding the simulations, Section \ref{sec:sim-results} explored the role of disk-orientation in altering the LMC's present-day HI truncation radius. Specifically, Figure \ref{fig:inclined} showed runs with intermediate angles between the disk-plane's normal and the wind vector (neither especially close to face-on nor edge-on wind) featured truncation radii $R_t$ that evolved in a nearly identical fashion for a given halo model. This suggests uncertainty in the LMC disk plane's orientation and the direction of the LMC's pericentric velocity vector inferred from this via backwards-integrated orbits ought to have a negligible impact on the uncertainty in our model's mapping between pericentric halo density and observed truncation radius. Regarding the observed HI distribution, the disk plane's orientation also enters into our calculations when we fit a constrained ellipse to the leading edge column contour, to infer a truncation radius (specifically, it determines the ellipse's axis ratio and orientation). vdM02 places the uncertainty in the position angle at $10^{\circ}$ whereas the inclination angle's uncertainty is roughly half that. The spread in inferred truncation radius introduced by this is significantly smaller than uncertainty introduced by the LMC COM distance measurement, which is already in turn much smaller than the $\sigma = .25$ kpc we introduced into $R_t$'s observed value due to the flocculent appearance of the leading edge.

\stdFig{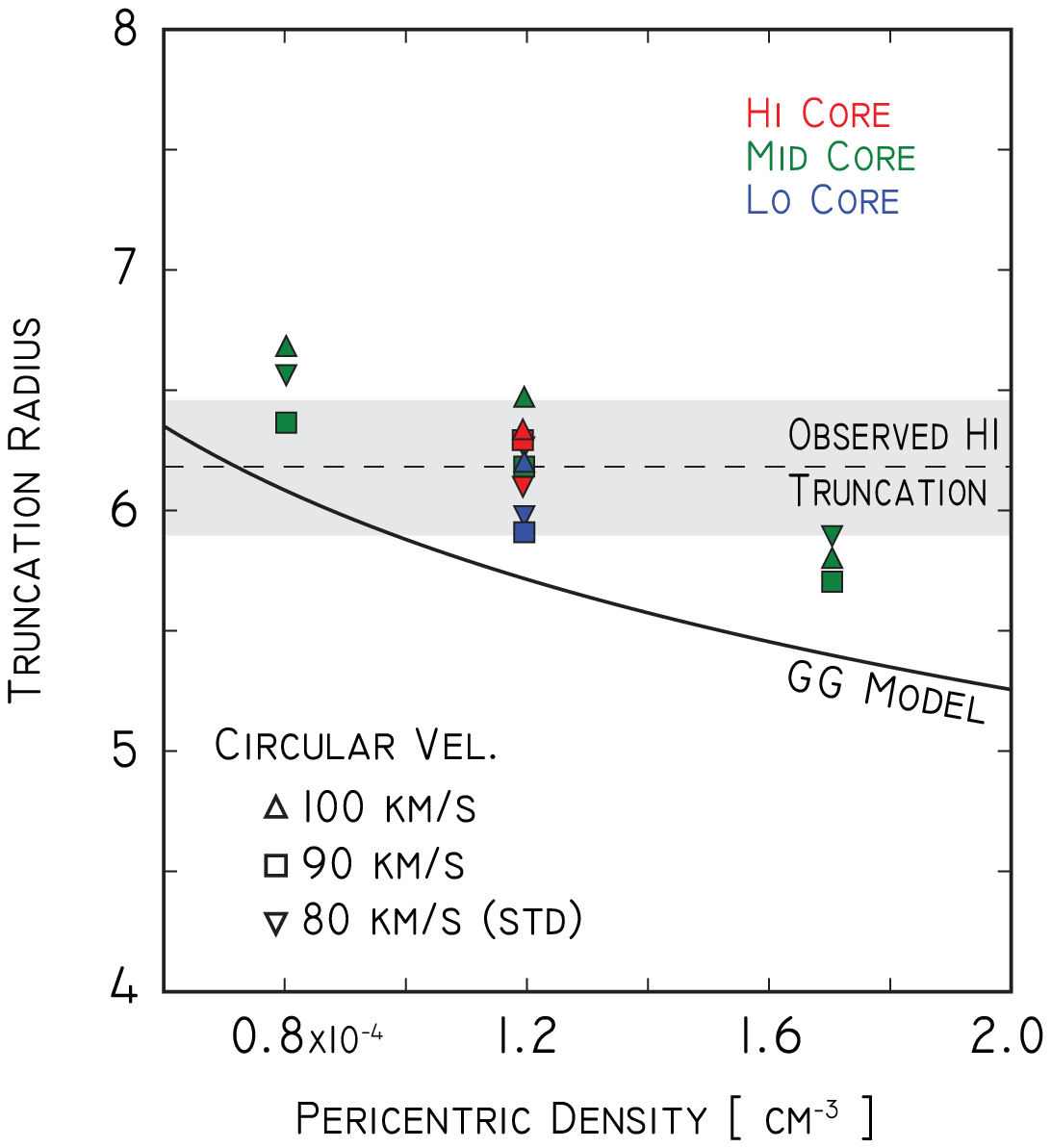}{A comparison of simulations with disparate LMC disk circular velocities (i.e. DM potentials), plotting quantities as in Figure \ref{fig:n-r}. Upward triangles, squares and downward triangles denote circular velocities of 100, 90 and 80 km/s, the final value representing our standard model throughout this work. Colors denote different MW gas halo core density models. Across $n_p$ and wind models, we fail to find a significant trend with $v_{\rm circ}$, and thus the DM potential does not appear to play a strong role in the setting the truncation radius.}{fig:dark-matter}{.7}

\paragraph{LMC Dark Matter Potential: } Our parameter choices for the DM halo were motivated by matching the observed LMC disk's rotation curve while also matching our Burkert profile closely to an NFW profile with concentration parameter $c=10$. Our simulated rotation curve peak of 80 km/s is admittedly on the lower end of the observations \citep[vdM14,][]{Olsen2011}. An alternative DM potential impacts the dynamics in a few notable ways. First, a more (less) massive LMC would make dynamical friction more (less) efficient, altering its orbital behavior. Our sampling of LMC/MW masses described above accounts for this uncertainty. Second, an enhanced (diminished) circular velocity  may make stripping more (less) efficient where the disk gas motion parallels the headwind (momentum-wise it has a ``head start''). And third, the gravitational restoring force of the DM potential may alter the required ram pressure for stripping at a given radius, an issue increasingly relevant in dwarf systems. The GG model was first developed to handle $\sim 10^{12} M_\odot$ late-type systems falling into clusters. For these systems, it's a good approximation to assume a (thin) stellar disk dominates the restoring force during RPS (see Section \ref{sec:gas-rich} for a discussion of the gas' self gravity). For the LMC, the vertical acceleration of the DM potential at the truncation radius $r \approx 6$ kpc is an appreciable fraction of the stellar surface density's acceleration field, which suggests DM could alter the predicted truncation radius. These latter two issues have thus far not been captured in our models nor our simulations, and could introduce additional uncertainty. To explore whether or not this was the case, we ran ten additional simulations: five with a rotation curve peaked at 90 km/s and five at 100 km/s. We chose three $n_p \in [ 0.8, 1.2, 1.7] \times 10^{-4}$ cm$^{-3}$ and at the median $n_p$ employed our three distinct halo core models. Figure \ref{fig:dark-matter} shows these results on our standard $R_{\rm trunc}$ vs. $n_p$ plot. For each of the five wind scenarios, the three disparate DM potential runs appear in a roughly random order, suggesting no strong trend with circular velocity (a trend would involve the symbols aligned as in the legend in the lower left-hand of the figure). Further, the scatter among these variations is less than the spread in observed truncation radius and that inherent to the simulations. Thus we ignore this uncertainty.

\begin{table}[tb]
\small
\begin{center}\begin{tabular}{llll}
\toprule
\multicolumn{4}{c}{\textbf{Inferred MW Halo Density, $n_p(48 \; {\rm kpc})$ }} \\
\midrule
Parameters Varied    	& Mean	& Median	& Std Dev   \\
\midrule
\midrule
Simulation Scatter 	& 1.066	$\times 10^{-4} {\rm cm}^{-3}$& 1.044	& .2214 	   \\
Stellar Surf Density	& 1.104	& 1.081	& .2294 		  \\
Truncation Radius  	& 1.087	& 1.047	& .2924 		  \\
Pericentric Velocity 	& 1.059	& 1.053	& .1270 		  \\
\midrule
Combined Errors    	& 1.118	& 1.030 	& .4950 \\
\bottomrule
\label{tab:errors}
\end{tabular}\end{center}
{ \small \textbf{Table 6:} Inferred MW gaseous halo density, $n_p$ at $r_p = 48.2 \pm 2.5$ kpc from Sgr A$^*$, from our GG model with an offset calibrated from our suite of wind-tunnel simulations. The top row incorporates only the scatter in the relation between the HI truncation radius, $R_t$,  and $n_p$ found from the simulations; the next includes only the mean offset from the simulations and spread in parameters for our modeled stellar disk; the next includes only spread in the observed $R_t$ (assuming a Gaussian spread, although this choice does not strongly influence the results); the next row from including all various LMC pericentric orbital speeds from our 10,000 orbit samples; and the last combining all these uncertainties.}
\end{table}

Table 5.5 summarizes results of the sampled models, including a breakdown of spread in all the parameters considered. From these results, we find the uncertainty in the observed truncation radius of the leading edge accounts for the most scatter in our results: $\sim 40\%$ of the variance. We emphasize that our choice of variance for this observed truncation radius was set to a conservatively large value of $.25$ kpc. Next, the uncertainty introduced by scatter in the simulated truncation radius (mostly due to oscillations of the disk edge) and imprecision in the stellar surface density profile each contribute roughly equally to the variance, $\sim 25\%$ each. Finally, the spread in pericentric velocity from our sampled orbits contributes only $\sim 8\%$ to the uncertainty in our inferred value of $n_p$.

\section{Discussion}
\label{sec:discussion}

\subsection{MW Gaseous Halo Mass Estimate}
\label{sec:mass-estimate}

Baryonic mass accounts for a fraction $\Omega_b / \Omega_m = .17$ of matter in the universe. But stars and the ISM in the disks of $L \sim L^*$ galaxies account for a far smaller share of their host halo's total mass \citep[e.g.][]{Behroozi2010}. Although galaxies may efficiently disperse the remaining baryons into the intergalactic medium (IGM), observations of that material have thus far failed to account for a sufficient mass in baryons  \citep{Prochaska2011,Cen1999}. The MW's CGM has often been invoked as a reservoir of these so-called ``missing baryons''. MB13's results suggests the MW hot halo accounts for $4.3 \pm .9 \times 10^{10} M_\odot$ or roughly 50\% of these baryons, though \cite{Gupta2012} suggest a much larger mass of hot gas. 

Surveys of $\sim L^*$ galaxies at low-redshift suggest a similar role for X-ray gas \citep{Anderson2013}. The recent COS-halos results of QSO absorption in UV suggests the CGM of $\sim L^*$ systems is indeed rife with baryons, though at far lower temperatures: potentially half the galaxy's mean baryon budget may exist as diffuse $T<10^5$ K CGM gas \citep{Werk2014,Tumlinson2013}; while warm-hot $10^{5} - 10^{7}$ K material could likewise harbor a signifiant fraction of mass \citep{Peeples2014,Tumlinson2011}.

Our work thus far has focused on inferring a local density measurement at the LMC's last pericentric passage, and the analysis of Section \ref{sec:sim-results} showed this measurement is largely insensitive to the broader halo profile. In this section, we consider the range of total CGM mass estimates implied by this density measurement. Such an exercise is indeed an extrapolation; in addition to invoking a specific density profile, we'll be integrating well beyond LMC distances, assuming a consistent exponential falloff and spherical symmetry.

For the density distribution we again assume a $\beta$-profile. For each $n_p$ sampled as in Section \ref{sec:error-anal}, we also randomly sample a halo core density, $n_0 r_c^{3 \beta}$, drawing from a uniform distribution of $n_0 r_c^{3 \beta} \in [ .005 , .003 ]$ corresponding to the range in MB15 though with the upper bound set a factor of three higher, since their work fails to probe cooler phases of the CGM. We then infer $\beta$ from this core density and the sampled pericentric density.

\stdFullFig{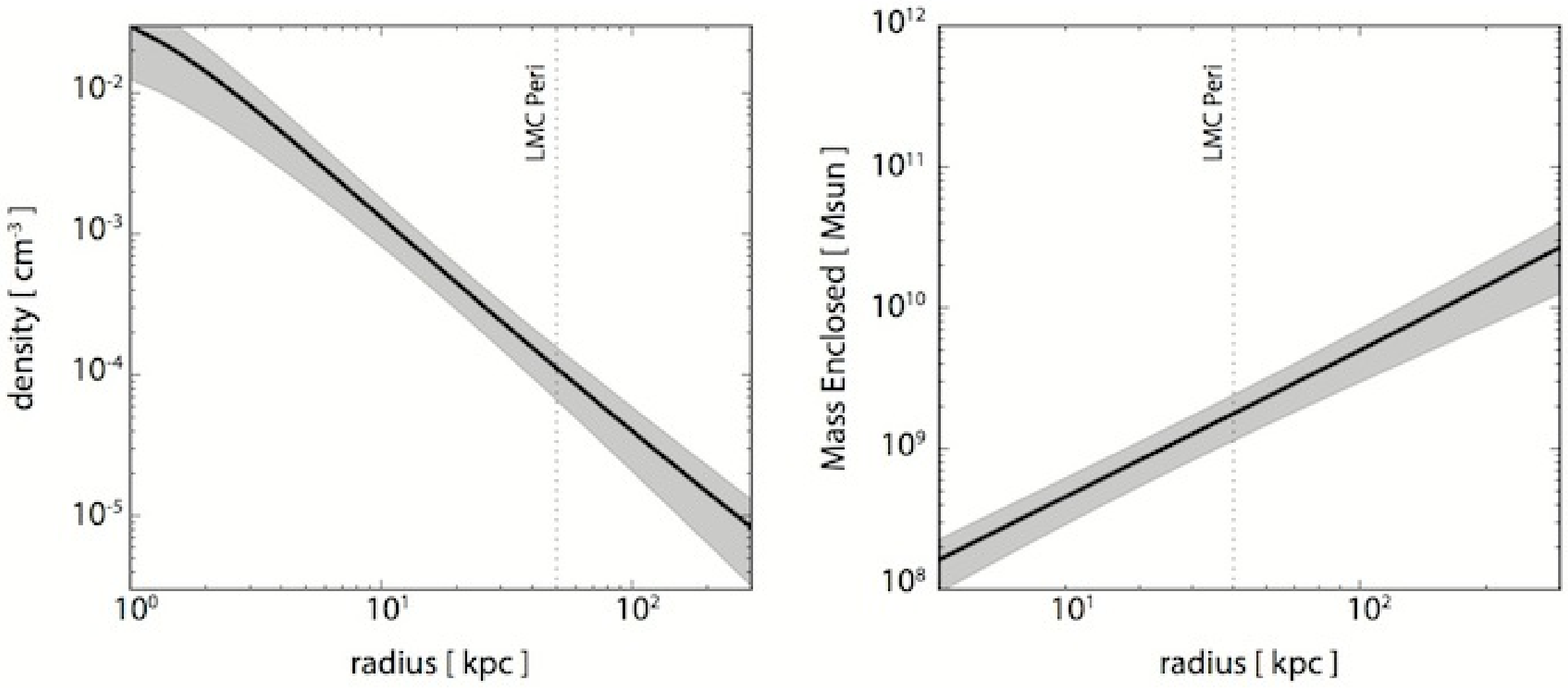}{Density and enclosed mass profiles for the MW CGM, generated by sampling $n_p$, the diffuse CGM density at the LMC's pericentric passage, from our stripping simulations, and then sampling $\beta$-profile core densities from $n_0 r_c^{3\beta} \in [.005,.03]$. Table 6 summarizes the mass enclosed and provides implied baryon fractions.}{fig:halo-profiles}{.65}

\begin{table*}[tbp]
\small
\begin{center}
\renewcommand{\tabcolsep}{6 pt}
	\begin{tabular}[b]{l|cccc}
		\hline \hline
		\multicolumn{5}{c}{\textbf{Extrapolated Gas Halo Mass}}	 \\
		\hline
										&	 \multicolumn{4}{c}{\textbf{$M_{\rm gas} / (.17 \cdot M_{\rm Halo})$}}							\\
	$M_{\rm gas}(300 \;{\rm kpc})$				&	$.7 \times 10^{12}$			&	$1 \times 10^{12}$			&	$1.5 \times 10^{12}$			&	$2 \times 10^{12}$	\\
		\hline
	$2.68^{+1.4}_{-1.4} \times 10^{10}M_\odot$	&    $21.58^{+10.9}_{-10.6}\%$ 		&	$15.11^{+7.6}_{-7.4}\%$		&	$10.07^{+5.1}_{-5.0}\%$		&    $7.55^{+3.8}_{-3.7}\%$	\\	
		\hline 
	\end{tabular}
\end{center}
{ \small \textbf{Table 6:} The total CGM mass enclosed within 300 kpc from our LMC stripping calculations. Also shown are the CGM's contribution to the Milky Way ``Baryon Budget'' defined as the halo's total mass multiplied by the cosmic mean baryon fraction, $\Omega_b / \Omega_m = .17$. From here we find the CGM likely contributes $\approx 5\% - 30\%$ of the MW's expected baryon mass. These results assume a spherical $\beta$-profile calibrated to the CGM density found at LMC pericenter.}
\label{tab:halo-mass}
\end{table*}

The left-hand panel of \ref{fig:halo-profiles} shows the average density profile and $1-\sigma$ bound at each radius from these samples. The right-hand panel integrates this profile to find the total mass enclosed by each radius. From here, we find the standard runs imply a total CGM mass of $2.68\pm 1.4 \times 10^{10} M_\odot$. To place this bound in context, we can normalize to the ``expected'' mass of baryons in the MW, by multiplying the MW's total halo mass by the mean cosmic baryon fraction, $\Omega_b / \Omega_M = .17$. For a MW mass of $1.5 \times 10^{12} M_\odot$ \citep[e.g.][]{Boylan-Kolchin2013}, our standard runs imply the CGM contains $\approx 10\%$ of the Galaxy's baryons. 

Table 6 quotes values at 300 kpc and also expresses these masses as a fraction of ``available'' baryons, i.e. $.17 M_{\rm halo}$ for a set of MW masses $M \in [ .7 , 1.0, 1.5, 2] \times 10^{12}$. From our work, the most reasonable estimate places the CGM baryon fraction at $\approx10\% - 20\%$ of the available baryons. In comparison, the COS-halos team finds anywhere from $25- 45\%$ in a cool, $<10^5$ K form, and warm-hot contribution of $5-37\%$. MB15 in contrast predicts a $\lesssim 50\%$ contribution from X-ray gas. Our result is on the conservative end of these predictions, despite adopting wide bounds in the halo model and probing the full temperature span of gas.

Although our work focused on a spherically symmetric $\beta$-profile, our analysis has shown that the currently observed truncation radius is sensitive only to the density at perigalacticon, where ram pressure is strongest, and thus our density bounds are consistent with a broad array of models for the diffuse CGM, so long as density tended to rise appreciably as the LMC moved deeper into the MW's potential. Alternatively, \cite{Tonnesen2008} examined RPS dynamics in a larger, forming cluster environment, where they found ram pressure experienced at a single radius from the cluster center can vary significantly. They found stripping to be a fast enough process in such an environment that truncation of the cluster galaxies was in fact a probe of dense pockets. If we assume our scaled down setting would obey similar dynamics, this implies our halo density measurement is in fact an upper limit, as the current LMC truncation radius could be set by the highest ram pressure pocket encountered by the cloud, rather than at perigalacticon.

\subsection{SMC Stripping Consistency Check}
\label{sec:smc-stripping}

\begin{table}[tbp]
\small
\renewcommand{\tabcolsep}{12 pt}
\begin{center}
\vspace{10 pt}
\begin{tabular}{lll}
\toprule
Quantity					&	Value			& Unit		\\	
\midrule
$M_\star(2\;{\rm kpc})$  		&	$3.1\times 10^{8}$	& $M_\odot$	\\
$a_\star$	 				&	$.65$		& kpc		\\
						&					&			\\
$M_{\rm gas}(2\;{\rm kpc})	$	&	$4.0\times10^8$	& $M_\odot$	\\
$a_{\rm gas}$				&	$.65$			& 			\\
$\Sigma_{\rm max, gas}$		&	$9.98\times10^{21}$	&$\rm cm^{-2}$	\\
						&					&			\\
$\sigma_{\rm DM}$			&	$27.5$ 			& km/s		\\
						&					&			\\
$r_{\rm SMC}$				&	$60$				& kpc 		\\
$v_{\rm SMC}$				&	$217$			& km/s		\\
\bottomrule
\label{tab:smc-params}
\end{tabular}
\end{center}
{\small \textbf{Table 7:} SMC model parameters, as explained in Section \ref{sec:smc-stripping}}
\end{table}

The LMC's companion galaxy, the Small Magellanic Cloud (SMC), provides another late-type dwarf traversing the CGM at high speeds at intermediate distances from the galactic plane. In this section we briefly verify that our results are consistent with the survival of the SMC's observed HI distribution.

The SMC system's position, orientation and motion are considerably less well-constrained than the LMC, rendering conclusions we arrive at here necessarily more speculative. 
In particular, the past pericentric approach of the SMC to the MW is highly uncertain as its orbit is strongly perturbed by the LMC.  
Furthermore, the system is surrounded by HI at appreciable column along all sight-lines, with no leading edge-analogue where RPS would be an obvious shepherd of HI (i.e. there is no location where gas is truncated relative to the stellar distribution as for the LMC).  
Thus RPS arguments are only useful in verifying that our density bounds allow for the survival of the observed SMC gaseous system. 

We employ the same analytic GG model as for the LMC. Table 7 summarizes all inputs for this model. We once again employ Equations 1 and 2 to describe exponential profiles for the stellar and gaseous disks. We adopt a gaseous mass $M_{\rm gas}(2 \; {\rm kpc}) = 4 \times 10^8 M_\odot$ and central surface brightness of $\Sigma_{\rm mas,gas} = 9.98 \times 10^{21}$ cm$^{-2}$ \citep{Bruns2005}, which for an exponential profile implies a scale radius $a_{\rm gas} = .65$ kpc. For the stars, we adopt a mass within the same radius of $3.1 \times 10^{8} M_\odot$  \citep{Stanimirovic2004,vanderMarel2009}. An accurate account of the gas' restoring force must include the SMC's DM mass, which is relatively large and concentrated. We adopt a dispersion velocity $\sigma_{\rm DM} = 27.5$ km/s \citep{Harris2006}, and follow \cite{Grcevich2009} by modeling its affect via
\begin{equation}
n_{p} v_{p}^2 \approx \sigma_{\rm DM}^2 n_{\rm gas}(R_{\rm Trunc})
\label{eq:dm-stripping}
\end{equation}
where the ambient gas' ram pressure is formed from $n_{p}$ and $v_{p}$, the CGM density and relative velocity at the SMC's pericenteric passage.
We examine the present properties of the SMC, which is moving with a Galactocentric velocity of $217 \pm 26$ km/s (K13) at a distance of $r_{\rm SMC} \approx 60$ kpc. 
For $n_{\rm gas}$ we take the SMC ISM's gas density at the disk's mid-plane (i.e. an upper bound). The final input is a truncation radius for the SMC's gaseous extent, 
which we set to 3.5 kpc, taken from the HI disk's observed angular extent \citep{Bruns2005}. This analysis implies a density upper bound just above $10^{-4}$ cm$^{-3}$.
This limit is consistent with the LMC fiducial case we've considered throughout this paper, but excludes the gas-rich scenario of Section \ref{sec:gas-rich}.

\subsection{Contributions to and Constraints from the Magellanic Stream}
\label{sec:mag-stream}
\stdFullFig{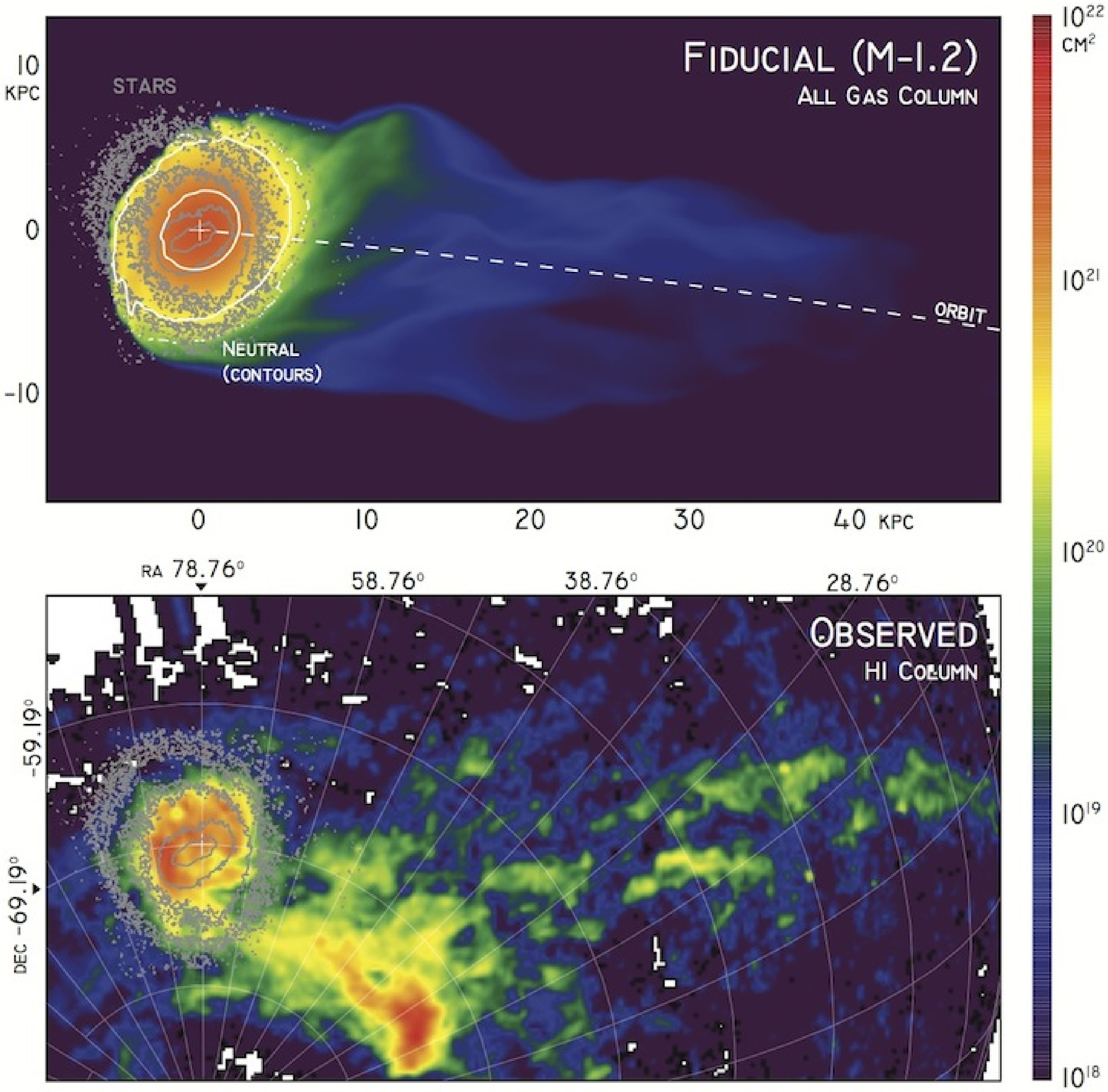}{Upper Panel: gas column density distribution of the simulated LMC after 1 Gyr of evolution. No density or temperature cuts are applied. 
Lower Panel: the true distribution of HI gas
in the Magellanic Clouds, Bridge and Stream in the same viewing perspective. Overplotted in grey scale is the distribution of stars in the LMC disk. 
The simulated material stripped from the LMC is of much lower column density ($< 10^{19}$cm$^{-2}$) than the bulk of the observed stream and follows 
the past orbit of the LMC on the plane of the sky, which does not trace the location of the Magellanic Stream.  The exception is material in the North-West quadrant of the LMC (upper right of
the LMC disk), which may be higher column. There is HI at high column observed at that location, which we predict should have originated exclusively from the outer disk of the LMC. }{fig:mag-stream}{.7}

The Magellanic Stream has itself been used to place constraints on the local halo gas density.  Observations of the radius and velocity width of clumps of HI in the 
Stream can be used to derive their internal pressure. If the clumps are assumed to be entirely confined by gas pressure (no contribution from magnetic fields or turbulence),
the necessary halo gas density can be estimated \citep{Stanimirovic2002, Murali2000}.  Assuming a halo temperature to $3.5\times 10^6$K, \citet{Anderson2010} place an upper limit on the 
hot halo electron density of $n(45 \rm{kpc}) < 9 \times 10^{-5}$ cm$^{-3}$.  Note, however, that the true distance to clumps in the Stream is likely 
significantly larger than 45 kpc \citep{Besla2012}.  \cite{Hsu2011} conduct a similar analysis for clumps at the tip of the Magellanic Stream assuming a distance of 120 kpc. 
They find that pressure confinement arguments require halo densities from $10^{-4.4}$ to $10^{-4.0}$ cm$^3$.  Such results imply a substantial flattening 
of the density profile at radii beyond 50 kpc, inconsistent with a $\beta$-profile. 

A pure ram pressure stripping origin for the Magellanic Stream has been explored by many authors \citep[e.g.,][]{Meurer1985,Moore1994,Heller1994,Mastropietro2005}. Such models naturally explain the apparent lack of stars in the trailing Magellanic Stream, but do not explain the existence of the Magellanic Bridge connecting the LMC and SMC or the existence of the Leading Arm (which leads the entire system).  
It is thus clear that tidal interactions between the LMC and SMC must play a dominant role in shaping the gaseous Magellanic System. However, this does not imply that ram pressure stripping provides a negligible contribution to the Stream.  Indeed our analysis illustrates that ram pressure will alter the structure of the LMC's gaseous disk. 
In particular, while the bulk of the stream has a very low metallicity (0.1 Z$_\odot$), consistent with an SMC origin \citep{Fox2013}, closer to the Clouds the metallicity increases to 0.5 Z$_\odot$ \citep{Richter2013}. These observations suggests that the Stream is being enriched, possibly from gas recently ram pressure stripped from the LMC.

We do not explicitly track the multiphase structure of the ISM in our simulations, instead we identify neutral HI by following only the evolution of gas in cells with densities above $\rho = .03$ cm$^{-3}$ and column densities of 10$^{19}$ cm$^{-2}$. 
Gas obeying these conditions is not found at large radii beyond the LMC disk, indicating that the LMC is not a dominant contributor to the Magellanic Stream.  
However, it is certainly possible that gas disobeying such conditions may in fact cool radiatively to form neutral HI in a full treatment of the multiphase ISM. 
As such, in Figure \ref{fig:mag-stream} we plot the column of {\bf all} gas that was once in the LMC disk in our fiducial simulation after 1 Gyr of evolution (regardless of density or temperature).  
The bottom panel illustrates the observed distribution of HI in the Magellanic System.
We find that the bulk of the gas removed from the LMC exists at low column densities ($< 10^{19}$ cm$^{-2}$). Furthermore, this material almost exclusively follows the past orbital trajectory of the LMC on the plane of the sky. 
As pointed out by \cite{Besla2007} and K13, HST proper motions rule out scenarios where the past orbital trajectory of the LMC is coincident with the location of the Magellanic Stream on the plane of the sky. It is instead offset by $\sim 10^\circ$.
Such offsets are natural consequences of the tidal stripping of rotating disks; since MW tides are too weak to affect the LMC, this results favors a scenario in which the Stream forms through tidal stripping of the SMC  \cite{Besla2012,Besla2010}.  

The exact contribution of this low column density material to the mass budget of the Stream is speculative, as it depends sensitively on the original extent of the LMC's gaseous disk and also the dark matter mass of the LMC, which controls how much gas can escape its potential well. 
For our fiducial simulation presented in Figure \ref{fig:mag-stream}, we find that $\sim 7 \times 10^6 M_\odot$ of gas (at any temperature or density) is removed from the LMC's gas disk outside a radius of 9 kpc (the nominal extent of the LMC's stellar disk). 
As such, we estimate that the LMC may contribute less than 1\% to the total mass budget of gas outside the Magellanic Clouds (the total gas mass, including the ionized component, is $2\times 10^{9} M_\odot$, assuming an average distance of 55 kpc \cite{Fox2014}).
If all of this material were neutral HI, it would contribute no more than 6.4\% of the mass budget of the head of the HI Magellanic Stream \citep[$1.1 \times 10^8 M_\odot$, assuming a distance of 55 kpc][]{Putman2003}. 

Note that, in our fiducial model, the total LMC mass within a radius of 100 kpc is $\sim 5 \times 10^{10} M_\odot$. Given the observed gas and stellar mass of the LMC, a reasonable baryon fraction (3-5\%) would imply a total halo mass of order $2\times 10^{11} M_\odot$
\citep[i.e., expectations from abundance matching][]{BoylanKolchin2011}. In such a massive halo it is likely that much of the stripped material would remain bound to the LMC. Our fiducial model is thus an optimistic estimate
of the contribution of material ram pressure stripped from the LMC to the Magellanic Stream. 
Even in our lower LMC halo model, material that is removed from the LMC does not travel to large distances (the Stream spans more than 100$^\circ$ on the sky). 
As such, the vast majority of the Stream is not likely to be ram pressure stripped material from the LMC and this material would only be found close to the clouds consistent with the higher metallicity material found in observations
 \citep{Fox2013,Richter2013}. 

In all simulations we find that stripped gas is initially removed asymmetrically, building up in the North-West quadrant (upper right) of the LMC disk. This occurs because the clockwise sense of rotation of the LMC disk 
allows material to be removed more easily in the north, where disk gas is already moving in the same direction as the headwind \citep[e.g., ][]{Roediger2006}. 
This leads to a very specific prediction: gas in this region should have metallicities consistent with that of the outer disk of the LMC and higher than that of the tip of the Stream. 
This can be tested with QSO absorption line studies such as that of \cite{Fox2013} and \cite{Lehner2012}. Further observations to confirm the metallicity break in the Stream 
observed with one QSO sightline near the Clouds by \cite{Richter2013} are also warranted.

\section{Summary}
\label{sec:summary}
We have provided a density measurement of the Milky Way's circumgalactic medium at $r = 48$ kpc by exploring the dynamics of ram pressure stripping of the Large Magellanic Cloud via analytic arguments and a large suite of three-dimensional hydrodynamic simulations. RPS probes the diffuse CGM in all its phases, providing a robust measurement independent of line studies in emission and absorption. We adopted a simple model for the LMC's stars, gas and DM profiles and adopted the model of \cite{Gunn1972} (GG model) to explore how a ram pressure headwind truncates the HI profile of disk systems, providing a wind computed from models for the LMC's orbit through the MW potential and sampling from a $\beta$-profile for the CGM's density. Our main conclusions are:
\begin{enumerate}

\item The presence of stars well beyond the HI truncation radius of the Large Magellanic Cloud's leading edge suggests ram pressure stripping played a strong role during the LMC's recent pericentric passage truncating its gaseous disk.

\item From analytic arguments, assuming the LMC's disk was truncated by the peak ram pressure at pericentric passage, we found the GG model provides a direct map from the LMC's observed truncation radius along its leading edge to a CGM density at $r \approx 48$ kpc from the Galactic Center. This toy model implied a halo density $\approx 5 \times 10^{-5}$ cm$^{-3}$.

\item From our suite of simulations sampling a broad range of halo models, we found the LMC's leading edge truncation radius was indeed tightly correlated to the MW CGM density at pericentric passage, as predicted in the GG model, and roughly independent of the following considerations:
\begin{enumerate}
\item The broader halo profile's core density and power-law falloff --- i.e. the density measurement was very localized to pericentric passage, thanks to the steep rise in ram pressure at this point. This implies our measurement is \emph{independent} of our choice to use a $\beta$-profile and applies to any distribution for the diffuse CGM. 
\item The disk orientation with respect to the wind, provided it was neither head-on nor precisely face-on, making our result insensitive to the precise orientation of the LMC disk at pericentric passage.
\item The DM potential and rotation curve peak of the LMC
\item The restoring force of the disk in such a gas-rich late-type dwarf is best described by including not only the stellar disk's gravitational potential but also the self-gravity of the gas.
\end{enumerate}

\item The simulations also elucidated some disagreement from/extensions of the GG model, including:
\begin{enumerate}
\item The presence of extra-planar material in the stripping zone renders the truncation radius seen in column viewer-dependent, as material in the process of liberation from the disk contributes gas at high column depending on where one views the galaxy from. This effect can introduce a substantial asymmetry in the observed extent of the disk in different quadrants of the system.
\item Our most relevant simulations included a significant $5\%$ offset from the GG model, i.e. our truncation radius for a given headwind ram pressure was 5\% larger than the analytic prediction with the gas' self-gravity. A portion of this offset was present \emph{regardless} of viewing perspective (from the solar line-of-site, face-on and edge-on projections of the disk), and attributable to time-varying oscillation's in the disk's extent.
\end{enumerate} 

\item Analysis of this suite of simulations, for our fiducial LMC disk model, yielded an implied CGM density substantially higher than the analytic toy model, with $n(48.2 \pm 2.5 \; {\rm kpc}) = 1.1 \pm .45 \times 10^{-4} \; {\rm cm^{-3}}$. The error bounds on this include a host of modeled uncertainties, including those related to the scatter in the simulations (mostly due to the oscillatory behavior of the disk edge), orbital uncertainties (including that of the LMC/MW mass ratio), the flocculent appearance of the LMC disk's leading edge (the dominant source of error) and uncertainty in the stellar profile.

\item From our simulated results, sampling parameters of a $\beta$-profile and integrating to $300$ kpc, our standard LMC model results yielded a total CGM gas mass of $2.68 \pm 1.4 \times 10^{10} M_\odot$, which could account for roughly 15\% of a 10$^{12} M_\odot$ MW's expected baryon content.

\item It is unlikely that material ram pressure stripped from the LMC can account for more than a few percent of the mass budget of the Magellanic Stream. 
The simulations suggest that material to the North-West of the LMC's disk consists of gas stripped from the LMC (but not necessarily unbound).  This material should have metallicities consistent with the outskirts of the 
LMC's disk, a prediction that is testable through absorption line studies towards background QSOs. 

\end{enumerate}

Movies for our fiducial simulation can be found at https://lavinia.as.arizona.edu/~gbesla/MWHotHalo.html

\section{Acknowledgements}

This research was funded through HST AR grant \#12632.  Support for program \#12632 was provided by NASA through a grant from the Space Telescope Science Institute, which is operated by the Association of Universities for Research in Astronomy, Inc., under NASA contract NAS 5-26555. Additional support was provided by NSF grants AST-0908390 and AST-1008134, and NASA grant NNX12AH41G, as well as computational resources from NSF XSEDE, and Columbia University's Yeti cluster. We benefited from helpful discussions with Susan Clark, Joshua Peek, Adrian Price-Whelan and Yong Zheng.

\bibliography{references}{}
\bibliographystyle{apj}

\appendix 

\section{LMC Disk Plane Velocity Corrections}
\label{sec:velocity-transforms}
To produce the HI velocity map in Section \ref{sec:hi-velocity-map}, we needed to subtract motion of the gas due to the LMC's COM motion and its solid body rotation. To do so, we employ the machinery laid out in vdM02 and \cite{vanderMarel2001}. This appendix briefly describes the necessary equations along with a brief motivation of our parameter choices. 

This prescription takes as inputs the LMC's COM position, $(\alpha_0,\delta_0, D_0)$; disk orientation $(i,\theta)$; COM proper motion in the LOS frame,  $v_x,v_y,v_z$, (see Section \ref{sec:coordinates}); and the change in the disk plane's inclination angle, $di/dt$. Note the change in position angle $d\theta/dt$ does not contribute to the radial velocity. Values for all these quantities were taken from the aforementioned sources, and are summarized, for convenience, in Table \ref{tab:position-params}. With these values in place, a set of angular coordinates, $\rho$ and $\phi$, can be computed at every point in the projected image in terms of celestial coordinates:
\begin{eqnarray}
\cos \rho 			&=&		\cos \delta \cos \delta_0 \cos ( \alpha - \alpha_0 ) + \sin \delta \sin \delta_0		\nonumber \\
\sin \rho \cos \phi 	&=&		-\cos \delta \sin (\alpha - \alpha_0)	\nonumber \\
\sin \rho \sin \phi 	&=&		\sin \delta \cos \delta_0 - \cos \delta \sin \delta_0 \cos ( \alpha - \alpha_0 ) \; .
\end{eqnarray} 
although these are three equations in two unknowns, the system is not overdetermined since trigonometric functions are not one-to-one. From there, we can recast the LMC's cartesian velocity vector into an angular form, still in the LOS frame:
\begin{eqnarray}
\left( \begin{array}{c} \vsys \\ v_t \\ \Theta_t  \end{array}\right) 
= \left( \begin{array}{c} -v_z \\ (v_x^2 + v_y^2)^{1/2} \\ \tan^{-1}(v_y/v_x) - 90^\circ  \end{array} \right) \; .
\label{eq:celestial-velocity}
\end{eqnarray}
Here $\vsys$ is motion \emph{away} form the observer, $v_t$ is speed of the motion in the plane of the image (transverse), and $\Theta_t$ is the position angle of this transverse motion, measured counterclockwise from the $\hat{y}$ north-vector. Finally we're in a position to compute the radial velocity correction, which must be subtracted from the observed velocities:
\begin{equation}
v_{\rm los}(\rho, \Phi) = \vsys \cos \rho + v_t \sin \rho \cos ( \Phi - \Theta_t )
	+ D_0 ( di/dt) \sin \rho \sin ( \Phi - \Theta ) \; ,
\label{equ:correction}
\end{equation}
where $\Theta = \theta - 90$ and $\Phi = \phi - 90$. The first term in this expression corresponds to the COM proper motion along the line of site, and differers from a na\"ive subtraction of the bulk motion by the cosine factor. The second term is more pernicious, introducing a velocity gradient across the image plane of amplitude $\sim 100$ km/s, which applies a systematic twist to the observed velocity field. The final term is related to any rotation of the LMC disk plane about the line of nodes, which all evidence at present has failed to detect at a significant level \citep{vanderMarel2014}.

\end{document}